\documentclass[draft,tightenlines,nofootinbib,preprint,aps,eqsecnum,amsmath,amssymb]{revtex4}
\begin{document}
% \tightenlines

\newcommand{\beq}{\begin{equation}}
\newcommand{\eeq}{\end{equation}}
\newcommand{\bea}{\begin{eqnarray}}
\newcommand{\eea}{\end{eqnarray}}
\newcommand{\cir}{{\buildrel \circ \over =}}

\title{ Multipolar Expansions for Closed and Open Systems of Relativistic
Particles.}

\author{David Alba}

\affiliation
 {Dipartimento di Fisica\\ Universita' di Firenze\\
L.go E.Fermi 2 (Arcetri)\\ 50125 Firenze, Italy\\ E-mail:
ALBA@FI.INFN.IT}

\author{Luca Lusanna}

\affiliation
 {Sezione INFN di Firenze\\ L.go E.Fermi 2 (Arcetri)\\
50125 Firenze, Italy\\ E-mail: LUSANNA@FI.INFN.IT}

\author{Massimo Pauri}

\affiliation
{Dipartimento di Fisica\\ Universita' di Parma\\
 Parco Area Scienze 7/A\\
  43100 Parma, Italy\\
   E-mail: PAURI@PR.INFN.IT}

\begin{abstract}
Dixon's multipoles for a system of N relativistic positive-energy
scalar particles are evaluated in the rest-frame instant form of
dynamics. The Wigner hyper-planes (intrinsic rest frame of the
isolated system) turn out to be the natural framework for
describing multipole kinematics. Classical concepts like the {\it
barycentric tensor of inertia} turn out to be extensible to
special relativity only by means of the quadrupole moments of the
isolated system. Two new applications of the multipole technique
are worked out for systems of interacting particles and fields. In
the rest-frame of the isolated system of either free or
interacting positive energy particles it is possible to define a
unique world-line which embodies the properties of the most
relevant centroids introduced in the literature as candidates for
the collective motion of the system. This is no longer true,
however, in the case of open subsystems of the isolated system.
While effective mass, 3-momentum and angular momentum in the rest
frame can be calculated from the definition of the {\it subsystem
energy-momentum tensor}, the definitions of effective center of
motion and effective intrinsic spin of the subsystem are not
unique. Actually, each of the previously considered centroids
corresponds to a different world-line in the case of open systems.
The pole-dipole description of open subsystems is compared to
their description as effective extended objects. Hopefully, the
technique developed here could be instrumental for the
relativistic treatment of binary star systems in metric gravity.

\vskip 1truecm

\today

\vskip 1truecm

\end{abstract}
\pacs{} \vfill\eject

\maketitle

\vfill\eject

\section{Introduction.}

An important area of research is nowadays the construction of
templates for gravitational waves emitted by binary systems.
Analytically, this can be done within the framework of PN
approximations by means of essentially non-relativistic multipole
expansions for compact bodies. On the other hand, since the main
emission are supposed to take place in a region where the PN
approximation fails, it would be desirable to have at disposal a
relativistic treatment of multipolar expansions as a preliminary
kinematical tool for dealing with open gravitating systems. This
paper focuses just on the construction of a suitable relativistic
kinematical background by an N-body system as a tool. A
preliminary extension of the results of this paper to relativistic
fluids is given in \cite{92}.
\bigskip

Our starting point will be the result recently
obtained\cite{iten1} concerning a complete treatment of the
kinematics of the relativistic N-body problem in the {\it
rest-frame instant form of dynamics}
\cite{lus,albad,crater,india}. This program required the
re-formulation of the theory of isolated relativistic systems on
arbitrary space-like hyper-surfaces \cite{c1}. and has been based
essentially upon Dirac's reformulation\cite{dirac} of classical
field theory (suitably extended to particles) on arbitrary
space-like hyper-surfaces ({\it equal time and Cauchy} surfaces).
For each isolated system (containing any combination of particles,
strings and fields) one obtains in this way a re-formulation of
the standard theory as a parametrized Minkowski theory\cite{lus}.
This program shows the extra bonus of being naturally prepared for
the coupling to gravity in its ADM formulation. The price to be
paid is that the functions $z^{\mu}(\tau ,\vec \sigma )$
describing the embedding of the space-like hyper-surface in
Minkowski space-time become configuration variables in the action
principle. Since the action is invariant under separate
$\tau$-reparametrizations and space-diffeomorphisms, first class
constraints emerge ensuring the independence of the description
from the choice of the 3+1 splitting. The embedding configuration
variables $z^{\mu}(\tau ,\vec \sigma )$ are thus the {\it gauge}
variables associated with this particular kind of general
covariance.
\bigskip

We summarize here the main results that are necessary to
understand all the subsequent technical developments. First of
all, recall that since the intersection of a time-like world-line
with a space-like hyper-surface corresponding to a value $\tau$ of
the time parameter is identified by 3 numbers $\vec \sigma =\vec
\eta (\tau )$ instead of four, in parametrized Minkowski theories
each particle must have {\it a well defined sign of the energy}:
therefore the two topologically disjoint branches of the mass
hyperboloid cannot be described simultaneously as in the standard
manifestly Lorentz-covariant theory, and there are no more
mass-shell constraints. Each particle with a definite sign of the
energy is described by the canonical coordinates ${\vec
\eta}_i(\tau )$, ${\vec \kappa}_i(\tau )$ while the 4-position of
the particles are given by $x^{\mu}_i(\tau )=z^{\mu}(\tau ,{\vec
\eta}_i(\tau ))$. The following 4-momenta $p^{\mu}_i(\tau )$ are
${\vec \kappa}_i$-dependent solutions of $p^2_i -  m^2_i =0$ with
the chosen sign of the energy.
\medskip

In order to exploit the separate spatial and time
reparametrization invariances of parametrized Minkowski theories,
we can first of all restrict the foliation to space-like {\it
hyper-planes} as leaves. For each configuration of the isolated
system with time-like 4-momentum, we further restrict to the
special leaves defined by hyper-planes orthogonal to the conserved
system 4-momentum ({\it Wigner hyper-planes}). This foliation is
fully determined by the configuration of the isolated system. One
gets in this way\cite{lus} the definition of the {\it
Wigner-covariant rest-frame instant form of dynamics} for any
isolated system whose configurations have well defined and finite
Poincar\'e generators with time-like total 4-momentum (see
Ref.\cite{dir} for the traditional forms of dynamics). Finally,
this formulation casts some light on the long standing problem of
defining a relativistic center of mass. As well known, no
definition of this concept can enjoy all the properties of its
non-relativistic counterpart. See
Refs.\cite{pau,com1,com2,com3,com4} for a partial bibliography of
all the existing attempts.
\medskip

As shown in Appendix A of Ref.\cite{iten1} only four first class
constraints survive in the rest-frame instant form on Wigner
hyper-planes. The original configuration variables $z^{\mu}(\tau
,\vec \sigma )$, ${\vec \eta}_i(\tau )$ and their conjugate
momenta $\rho_{\mu}(\tau ,\vec \sigma )$, ${\vec \kappa}_i(\tau )$
are reduced to:\medskip

i) {\it a decoupled particle} ${\tilde x}^{\mu}_s(\tau )$,
$p^{\mu}_s$ (the only remnant of the space-like hyper-surface)
with a positive mass $\epsilon_s=\sqrt{ p^2_s}$ determined by the
first class constraint $\epsilon_s-M_{sys} \approx 0$, $M_{sys}$
being the invariant mass of the isolated system. As a gauge fixing
to the constraint, the rest-frame Lorentz scalar time
$T_s={{{\tilde x}_s\cdot p_s}\over {\epsilon_s}}$ is put equal to
the mathematical time: $T_s-\tau \approx 0$. Here, ${\tilde
x}^{\mu}_s(\tau )$ is a {\it non-covariant canonical} variable for
the {\it 4-center of mass}. After the elimination of $T_s$ and
$\epsilon_s$ by means of such pair of second class constraints, we
are left with a decoupled free point ({\it point particle clock})
of mass $M_{sys}$ and canonical 3-coordinates ${\vec
z}_s=\epsilon_s ({\vec {\tilde x}}_s-{{{\vec p}_s}\over {p^o_s}}
{\tilde x}^o)$, ${\vec k}_s={{{\vec p}_s}\over {\epsilon_s}}$
\cite{c2}. The non-covariant canonical ${\tilde x}^{\mu}_s(\tau )$
must not be confused with the 4-vector $x^{\mu}_s(\tau
)=z^{\mu}(\tau ,\vec \sigma =0)$ identifying the origin of the
3-coordinates $\vec \sigma$ inside the Wigner hyper-planes. The
world-line $x^{\mu}_s(\tau )$ is arbitrary because it depends on
$x^{\mu}_s(0)$ while its 4-velocity ${\dot x}^{\mu}_s(\tau )$
depends on the Dirac multipliers associated with the remaining 4
(or 3 after having imposed $T_s - \tau \approx 0$) first class
constraints (see Section II). Correspondingly, this world-line may
be considered as an arbitrary {\it centroid} for the isolated
system. The unit time-like 4-vector $u^{\mu}(p_s) =
p_s^{\mu}/\epsilon_s\,\,$ ($= {\dot {\tilde x}}^{\mu}_s(\tau )$
after having imposed $T_s - \tau \approx 0$) is orthogonal to the
Wigner hyper-planes and describes their orientation in the chosen
inertial frame.

ii) {\it the particle canonical variables} ${\vec \eta}_i(\tau )$,
${\vec \kappa}_i(\tau )$  inside the Wigner hyper-planes. They are
Wigner spin-1 3-vectors, like the coordinates $\vec \sigma$. They
are restricted by the three first class constraints (the {\it
rest-frame conditions}) ${\vec \kappa}_{+}=\sum_{i=1}^N {\vec
\kappa}_i \approx 0$. Since the role of the relativistic decoupled
{\it 4-center of mass} is taken by ${\tilde x}^{\mu}_s(\tau )$
(or, after the gauge fixing $T_s-\tau \approx 0$, by an {\it
external 3-center of mass} ${\vec z}_s$, defined in terms of
${\tilde x}^{\mu}_s$ and $p^{\mu}_s$ \cite{iten1}), the rest-frame
conditions imply that the {\it internal} canonical 3-center of
mass ${\vec q}_{+}={\vec \sigma}_{com}$ is a {\it gauge variable}
that can be eliminated through a gauge fixing \cite{c3}. This
amounts in turn to a definite choice of the world-line
$x^{\mu}_s(\tau )$ of the centroid.
\bigskip

All this leads to a {\it doubling of viewpoints and concepts}:
\medskip

1) The {\it external} viewpoint, taken by an arbitrary inertial
Lorentz observer, who describes the Wigner hyper-planes determined
by the time-like configurations of the isolated system. A change
of inertial observer by means of a Lorentz transformation rotates
the Wigner hyper-planes and induces a Wigner rotation of the
3-vectors inside each Wigner hyperplane. Every such hyperplane
inherits an induced {\it internal Euclidean structure} while an
{\it external} realization of the Poincar\'e group induces an {\it
internal} Euclidean action.
\medskip

Then, three {\it external} concepts of 4-center of mass can be
defined by using the {\it external} realization of the Poincar\'e
algebra (each one corresponding to a different 3-location inside
the Wigner hyper-planes):\medskip

a) the {\it external} non-covariant canonical {\it 4-center of
mass} (also named {\it 4-center of spin}) ${\tilde x}^{\mu}_s$
(with 3-location ${\vec {\tilde \sigma}}$),

b) the {\it external} non-covariant non-canonical M\o ller {\it
4-center of energy} $R^{\mu}_s$ (with 3-location ${\vec
\sigma}_R$),

c) the {\it external} covariant non-canonical Fokker-Pryce {\it
4-center of inertia} $Y^{\mu}_s$ (with 3-location ${\vec
\sigma}_Y$).
\bigskip

Only the canonical non-covariant center of mass ${\tilde
x}^{\mu}_s(\tau )$ is relevant to the Hamiltonian treatment with
Dirac constraints, while only the Fokker-Pryce $Y^{\mu}_s$ is a
4-vector by construction. See Ref.\cite{iten1} for the
construction of the {\it 4-centers} starting from the
corresponding {\it 3-centers} (3-center of spin\cite{com3},
3-center of energy \cite{mol}, 3-center of
inertia\cite{com2,com3}), which are group-theoretically defined in
terms of generators of the external Poincare' group.
\medskip

2) The {\it internal} viewpoint, taken by an observer inside the
Wigner hyper-planes. This viewpoint is associated to a unfaithful
{\it internal} realization of the Poincar\'e algebra: the total
{\it internal} 3-momentum of the isolated system vanishes due to
the rest-frame conditions. The {\it internal} energy and angular
momentum are the invariant mass $M_{sys}$ and the spin (the
angular momentum with respect to ${\tilde x}^{\mu}_s(\tau )$) of
the isolated system, respectively. By means of the {\it internal}
realization of the Poincar\'e algebra we can define three {\it
internal} 3-centers of mass: the {\it internal} canonical 3-center
of mass (or 3-center of spin) ${\vec q}_{+}$, the {\it internal}
M\o ller 3-center of energy ${\vec R}_{+}$ and the {\it internal}
Fokker-Pryce 3-center of inertia ${\vec y}_{+}$. However, due to
the rest-frame condition ${\vec \kappa}_{+} \approx 0$, {\it they
all coincide}: ${\vec q}_{+} \approx {\vec R}_{+} \approx {\vec
y}_{+}$. As a natural gauge fixing to the rest-frame conditions,
we can add the vanishing of the {\it internal} Lorentz boosts
$\vec K$ (recall that they are equal to $- {\vec R}_{+} /
M_{sys}$). This is equivalent to locate the internal canonical
3-center of mass ${\vec q}_{+}$ in $\vec \sigma =0$, i.e. in the
origin $x^{\mu}_s(\tau )=z^{\mu}(\tau ,\vec 0)$. Upon such gauge
fixings and with $T_s-\tau \approx 0$, the world-line
$x^{\mu}_s(\tau )$ becomes uniquely determined except for the
arbitrariness in the choice of $x^{\mu}_s(0)$ [$\,
u^{\mu}(p_s)=p^{\mu}_s/\epsilon_s$]

\beq
 x^{\mu}_s(\tau )=x^{\mu}_s(0) + u^{\mu}(p_s) T_s,
  \label{I1}
 \eeq

\noindent   and, modulo $x^{\mu}_s(0)$, it coincides with the {\it
external} covariant non-canonical Fokker-Pryce 4-center of
inertia, $x^{\mu}_s(\tau ) = x^{\mu}_s(0) +
Y^{\mu}_s$\cite{iten1}.
\bigskip

The doubling of concepts, the {\it external} non-covariant
canonical 4-center of mass ${\tilde x}^{\mu}_s(\tau )$ (or the
{\it external} 3-center of mass ${\vec z}_s$ when $T_s-\tau
\approx 0$) and the {\it internal} canonical 3-center of mass
${\vec q}_{+}\approx 0$ {\it replaces} the separation of the
non-relativistic 3-center of mass due to the Abelian translation
symmetry. Correspondingly, the non-relativistic conserved
3-momentum is replaced by the {\it external} ${\vec
p}_s=\epsilon_s {\vec k}_s$, while, as already said, the {\it
internal} 3-momentum vanishes, ${\vec \kappa}_{+}\approx 0$, as a
definition of rest frame.\medskip

In the gauge where $\epsilon_s \equiv M_{sys}$, $T_s \equiv \tau
$, the canonical basis ${\vec z}_s$, ${\vec k}_s$, ${\vec
\eta}_i$, ${\vec \kappa}_i$ is restricted by the three pairs of
second class constraints ${\vec \kappa}_{+}=\sum_{i=1}^N{\vec
\kappa}_i \approx 0$, ${\vec q}_{+} \approx 0$, so that 6N
canonical variables describe the N particles like in the
non-relativistic case. We still need a canonical transformation
${\vec \eta}_i$, ${\vec \kappa}_i$ $\,\, \mapsto \,\,$ ${\vec
q}_{+} [\approx 0]$, ${\vec \kappa}_{+} [\approx 0]$, ${\vec
\rho}_a$, ${\vec \pi}_a$ [$a=1,..,N-1$] identifying a set of
relative canonical variables. The final 6N-dimensional canonical
basis is ${\vec z}_s$, ${\vec k}_s$, ${\vec \rho}_a$, ${\vec
\pi}_a$. To get this result  we need a highly non-linear canonical
transformation\cite{iten1}, which can be obtained by exploiting
the Gartenhaus-Schwartz singular transformation \cite{garten}.
\medskip

At the end we obtain {\it the Hamiltonian for the relative motions
as a sum of N square roots}, each one containing a squared mass
and a quadratic form in the relative momenta, which goes into the
non-relativistic Hamiltonian for relative motions in the limit
$c\, \rightarrow \infty$. This fact has the following
implications:\medskip

a) if one tries to make the inverse Legendre transformation to
find the associated Lagrangian, it turns out that, due to  the
presence of square roots, the Lagrangian is a hyperelliptic
function of ${\dot {\vec \rho}}_a$ already in the free case. A
closed form exists only for N=2, $m_1=m_2=m$: $L = -  m
\sqrt{4-{\dot {\vec \rho}}{}^2}$. This exceptional case already
shows that the existence of the limiting velocity $c$ forbids a
linear relation between the spin (center-of-mass angular momentum)
and the angular velocity.

b) the N quadratic forms in the relative momenta appearing in the
relative Hamiltonian {\it cannot be simultaneously diagonalized}.
In any case the Hamiltonian is a sum of square roots, so that
concepts like {\it reduced masses}, {\it Jacobi normal relative
coordinates} and {\it tensor of inertia} cannot be extended to
special relativity. As a consequence, for example, a relativistic
static orientation-shape SO(3) principal bundle approach
\cite{little,c4} can be implemented only by using non-Jacobi
relative coordinates.

c) the best way of studying rotational kinematics, {\it viz} the
non-Abelian rotational symmetry associated with the conserved {\it
internal} spin, is based on the {\it canonical spin bases} with
the associated concepts of {\it spin frames} and {\it dynamical
body frames} introduced in Ref.\cite{iten1}. It is remarkable that
they can be build as in the non-relativistic case \cite{iten2}
starting from the canonical basis ${\vec \rho}_a$, ${\vec \pi}_a$.
\medskip

Let us clarify this important point. In the non-relativistic
N-body problem it is easy to make the separation of the absolute
translational motion of the center of mass from the relative
motions, due to the Abelian nature of the translation symmetry
group. This implies that the associated Noether constants of
motion (the conserved total 3-momentum) are in involution, so that
the center-of-mass degrees of freedom decouple. Moreover, the fact
that the non-relativistic kinetic energy of the relative motions
is a quadratic form in the relative velocities allows the
introduction of special sets of relative coordinates, the {\it
Jacobi normal relative coordinates} that diagonalize the quadratic
form and correspond to different patterns of clustering of the
centers of mass of the particles. Each set of Jacobi normal
relative coordinates organizes the N particles into a {\it
hierarchy of clusters}, in which each cluster of two or more
particles has a mass given by an eigenvalue ({\it reduced masses})
of the quadratic form; Jacobi normal coordinates join the centers
of mass of pairs of clusters.\medskip

However, the non-Abelian nature of the rotation symmetry group
whose associated Noether constants of motion (the conserved total
angular momentum) are not in involution, prevents the possibility
of a global separation of absolute rotations from the relative
motions, so that there is no global definition of absolute {\it
vibrations}. Consequently, an {\it isolated} deformable body can
undergo rotations by changing its own shape (see the examples of
the {\it falling cat} and of the {\it diver}). It was just to deal
with these problems that the theory of the orientation-shape SO(3)
principal bundle approach\cite{little} has been developed. Its
essential content is that any {\it static} (i.e.
velocity-independent) definition of {\it body frame} for a
deformable body must be interpreted as a gauge fixing in the
context of a SO(3) {\it gauge} theory. Both the laboratory and the
body frame angular velocities, as well as the orientational
variables of the static body frame, become thereby {\it
unobservable gauge} variables. This approach is associated with a
set of {\it point} canonical transformations, which allow to
define the body frame components of relative motions in a
velocity-independent way. \bigskip

Since in many physical applications (e.g. nuclear physics,
rotating stars,...) angular velocities are viewed as {\it
measurable} quantities, one would like to have an alternative
formulation complying with this requirement and possibly
generalizable to special relativity. This program has been first
accomplished in a previous paper \cite{iten2} and then
relativistically extended (Ref.\cite{iten1}). Let us summarize,
therefore, the main points of our formulation.

First of all, for $N \geq 3$, we have constructed (see
Ref.\cite{iten2}) a class of {\it non-point} canonical
transformations which allow to build the already quoted {\it
canonical spin bases}: they are connected to the patterns of the
possible {\it clusterings of the spins} associated with relative
motions. The definition of these {\it spin bases} is independent
of Jacobi normal relative coordinates, just as the patterns of
spin clustering are independent of the patterns of center-of-mass
Jacobi clustering. We have  found two basic frames associated to
each spin basis: the {\it spin frame} and the {\it dynamical body
frame}. Their construction is guaranteed by the fact that in the
relative phase space, besides the natural existence of a
Hamiltonian symmetry {\it left} action of SO(3) \cite{c5,c6}, it
is possible to define as many Hamiltonian non-symmetry  {\it
right} actions of SO(3) \cite{c7} as the possible patterns of spin
clustering. While for N=3 the unique canonical spin basis
coincides with a special class of global cross sections of the
trivial orientation-shape SO(3) principal bundle, for $N \geq 4$
the existing {\it spin bases} and {\it dynamical body frames} turn
out to be unrelated to the local cross sections of the {\it
static} non-trivial orientation-shape SO(3) principal bundle, and
{\it evolve} in a dynamical way dictated by the equations of
motion. In this new formulation {\it both} the orientation
variables and the angular velocities become, by construction, {\it
measurable} quantities in each canonical spin basis.\medskip

For each N, every allowed spin basis provides a physically
well-defined separation between {\it rotational} and {\it
vibrational} degrees of freedom. The non-Abelian nature of the
rotational symmetry implies that there is no unique separation of
{\it absolute rotations} and {\it relative motions}. The unique
{\it body frame} of rigid bodies is replaced here by a discrete
number of {\it evolving dynamical body frames} and of {\it spin
canonical bases}, both of which are grounded in patterns of spin
couplings, direct analog of the coupling  of quantum angular
momenta.
\bigskip

As anticipated at the outset, we want to complete our study of
relativistic kinematics for the N-body system by first evaluating
the rest-frame Dixon multipoles \cite{dixon,c8} and then by
analyzing the role of Dixon's multipoles for open subsystems. The
basic technical tool will be the standard definition of the {\it
energy momentum tensor} of the N positive-energy {\it free}
particles on the Wigner hyperplane. It will be seen, however, that
in order to get a sensible extension of this definition to open
subsystems, a physically significant convention is required. On
the whole, it turns out that the Wigner hyperplane is the natural
framework for reorganizing a lot of kinematics connected with
multipoles. Only in this way, moreover, a concept like the {\it
barycentric tensor of inertia} can be introduced in special
relativity, specifically by means of the quadrupole
moments.\medskip

A first application of the formalism is done for an isolated
system of $N$ positive-energy particles with {\it mutual
action-at-a-distance interactions}. Then the formalism is applied
to the case of an {\it open $n < N$ particle subsystem} of an
isolated system consisting of $N$ charged positive-energy
particles (with Grassmann-valued electric charges to regularize
the Coulomb self-energies) plus the electro-magnetic field
\cite{crater}. In the rest frame of the isolated system a {\it
suitable} definition of the energy-momentum tensor of the open
subsystem allows to define its effective mass, 3-momentum and
angular momentum.

Then we evaluate the rest-frame Dixon multipoles of the
energy-momentum tensor of the open subsystem with respect to
various centroids describing possible {\it collective centers of
motion}. Unlike the case of isolated systems, each centroid
generates a different world-line and there are many candidates for
the effective center of motion and the effective intrinsic spin.
Two centroids (namely the center of energy and Tulczyjew centroid)
appear to be preferable due to their properties. The case $n = 2$
is studied explicitly. It is also shown that the pole-dipole
description of the 2-particle cluster can be replaced by a
description of the cluster as an extended system (its effective
spin frame can be evaluated). This can be done, however, at the
price of introducing an explicit dependence on the action of the
external electro-magnetic field upon the cluster. By comparing the
effective parameters of an open cluster of $n_1 + n_2$ particles
to the effective parameters of the two clusters with $n_1$ and
$n_2$ particles, it turns out in particular that only the
effective {\it center of energy} appears to be a viable center of
motion for studying the interactions of open subsystems.

\bigskip

A review of the rest-frame instant form of dynamics for N scalar
free positive-energy particles and some new original results on
the canonical transformation to the internal center of mass and
relative variables are given in Section II.

In Section III we evaluate the energy momentum tensor in the
Wigner hyper-planes.

Dixon's multipoles are introduced in Section IV. A special study
of monopole, dipole and quadrupole moments is given and the
multipolar expansion is defined.

After the extension of the previous results to isolated systems
with mutual action-at-a-distance interactions, we study in Section
V the behavior of open subsystems of isolated systems, the
centroids which are good candidates for the description the
collective center of motion, and discuss the determination of the
effective parameters (mass, spin, momentum, variables relative to
the center of motion) for the open subsystem.

Some comments about standing problems are given in the
Conclusions.

The non-relativistic N-particle multipolar expansion is given in
Appendix A, while in Appendix B contains is a review of symmetric
trace-free (STF) tensors. The Gartenhaus-Schwartz transformation
is summarized in Appendix C. Finally, other properties of Dixon's
multipoles are reported in Appendix D.

\vfill\eject

\section{Review of the Rest-Frame Instant Form.}

We briefly review the treatment of N free scalar positive-energy
particles in the framework of parametrized Minkowski theory (see
Appendices A and B and Section II  of Ref.\cite{iten1}). Each
particle is described by a configuration 3-vector ${\vec
\eta}_i(\tau )$. The particle world-line is $x^{\mu }_i(\tau
)=z^{\mu }(\tau ,{\vec \eta}_i(\tau ))$, where $z^{\mu}(\tau ,\vec
\sigma )$ are the embedding configuration variables describing the
space-like hyper-surface $\Sigma_{\tau}$.
\bigskip

The foliation is defined by an embedding $R\times \Sigma
\rightarrow M^4$, $(\tau ,\vec \sigma ) \mapsto z^{\mu}(\tau ,\vec
\sigma )$, with $\Sigma$ an abstract 3-surface diffeomorphic to
$R^3$. $\Sigma_{\tau}$ is the Cauchy surface of {\it equal time}.
The metric induced on $\Sigma_{\tau}$ is $g_{ AB}[z]
=z^{\mu}_{A}\eta_{\mu\nu}z^{\nu}_{B}$, a functional of $z^{\mu}$,
and the embedding coordinates $z^{\mu}(\tau ,\vec \sigma )$ are
considered as independent fields. We use the notation
$\sigma^{A}=(\tau ,\sigma^{\check r})$ of Refs.\cite{lus,crater}.
The $z^{\mu}_{A}(\sigma )= \partial z^{\mu}(\sigma )/\partial
\sigma^{A}$ are flat cotetrad fields on Minkowski space-time with
the $z^{\mu}_{\check r}$'s tangent to $\Sigma_{\tau}$. The dual
tetrad fields are $z^A_{\mu}(\sigma ) = {{\partial
\sigma^A(z)}\over {\partial z^{\mu}}}$, $z^A_{\mu}(\tau ,\vec
\sigma )\, z^{\mu}_B(\tau ,\vec \sigma ) = \delta^A_B$. While in
Ref.\cite{iten1} we used the metric convention $\eta_{\mu\nu} =
\epsilon (+---)$ with $\epsilon =\pm$, in this paper we shall use
$\epsilon =1$ like in Ref.\cite{lus}.
\medskip

If we put $\sqrt{g(\tau ,\vec \sigma )} = \sqrt{ - det\,
g_{AB}(\tau ,\vec \sigma )}$ and $\sqrt{\gamma (\tau ,\vec \sigma
)} = \sqrt{- det\, g_{\check r\check s}(\tau ,\vec \sigma )}$, we
have $z^{\mu}_{\tau}(\tau ,\vec \sigma ) = \Big(\sqrt{{{g}\over
{\gamma}} }\, l^{\mu} + g_{\tau \check r}\, \gamma^{\check r\check
s}\, z^{\mu}_{\check s} \Big)(\tau ,\vec \sigma )$
($\gamma^{\check r\check u}\, g_{\check u\check s} =
\delta^{\check r}_{\check s}$), $l^{\mu}(\tau ,\vec \sigma ) =
\Big({1\over {\sqrt{\gamma}}}\,
\epsilon^{\mu}{}_{\alpha\beta\gamma}\, z^{\alpha}_{\check 1}\,
z^{\beta}_{\check 2}\, z^{\gamma}_{\check 3} \Big)(\tau ,\vec
\sigma )$ (normal to $\Sigma_{\tau}$; $l^2(\tau ,\vec \sigma ) =
1$; $l_{\mu}(\tau ,\vec \sigma )\, z^{\mu}_{\check r}(\tau ,\vec
\sigma ) = 0$) and $d^4z = z^{\mu}_{\tau}(\tau ,\vec \sigma )\,
d\tau\, d^3\Sigma_{\mu} = d\tau\, [z^{\mu}_{\tau}(\tau ,\vec
\sigma )\, l_{\mu}(\tau ,\vec \sigma )]\, \sqrt{\gamma (\tau ,\vec
\sigma )}\, d^3\sigma = \sqrt{g(\tau ,\vec \sigma )}\, d\tau\,
d^3\sigma$.

\bigskip

The system is described by the action\cite{lus,albad,crater}

\begin{eqnarray}
S&=& \int d\tau d^3\sigma \, {\cal L}(\tau ,\vec \sigma )=\int
d\tau L(\tau ),\nonumber \\
 &&{\cal L}(\tau ,\vec \sigma )=-\sum_{i=1}^N\delta^3(\vec \sigma -{\vec \eta}_i
(\tau ))m_i\sqrt{ g_{\tau\tau}(\tau ,\vec \sigma )+2g_{\tau
{\check r}} (\tau ,\vec \sigma ){\dot \eta}^{\check r}_i(\tau
)+g_{{\check r}{\check s}} (\tau ,\vec \sigma ){\dot
\eta}_i^{\check r}(\tau ){\dot \eta}_i^{\check s} (\tau )
},\nonumber \\
 &&L(\tau
)=-\sum_{i=1}^Nm_i\sqrt{ g_{\tau\tau}(\tau ,{\vec \eta}_i (\tau
))+2g_{\tau {\check r}}(\tau ,{\vec \eta}_i(\tau )){\dot
\eta}^{\check r} _i(\tau )+g_{{\check r}{\check s}}(\tau ,{\vec
\eta}_i(\tau )){\dot \eta}_i ^{\check r}(\tau ){\dot
\eta}_i^{\check s}(\tau )  }. \label{II1}
\end{eqnarray}

\noindent The action is invariant under separate $\tau$- and $\vec
\sigma$-reparametrizations.
\medskip

The canonical momenta are

\begin{eqnarray}
\rho_{\mu}(\tau ,\vec \sigma )&=&-{ {\partial {\cal L}(\tau ,\vec
\sigma )} \over {\partial z^{\mu}_{\tau}(\tau ,\vec \sigma )}
}=\sum_{i=1}^N\delta^3 (\vec \sigma -{\vec \eta}_i(\tau
))m_i\nonumber \\
 &&{{z_{\tau\mu}(\tau ,\vec \sigma )+z_{{\check r}\mu}(\tau ,\vec \sigma )
{\dot \eta}_i^{\check r}(\tau )}\over {\sqrt{g_{\tau\tau}(\tau
,\vec \sigma )+ 2g_{\tau {\check r}}(\tau ,\vec \sigma ){\dot
\eta}_i^{\check r}(\tau )+ g_{{\check r}{\check s}}(\tau ,\vec
\sigma ){\dot \eta}_i^{\check r}(\tau ){\dot \eta}_i^{\check
s}(\tau ) }} }=\nonumber \\
&=&[(\rho_{\nu}l^{\nu})l_{\mu}+(\rho_{\nu}z^{\nu}_{\check
r})\gamma^{{\check r} {\check s}}z_{{\check s}\mu}](\tau ,\vec
\sigma ),\nonumber \\
 &&{}\nonumber \\
\kappa_{i{\check r}}(\tau )&=&-{ {\partial L(\tau )}\over
{\partial {\dot \eta}_i^{\check r}(\tau )} }=\nonumber \\ &=&m_i{
{g_{\tau {\check r}}(\tau ,{\vec \eta}_i(\tau ))+g_{{\check r}
{\check s}}(\tau ,{\vec \eta}_i(\tau )){\dot \eta}_i^{\check
s}(\tau )}\over { \sqrt{g_{\tau\tau}(\tau ,{\vec \eta}_i(\tau ))+
2g_{\tau {\check r}}(\tau ,{\vec \eta}_i(\tau )){\dot
\eta}_i^{\check r}(\tau )+ g_{{\check r}{\check s}}(\tau ,{\vec
\eta}_i(\tau )){\dot \eta}_i^{\check r} (\tau ){\dot
\eta}_i^{\check s}(\tau ) }} },\nonumber \\
 &&{}\nonumber \\
&&\lbrace z^{\mu}(\tau ,\vec \sigma ),\rho_{\nu}(\tau ,{\vec
\sigma}^{'}\rbrace =-\eta^{\mu}_{\nu}\delta^3(\vec \sigma -{\vec
\sigma}^{'}),\nonumber \\ &&\lbrace \eta^{\check r}_i(\tau
),\kappa_{j{\check s}}(\tau )\rbrace = -\delta_{ij}\delta^{\check
r}_{\check s}.
 \label{II2}
\end{eqnarray}
\medskip

The canonical Hamiltonian $H_{c}$ is zero, but there are the
primary first class constraints

\begin{eqnarray}
{\cal H}_{\mu}(\tau ,\vec \sigma )&=& \rho_{\mu}(\tau ,\vec \sigma
)-l_{\mu} (\tau ,\vec \sigma )\sum_{i=1}^N\delta^3(\vec \sigma
-{\vec \eta}_i(\tau )) \sqrt{ m^2_i-\gamma^{{\check r}{\check
s}}(\tau ,\vec \sigma ) \kappa_{i{\check r}}(\tau
)\kappa_{i{\check s}}(\tau ) }-\nonumber \\ &-&z_{{\check r}\mu}
(\tau ,\vec \sigma )\gamma^{{\check r}{\check s}}(\tau ,\vec
\sigma ) \sum_{i=1}^N\delta^3(\vec \sigma -{\vec \eta}_i(\tau
))\kappa_{i{\check s}} \approx 0,
 \label{II3}
\end{eqnarray}

\noindent so that the Dirac Hamiltonian is $H_D=\int d^3\sigma
\lambda^{\mu}(\tau ,\vec \sigma ) {\cal H}_{\mu}(\tau ,\vec \sigma
)$, where $\lambda^{\mu}(\tau ,\vec \sigma )$ are Dirac
multipliers.
\medskip

The conserved Poincar\'e generators are (the suffix ``s" denotes
the hypersurface $\Sigma_{\tau}$)

\begin{eqnarray}
&&p^{\mu}_s=\int d^3\sigma \rho^{\mu}(\tau ,\vec \sigma
),\nonumber \\ &&J_s^{\mu\nu}=\int d^3\sigma [z^{\mu}(\tau ,\vec
\sigma )\rho^{\nu}(\tau , \vec \sigma )-z^{\nu}(\tau ,\vec \sigma
)\rho^{\mu}(\tau ,\vec \sigma )].
 \label{II4}
\end{eqnarray}

\bigskip

After the restriction to space-like {\it hyper-planes},
$z^{\mu}(\tau ,\vec \sigma ) = x^{\mu}_s(\tau ) + b^{\mu}_{\check
s}(\tau )\, \sigma^{\check s}$, the Dirac Hamiltonian is reduced
to $H_D=\lambda_{\mu}(\tau ){\tilde {\cal H}}^{\mu}(\tau
)+\lambda_{\mu\nu}(\tau ){\tilde {\cal H}}^{\mu\nu}(\tau )$  (only
ten Dirac multipliers survive) while the remaining ten constraints
are given by

\begin{eqnarray}
{\tilde {\cal H}}^{\mu}(\tau )&=&\int d^3\sigma {\cal
H}^{\mu}(\tau ,\vec \sigma )=\,
p^{\mu}_s-l^{\mu}\sum_{i=1}^N\sqrt{m^2_i+{\vec \kappa}^2_i (\tau
)}+b^{\mu}_{\check r}(\tau )\sum_{i=1}^N\kappa_{i{\check r}}(\tau
) \approx 0,\nonumber \\ {\tilde {\cal H}}^{\mu\nu}(\tau
)&=&b^{\mu}_{\check r}(\tau )\int d^3\sigma \sigma^{\check r}\,
{\cal H}^{\nu}(\tau ,\vec \sigma )-b^{\nu}_{\check r}(\tau ) \int
d^3\sigma \sigma^{\check r}\, {\cal H}^{\mu}(\tau ,\vec \sigma )=
\nonumber \\ &=&S_s^{\mu\nu}(\tau )-[b^{\mu}_{\check r}(\tau
)b^{\nu}_{\tau}-b^{\nu}_{\check r}(\tau
)b^{\mu}_{\tau}]\sum_{i=1}^N\eta_i^{\check r}(\tau )
\sqrt{m^2_i+{\vec \kappa}^2_i(\tau )}-\nonumber \\
&-&[b^{\mu}_{\check r}(\tau )b^{\nu}_{\check s}(\tau
)-b^{\nu}_{\check r}(\tau ) b^{\mu}_{\check s}(\tau
)]\sum_{i=1}^N\eta_i^{\check r}(\tau )\kappa_i^{\check s}(\tau
)\approx 0.
 \label{II5}
\end{eqnarray}

Here $S^{\mu\nu}_s$ is the spin part of the Lorentz generators

\begin{eqnarray}
J^{\mu\nu}_s&=&x^{\mu}_sp^{\nu}_s-x^{\nu}_sp^{\mu}_s+S^{\mu\nu}_s,\nonumber
\\ &&S^{\mu\nu}_s=b^{\mu}_{\check r}(\tau )\int d^3\sigma
\sigma^{\check r} \rho^{\nu}(\tau ,\vec \sigma )-b^{\nu}_{\check
r}(\tau )\int d^3\sigma \sigma^{\check r}\rho^{\mu}(\tau ,\vec
\sigma ).
 \label{II6}
\end{eqnarray}

\medskip

The condition ${\dot l}^{\mu} = 0$ that the unit normal $l^{\mu} =
\epsilon^{\mu}{}_{\alpha\beta\gamma}\, b^{\alpha}_{\check 1}(\tau
)\, b^{\beta}_{\check 2}(\tau )\, b^{\gamma}_{\check 3}(\tau )$,
$l^2 = 1$, to the hyper-planes be $\tau$-indepedent, implies three
gauge fixing constraints $l^{\mu} = const.$ for the first class
constraints ${\tilde {\cal H}}^{\mu\nu}(\tau ) \approx 0$. These
imply in turn that the $b^{\mu}_{\check r}(\tau )$'s depend only
on three Euler angles describing a rotating spatial frame.
Therefore foliations with parallel hyper-planes have only 7
independent first class constraints.
\bigskip

When $p^2_s > 0$, the restriction to Wigner hyper-planes is done
by imposing $b^{\mu}_{\check A} = ( b^{\mu}_{\tau} = l^{\mu};
b^{\mu}_{\check r}) \approx b^{\mu}_A = L^{\mu}{}_{\nu = A}(p_s,
{{\buildrel \circ \over =}}_s)$, where $ L^{\mu}{}_{\nu}(p_s,
{{\buildrel \circ \over =}}_s)$ is the standard Wigner boost for
time-like Poincare' orbits. The indices $\check r = r$ are Wigner
spin 1 indices and we have $b^{\mu}_{\tau} =  L^{\mu}{}_o(p_s,
{{\buildrel \circ \over =}}_s) = u^{\mu}(p_s) = p^{\mu}_s /
\sqrt{p^2_s}$, $b^{\mu}_r = \epsilon^{\mu}_r(u(p_s)) =
L^{\mu}{}_r(p_s, {{\buildrel \circ \over =}}_s)$.

\bigskip

On the {\it Wigner hyperplane} \cite{c9}, $z^{\mu}(\tau ,\vec
\sigma ) = x^{\mu}_s(\tau ) + \epsilon^{\mu}_r(u(p_s))\,
\sigma^r$, we have the following constraints and Dirac
Hamiltonian\cite{lus,crater}

\begin{eqnarray}
{\tilde {\cal H}}^{\mu}(\tau )&=&p^{\mu}_s-u^{\mu}(p_s)
\sum_{i=1}^N \sqrt{m_i ^2+{\vec \kappa}_i^2}
+\epsilon^{\mu}_r(u(p_s)) \sum_{i=1}^N\kappa_{ir}= \nonumber \\
 &=&u^{\mu}(p_s) \Big[\epsilon_s-\sum_{i=1}^N\sqrt{m_i^2+ {\vec
\kappa}_i^2}\Big] +\epsilon^{\mu}_r(u(p_s)) \sum_{i=1}^N
\kappa_{ir} \approx 0,\nonumber \\
 &&{}\nonumber \\
  &&or\nonumber \\
 &&{}\nonumber \\
 &&\epsilon_s-M_{sys}\approx 0,\quad\quad
M_{sys}=\sum_{i=1}^N\sqrt{m_i^2+{\vec \kappa}_i^2},\nonumber \\
 &&{\vec p}_{sys} = {\vec
\kappa}_{+}=\sum_{i=1}^N {\vec \kappa}_i \approx 0,\nonumber \\
 &&{}\nonumber \\
 H_D&=&
\lambda^{\mu}(\tau ) {\tilde {\cal H}}_{\mu}(\tau )= \lambda (\tau
) [\epsilon_s-M_{sys}]-\vec \lambda (\tau ) \sum_{i=1}^N {\vec
\kappa}_i, \nonumber \\
 &&{}\nonumber \\
  &&\lambda (\tau ) \approx
-{\dot x}_{s \mu}(\tau )u^{\mu}(p_s),\qquad \lambda_r(\tau
)\approx -{\dot x}_{s \mu}(\tau )\epsilon^{\mu}_r(u(p_s)),
\nonumber \\
 &&{}\nonumber \\
 {\dot {\tilde x}}^{\mu}_s(\tau ) &=&-\lambda (\tau )
u^{\mu}(p_s),\nonumber \\
 &&{}\nonumber \\
 {\dot x}_s^{\mu}(\tau )  &{\buildrel \circ \over =}\,& \{
x^{\mu}_s(\tau ), H_D \} ={\lambda}_{\nu}(\tau ) \{ x^{\mu}_s(\tau
), {\cal H}^{\mu}(\tau ) \} \approx\nonumber \\
 &\approx& -{\lambda}^{\mu}(\tau )=-\lambda (\tau )
 u^{\mu}(p_s)+\epsilon^{\mu}_r(u(p_s)) \lambda_r(\tau ),\nonumber \\
 &&{}\nonumber \\
 {\dot x}^2_s(\tau )&=& \lambda^2(\tau )-{\vec \lambda}^2(\tau ) >
0,\quad\quad {\dot x}_s\cdot u(p_s)=-\lambda (\tau ),\nonumber \\
 &&{}\nonumber \\
 U^{\mu}_s(\tau )&=& {{ {\dot x}^{\mu}_s(\tau )}\over { \sqrt{{\dot
x}^2 _s(\tau )} }}={{-\lambda (\tau )u^{\mu}(p_s)+\lambda_r(\tau
)\epsilon^{\mu}_r (u(p_s))}\over {\sqrt{\lambda^2(\tau )-{\vec
\lambda}^2(\tau )} }},\nonumber \\
 &&{}\nonumber \\
  \Rightarrow&& x^{\mu}_s(\tau )=x^{\mu}_s(0)-u^{\mu}(p_s)\int_0^{\tau}d\tau_1
\lambda (\tau_1) + \epsilon^{\mu}_r(u(p_s))\,
\int_0^{\tau}d\tau_1\lambda_r(\tau_1).
 \label{II7}
 \end{eqnarray}
\medskip

While the Dirac multiplier $\lambda (\tau )$  is determined by the
gauge fixing $T_s-\tau \approx 0$, the 3 Dirac's multipliers $\vec
\lambda (\tau )$ describe the classical zitterbewegung of the
centroid $x^{\mu}_s(\tau )=z^{\mu}(\tau ,\vec 0)$ which is the
origin of the 3-coordinates on the Wigner hyperplane. Each
gauge-fixing $\vec \chi (\tau )\approx 0$ to the 3 first class
constraints ${\vec \kappa}_{+}\approx 0$ (defining the {\it
internal rest-frame} 3-center of mass ${\vec \sigma}_{cm}$) gives
a different determination of the multipliers $\vec \lambda (\tau
)$ \cite{c10}. Therefore each gauge-fixing identifies a different
world-line for the covariant non-canonical centroid $x^{(\vec \chi
)\mu}_s(\tau )$. Of course, inside the Wigner hyperplane, three
degrees of freedom of the isolated system \cite{c13} become gauge
variables. The natural gauge fixing for eliminating the first
class constraints ${\vec \kappa}_{+} \approx 0$ is $\vec \chi
(\tau )={\vec \sigma}_{com} ={\vec q}_{+}\approx 0$ [vanishing of
the location of the internal canonical 3-center of mass, see after
Eq.(\ref{II16})]. We have that ${\vec q}_{+}\approx 0$ implies
$\lambda_{\check r}(\tau )=0$: in this way the {\it internal}
3-center of mass is located in a unique centroid $x^{({\vec q}_{+}
)\mu}_s(\tau )=z^{\mu}(\tau ,\vec \sigma =0)$ [${\dot x}_s^{({\vec
q}_{+} )\mu} = {\dot {\tilde x}}^{\mu}_s = u^{\mu}(p_s)$].

\medskip

Note that the constant $x^{\mu}_s(0)$ [and, therefore, also
${\tilde x} ^{\mu}_s(0)$] is arbitrary, reflecting the
arbitrariness in the absolute location of the origin of the {\it
internal} coordinates on each hyperplane in Minkowski space-time.
The centroid $x^{\mu}_s(\tau )$ corresponds to the unique special
relativistic center-of-mass-type world-line for isolated systems
of Refs. \cite{beig,c11}, which unifies previous proposals of
Synge, M$\o$ller and Pryce.
\medskip

The only remaining canonical variables describing the Wigner
hyperplane in the final Dirac brackets are a non-covariant
canonical coordinate ${\tilde x}^{\mu}_s(\tau )$ \cite{c12}  and
$p^{\mu}_s$. The point with coordinates ${\tilde x}^{\mu}_s(\tau
)$ is the decoupled canonical {\it external 4-center of mass} of
the isolated system, which can be interpreted as a decoupled
observer with his parametrized clock ({\it point particle clock}).
Its velocity ${\dot {\tilde x}}^{\mu}_s(\tau )$ is parallel to
$p^{\mu}_s$, so that it has no classical zitterbewegung. \bigskip

The relation between $x^{\mu}_s(\tau )$ and ${\tilde
x}_s^{\mu}(\tau )$ (${\vec {\tilde \sigma}}$ is its 3-location on
the Wigner hyperplane) is \cite{lus,iten1}

\begin{equation}
{\tilde x}^{\mu}_s(\tau )= ({\tilde x}^o_s(\tau ); {\vec {\tilde
x}}_s(\tau ) )=z^{\mu}(\tau ,{\vec {\tilde \sigma
}})=x^{\mu}_s(\tau )-{1\over {\epsilon_s(p^o_s+\epsilon_s)}}\Big[
p_{s\nu}S_s^{\nu\mu}+\epsilon_s(S^{o\mu}_s-S^{o\nu}_s{{p_{s\nu}p_s^{\mu}}\over
{\epsilon^2_s}}) \Big],
 \label{II8}
\end{equation}

\medskip

After the separation of the relativistic canonical non-covariant
{\it external} 4-center of mass ${\tilde x}_s^{\mu}(\tau )$, the N
particles are described on the Wigner hyperplane by the 6N Wigner
spin-1 3-vectors ${\vec \eta}_i(\tau )$, ${\vec \kappa}_i (\tau )$
restricted by the rest-frame condition ${\vec \kappa}_{+}=\sum^N
_{i=1} {\vec \kappa}_i \approx 0$. \bigskip

The various spin tensors and vectors are \cite{lus}

\begin{eqnarray}
J^{\mu\nu}_s&=&x^{\mu}_s p^{\nu}_s- x^{\nu}_s p^{\mu}_s+
S^{\mu\nu}_s= {\tilde x}^{\mu}_s p^{\nu}_s - {\tilde x}^{\nu}_s
p^{\mu}_s +{\tilde S} ^{\mu\nu}_s,\nonumber \\ &&{}\nonumber \\
S^{\mu\nu}_s&=&[u^{\mu}(p_s)\epsilon^{\nu}(u(p_s))-u^{\nu}(p_s)\epsilon^{\mu}
(u(p_s))] {\bar S}^{\tau
r}_s+\epsilon^{\mu}(u(p_s))\epsilon^{\nu}(u(p_s)) {\bar
S}^{rs}_s\equiv \nonumber \\ &\equiv& \Big[
\epsilon^{\mu}_r(u(p_s)) u^{\nu}(p_s)-\epsilon^{\nu} (u(p_s))
u^{\mu}(p_s)\Big] \sum_{i=1}^N \eta^r_i \sqrt{m^2_ic^2+{\vec
\kappa} _i^2}+\nonumber \\ &+&\Big[ \epsilon^{\mu}_r(u(p_s))
\epsilon^{\nu}_s(u(p_s))-\epsilon^{\nu}_r (u(p_s))
\epsilon^{\mu}_r(u(p_s))\Big] \sum_{i=1}^N \eta^r_i\kappa^s_i,
\nonumber \\ &&{}\nonumber \\ {\bar
S}^{AB}_s&=&\epsilon^A_{\mu}(u(p_s)) \epsilon^B_{\nu}(u(p_s))
S^{\mu\nu}_s ,\nonumber \\ &&{\bar S}^{rs}_s\equiv
\sum_{i=1}^N(\eta^r_i\kappa^s_i- \eta^s_i\kappa^r_i),\quad {\bar
S}^{\tau r}_s\equiv - \sum_{i=1}^N\eta^r_i\sqrt{m^2_ic^2+{\vec
\kappa}_i^2},\nonumber \\ &&{}\nonumber \\ {\tilde
S}^{\mu\nu}_s&=&S^{\mu\nu}_s+{1\over {\sqrt{\epsilon p^2_s}(p^o_s+
\sqrt{\epsilon p^2_s})}}\Big[ p_{s\beta}(S^{\beta\mu}_s
p^{\nu}_s-S^{\beta\nu}_s p^{\mu}_s)+\sqrt{p^2_s}(S^{o\mu}_s
p^{\nu}_s-S^{o\nu}_s p^{\mu}_s)\Big], \nonumber \\ &&{\tilde
S}^{ij}_s=\delta^{ir}\delta^{js} {\bar S}_s^{rs},\quad\quad
{\tilde S}^{oi}_s=-{{\delta^{ir} {\bar S}^{rs}_s\, p^s_s}\over
{p^o_s+ \sqrt{\epsilon p^2_s}}},\nonumber \\ &&{}\nonumber \\
{\vec  {\bar S}} &\equiv & {\vec {\bar S}}=\sum_{i+1}^N{\vec
\eta}_i\times {\vec \kappa}_i\approx \sum_{i=1}^N {\vec
\eta}_i\times {\vec \kappa}_i-{\vec \eta}_{+}\times {\vec
\kappa}_{+} = \sum_{a=1}^{N-1} {\vec \rho}_a\times {\vec \pi}_a.
 \label{II9}
\end{eqnarray}

Note that while
$L^{\mu\nu}_s=x^{\mu}_sp^{\nu}_s-x^{\nu}_sp^{\mu}_s$ and
$S^{\mu\nu}_s$ are not constants of the motion due to the
classical zitterbewung (when $\vec \lambda (\tau ) \not= 0$), both
${\tilde L}_s^{\mu\nu}={\tilde x}^{\mu}_sp_s^{\nu}-{\tilde
x}^{\nu}_sp^{\mu}_s$ and ${\tilde S}^{\mu\nu}_s$ are conserved.

\medskip

The canonical variables ${\tilde x}^{\mu}_s$, $p^{\mu}_s$ for the
{\it external} 4-center of mass, can be replaced by the canonical
pairs\cite{ll}

\begin{eqnarray}
T_s&=& {{p_s\cdot {\tilde x}_s}\over {\epsilon_s}}={{p_s
 \cdot x_s}\over {\epsilon_s}},\qquad
 \epsilon_s = \pm \sqrt{\epsilon p^2_s},\nonumber \\
 {\vec z}_s&=&\epsilon_s ({\vec {\tilde
x}}_s- {{ {\vec p}_s}\over {p^o_s}} {\tilde x}^o_s), \qquad
 {\vec k}_s = {{ {\vec p}_s}\over {\epsilon_s}},
 \label{II10}
\end{eqnarray}

\noindent which make explicit the interpretation of it as a {\it
point particle clock}. The inverse transformation is

\begin{eqnarray}
{\tilde x}^o_s&=&\sqrt{1+{\vec k} _s^2}(T_s+{{{\vec k}_s\cdot
{\vec z}_s}\over {\epsilon_s}}), \qquad
 {\vec {\tilde x}}_s = {{{\vec z}_s}\over {\epsilon_s}}+(T_s+{{{\vec
k}_s\cdot {\vec z}_s}\over {\epsilon_s}}){\vec k}_s, \qquad
 p^o_s = \epsilon_s \sqrt{1+{\vec k}_s^2}, \nonumber \\
 {\vec p}_s&=&\epsilon_s {\vec k}_s.
 \label{II11}
\end{eqnarray}

\medskip

This non-point canonical transformation can be summarized as
[$\epsilon_s - M_{sys} \approx 0$, ${\vec \kappa}_{+}=\sum_{i=1}^N
{\vec \kappa}_i \approx 0$]

\begin{equation}
\begin{minipage}[t]{5cm}
\begin{tabular}{|l|l|} \hline
${\tilde x}_s^{\mu}$& ${\vec \eta}_i$ \\  \hline
 $p^{\mu}_s$& ${\vec \kappa}_i$ \\ \hline
\end{tabular}
\end{minipage} \ {\longrightarrow \hspace{1cm}} \
\begin{minipage}[t]{5 cm}
\begin{tabular}{|l|l|l|} \hline
$\epsilon_s$& ${\vec z}_s$   & ${\vec \eta}_i$   \\ \hline
 $T_s$ & ${\vec k}_s$& ${\vec \kappa}_i$ \\ \hline
\end{tabular}
\end{minipage}.
 \label{II12}
\end{equation}

After the addition of the gauge-fixing $T_s-\tau \approx 0$
\cite{c14}, the invariant mass $M_{sys}$ of the system, which is
also the {\it internal} energy of the isolated system, replaces
the non-relativistic Hamiltonian $H_{rel}$ for the relative
degrees of freedom: this reminds of the frozen Hamilton-Jacobi
theory, in which the time evolution can be re-introduced by using
the energy generator of the Poincar\'e group as Hamiltonian
\cite{c15}.
\bigskip

After the gauge fixings $T_s-\tau \approx 0$ [implying $\lambda
(\tau ) = - 1$], the final Hamiltonian, the embedding of the
Wigner hyperplane into Minkowski space-time and the Hamilton
equations become

\bea
 H_D&=& M_{sys} -\vec \lambda (\tau ) \cdot {\vec \kappa}_{+} ,\nonumber \\
 &&{}\nonumber \\
z^{\mu}(\tau ,\vec \sigma ) &=& x^{\mu}_s(\tau ) + \epsilon^{\mu
}_r(u(p_s)) \sigma^r =\nonumber \\
 &=& x^{\mu}_s(0) + u^{\mu}(p_s)
\tau + \epsilon^{\mu}_r(u(p_s))\, \Big( \sigma^r + \int_o^{\tau}
d\tau_1\, \lambda_r(\tau_1 )\Big) ,\nonumber \\
 &&{}\nonumber \\
 &&with \nonumber \\
&&{}\nonumber \\
 {\dot x}^{\mu}_s(\tau )\, & =& u^{\mu}(p_s)+\epsilon^{\mu}_r(u(p_s))
\lambda_r(\tau ),\nonumber \\
 &&{}\nonumber \\
 &&{}\nonumber \\
 {\dot {\vec \eta}}_i(\tau ) &\cir& {{{\vec \kappa}_i(\tau )}\over
 {\sqrt{m^2_i + {\vec \kappa}^2_l(\tau )}}} - \vec \lambda
 (\tau ),\quad \Rightarrow\qquad {\vec \kappa}_i(\tau ) \cir m_i\,
 {{{\dot {\vec \eta}}_i(\tau ) + \vec \lambda (\tau )}\over
 {\sqrt{1 - [{\dot {\vec \eta}}_i^2(\tau ) + \vec \lambda (\tau )]^2}}},\nonumber \\
 {\dot {\vec \kappa}}_i(\tau ) &\cir& 0.
 \label{II13}
  \eea

The particles' world-lines in Minkowski space-time and the
associated momenta are

\begin{eqnarray}
x^{\mu}_i(\tau )&=&z^{\mu}(\tau ,{\vec \eta}_i(\tau
))=x^{\mu}_s(\tau )+\epsilon^{\mu}_r(u(p_s)) \eta^r_i(\tau
),\nonumber \\
 p^{\mu}_i(\tau )&=&\sqrt{m^2_i+{\vec \kappa}^2_i(\tau )} u^{\mu}(p_s) +
\epsilon^{\mu}_r(u(p_s)) \kappa_{ir}(\tau )\,\, \Rightarrow
p^2_i=m_i^2.
 \label{II14}
 \end{eqnarray}
\medskip

The {\it external} rest-frame instant form realization of the
Poincar\'e generators \cite{c16} with non-fixed invariants $ p^2_s
= \epsilon_s^2 \approx M^2_{sys}$, $- p^2_s {\vec {\bar S}}_s^2
\approx - M^2_{sys} {\vec {\bar S}}^2$, is obtained from
Eq.(\ref{II9}):

\begin{eqnarray}
p^{\mu}_s,&&\nonumber  \\
 J^{\mu\nu}_s&=&{\tilde x}^{\mu}_sp^{\nu}_s-{\tilde x}^{\nu}_s p^{\mu}_s
 + {\tilde S}^{\mu\nu}_s,\nonumber \\
 &&{}\nonumber \\
 p^o_s&=& \sqrt{\epsilon_s^2+{\vec p}_s^2}= \epsilon_s \sqrt{1+ {\vec
k}_s^2}\approx  \sqrt{M^2_{sys}+{\vec p}^2_s}=M_{sys}
\sqrt{1+{\vec k}_s^2},\nonumber \\
 {\vec p}_s&=& \epsilon_s{\vec k}_s\approx M_{sys} {\vec k}_s,
\nonumber \\
 J^{ij}_s&=&{\tilde x}^i_sp^j_s-{\tilde x}^j_sp^i_s +
\delta^{ir}\delta^{js}\sum_{i=1}^N(\eta^r_i\kappa^s_i-\eta^s_i\kappa^r_i)=
z^i_sk^j_s-z^j_sk^i_s+\delta^{ir}\delta^{js} \epsilon^{rsu}{\bar
S}^u_s,\nonumber \\
 K^i_s&=&J^{oi}_s= {\tilde x}^o_sp^i_s-{\tilde x}^i_s
\sqrt{\epsilon_s^2+{\vec p}_s^2}-{1\over
{\epsilon_s+\sqrt{\epsilon^2_s+ {\vec p}_s^2}}} \delta^{ir} p^s_s
\sum_{i=1}^N(\eta^r_i\kappa^s_i- \eta^s_i\kappa^r_i)=\nonumber \\
 &=&-\sqrt{1+{\vec k}_s^2} z^i_s-{{\delta^{ir} k^s_s\epsilon
^{rsu}{\bar S}^u_s}\over {1+\sqrt{1+{\vec k}_s^2} }}\approx
{\tilde x}^o_sp^i_s -{\tilde x}^i_s\sqrt{M^2_{sys}+{\vec
p}^2_s}-{{\delta^{ir}p^s_s\epsilon^{rsu}{\bar S}^u_s}\over
{M_{sys}+\sqrt{M_{sys}^2+{\vec p}^2_s}}}.
 \label{II15}
 \end{eqnarray}
\medskip

On the other hand the {\it internal} realization of the Poincar\'e
algebra is built inside the Wigner hyperplane by using the
expression of ${\bar S}_s^{AB}$ given by Eq.(\ref{II9}) \cite{c17}

\begin{eqnarray}
&&M_{sys}=H_M=\sum_{i=1}^N  \sqrt{m^2_i+{\vec
\kappa}_i^2},\nonumber \\
 &&{\vec \kappa}_{+}=\sum_{i=1}^N {\vec
\kappa}_i\, (\approx 0),\nonumber \\
 &&\vec J=\sum_{i=1}^N {\vec \eta}_i\times {\vec
\kappa}_i,\quad\quad J^r={\bar S}^r={1\over 2}\epsilon^{ruv}{\bar
S}^{uv} \equiv {\bar S}^r_s,\nonumber \\
 &&\vec K=- \sum_{i=1}^N
\sqrt{m^2_i+{\vec \kappa}_i^2}\,\, {\vec \eta}_i=-M_{sys}{\vec
R}_{+},
 \quad\quad K^r=J^{or}={\bar S}_s^{\tau r}, \nonumber \\
&&{}\nonumber \\
 &&\Pi = M^2_{sys}-{\vec \kappa}_{+}^2 \approx M^2_{sys} > 0,\nonumber \\
 &&W^2=-\epsilon (M^2_{sys}-{\vec \kappa}^2_{+}) {\vec {\bar
S}}^2_s \approx -\epsilon M^2_{sys} {\vec {\bar S}}^2_s.
 \label{II16}
\end{eqnarray}
\medskip

The constraints $\epsilon_s-M_{sys}\approx 0$, ${\vec
\kappa}_{+}\approx 0$ have the following meaning:

 i) the constraint $\epsilon_s-M_{sys}\approx 0$ is the bridge connecting
the {\it external} and {\it internal} realizations \cite{c18};

 ii)the constraints ${\vec \kappa}_{+}\approx 0$, together with $\vec
K\approx 0$ (i.e. ${\vec R}_{+} \approx {\vec q}_{+} \approx {\vec
y}_{+} \approx 0$) \cite{c19}, imply an unfaithful {\it internal}
realization, in which the only non-zero generators are the
conserved energy and spin of an isolated system.

\bigskip

The determination of ${\vec q}_{+}$  for the N particle system has
been carried out by the group theoretical methods of
Ref.\cite{pauri} in Section III of Ref.\cite{iten1}. Given a
realization  of the ten Poincar\'e generators on the phase space,
one can build three 3-position variables in terms of them only.
For N free scalar relativistic particles on the Wigner hyperplane
with ${\vec p}_{sys}={\vec \kappa}_{+}\approx 0$ within the {\it
internal} realization (\ref{II16}) they are:\hfill\break

 i) a canonical {\it internal} 3-center of mass (or 3-center of spin)  ${\vec q}_{+}$;
\hfill\break

 ii) a non-canonical {\it internal}  M\o ller 3-center
of energy ${\vec R}_{+}$;\hfill\break

 iii) a non-canonical {\it internal} Fokker-Pryce
3-center of inertia ${\vec y}_{+}$. \hfill\break
 It can be shown\cite{iten1} that, due to ${\vec \kappa}_{+}\approx 0$,
 they all {\it coincide}: ${\vec q}_{+} \approx {\vec R}_{+} \approx {\vec y}_{+}$.
\medskip

Therefore the gauge fixings $\vec \chi (\tau )={\vec q}_{+}\approx
{\vec R}_{+}\approx {\vec y} _{+}\approx 0$ imply $\vec \lambda
(\tau )\approx 0$ and force the three {\it internal} collective
variables to coincide with the origin of the coordinates, which
now becomes

\beq
 x^{({\vec q}_{+})\mu}_s(T_s)=x_s^{\mu}(0) +u^{\mu}(p_s) T_s.
 \label{II17}
  \eeq

\noindent As  shown in Section IV, the addition of the gauge
fixings $\vec \chi (\tau )={\vec q}_{+}\approx {\vec R}_{+}\approx
{\vec y}_{+} \approx 0$ also implies that the Dixon center of mass
of an extended object\cite{dixon1} and  the Pirani\cite{pirani}
and Tulczyjew\cite{mul4,ehlers,mul8} centroids \cite{c20}  all
simultaneously coincide with  the centroid $x_s^{({\vec q}_{+}(\,
\mu }(\tau )$.
\bigskip

The {\it external} realization (\ref{II15}) allows to build the
analogous {\it external} 3-variables ${\vec q}_s$, ${\vec R}_s$,
${\vec Y}_s$. Eq.(4.4) of Ref.\cite{iten1} shows the construction
of the associated {\it external} 4-variables ${\tilde x}^{\mu}_s$,
$Y^{\mu}_s$, $R^{\mu}_s$ and their locations ${\vec {\tilde
\sigma}}$, ${\vec \sigma}_Y$, ${\vec \sigma}_R$ on the Wigner
hyperplane. It appears that the {\it external} Fokker-Pryce
non-canonical covariant 4-center of inertia $Y^{\mu}_s$ coincides
with the centroid (\ref{II17}).
\bigskip

In Ref.\cite{iten1} there is the definition of the following
sequence of canonical transformations

\beq
 \begin{minipage}[t]{3cm}
\begin{tabular}{|l|} \hline
${\vec \eta}_i$  \\ \hline
 ${\vec \kappa}_i$  \\ \hline
\end{tabular}
\end{minipage}
\ {\longrightarrow \hspace{.2cm}}\
\begin{minipage}[t]{3cm}
\begin{tabular}{|l|l|} \hline
${\vec \eta}_{+}$ & ${\vec \rho}_a$ \\ \hline
 ${\vec \kappa}_{+}$& ${\vec \pi}_a$ \\ \hline
\end{tabular}
\end{minipage}
\ {{\longrightarrow} \hspace{.2cm}}\
\begin{minipage}[b]{3cm}
\begin{tabular}{|l|l|} \hline
${\vec q}_{+}$ & ${\vec \rho}_{qa}$ \\ \hline
 ${\vec \kappa}_{+}$ & ${\vec \pi}_{qa}$  \\
 \hline
 \end{tabular}
 \end{minipage},\,\, a = 1,..,N-1,
 \label{II18}
 \eeq
\medskip

\noindent leading to the canonical separation of the internal
3-center of mass (${\vec q}_{+}$, ${\vec \kappa}_{+}$) from the
internal relative variables ${\vec \rho}_{qa}$, ${\vec \pi}_{qa}$.
Since the rest-frame condition ${\vec \kappa}_{+} \approx 0$
implies \cite{iten1} ${\vec \rho}_{qa} \approx {\vec \rho}_a$,
${\vec \pi}_{qa} \approx {\vec \pi}_a$, in the gauge ${\vec q}_{+}
\approx 0$ and in terms of the associated Dirac brackets we get an
internal reduced phase space whose canonical basis is ${\vec
\rho}_{qa} \equiv {\vec \rho}_a$, ${\vec \pi}_{qa} \equiv {\vec
\pi}_a$, $a=1,..,N-1$.
\medskip

The intermediate linear point canonical transformation in
(\ref{II18}) is [actually this is a family of canonical
transformations, since the $\gamma_{ai}$'s are any set of numbers
satisfying $\sum_i\, \gamma_{ai} = 0$, $\sum_a\, \gamma_{ai}\,
\gamma_{aj} = \delta_{ij} - {1\over N}$, $\sum_i\, \gamma_{ai}\,
\gamma_{bi} = \delta_{ab}$]

\bea
 {\vec \eta}_i &=& {\vec \eta}_{+} + {1\over {\sqrt{N}}}\,
 \sum_a\, \gamma_{ai}\, {\vec \rho}_a,\nonumber \\
 {\vec \kappa}_i &=& {1\over N}\, {\vec \kappa}_{+} + \sqrt{N}\,
 \sum_a\, \gamma_{ai}\, {\vec \pi}_a,\nonumber \\
 &&{}\nonumber \\
 {\vec \eta}_{+} &=& {1\over N}\, \sum_i\, {\vec \eta}_i,\qquad
 {\vec \rho}_a = \sqrt{N}\, \sum_i\, \gamma_{ai}\, {\vec
 \eta}_i,\nonumber \\
 {\vec \kappa}_{+} &=& \sum_i\, {\vec \kappa}_i,\,\,\,\qquad
 {\vec \pi}_a = {1\over {\sqrt{N}}}\, \sum_i\, \gamma_{ai}\, {\vec
 \kappa}_i.
 \label{II19}
 \eea
\medskip

The second canonical transformation has been defined in Section V
of Ref.\cite{iten1} by using a singular Gartenhaus-Schwartz
transformation (see Appendix C), but it was not written explicitly
for ${\vec \kappa}_{+} \not= 0$. By using Eqs. (\ref{c14}) and
(\ref{c15}) we get the following results:
\medskip

1) For $N = 2$ ($\gamma_{1\, 1} = - \gamma_{1\, 2} = {1\over
{\sqrt{2}}}$) we have

\bea
 &&{\vec \eta}_1 = {\vec \eta}_{+} + {1\over 2}\, \vec \rho ,\qquad
 {\vec \kappa}_1 = {1\over 2}\, {\vec \kappa}_{+} + \vec
 \pi ,\nonumber \\
 &&{\vec \eta}_2 = {\vec \eta}_{+} - {1\over 2}\, \vec \rho
 ,\qquad {\vec \kappa}_2 = {1\over 2}\, {\vec \kappa}_{+} - \vec
 \pi ,\nonumber \\
 &&{}\nonumber \\
 &&{\vec \eta}_{+} = {1\over 2}\, ({\vec \eta}_1 + {\vec
 \eta}_2),\qquad {\vec \kappa}_{+} = {\vec \kappa}_1 + {\vec
 \kappa}_2,\nonumber \\
 &&\vec \rho =\,\,\, {\vec \eta}_1 - {\vec \eta}_2,\qquad \vec \pi
 = {1\over 2}\, ({\vec \kappa}_1 - {\vec \kappa}_2),\nonumber \\
 &&{}\nonumber \\
 &&{}\nonumber \\
 &&\vec J = {\vec \eta}_1 \times {\vec \kappa}_1 + {\vec \eta}_2
 \times {\vec \kappa}_2 = {\vec \eta}_{+} \times {\vec \kappa}_{+}
 + \vec S = {\vec q}_{+} \times {\vec \kappa}_{+} + {\vec S}_q,\qquad
 \vec S = \vec \rho \times \vec \pi,\quad {\vec S}_q = {\vec
 \rho}_q \times {\vec \pi}_q,\nonumber \\
 &&{}\nonumber \\
 &&{}\nonumber \\
 &&{\vec R}_{+} = {{\sqrt{m^2_1 + {\vec \kappa}_1^2}\, {\vec \eta}_1 +
 \sqrt{m^2_2 + {\vec \kappa}_2^2}\, {\vec \eta}_2}\over {\sqrt{m^2_1 + {\vec \kappa}_1^2}
 + \sqrt{m^2_2 + {\vec \kappa}_2^2}}} = {\vec \eta}_{+} + {1\over
 2}\, {{\sqrt{m^2_1 + {\vec \kappa}_1^2} - \sqrt{m^2_2 + {\vec \kappa}_2^2}}\over
 {\sqrt{m^2_1 + {\vec \kappa}_1^2} + \sqrt{m^2_2 + {\vec
 \kappa}_2^2}}}\, \vec \rho ,\nonumber \\
 &&{}\nonumber \\
 &&{\vec q}_{+} = {\vec R}_{+} +\nonumber \\
  &&+ {{{\vec S}_q \times {\vec \kappa}_{+}}\over
 {(\sqrt{m^2_1 + {\vec \kappa}_1^2} + \sqrt{m^2_2 + {\vec \kappa}_2^2})\,
 (\sqrt{m^2_1 + {\vec \kappa}_1^2} + \sqrt{m^2_2 + {\vec \kappa}_2^2} +
 \sqrt{(\sqrt{m^2_1 + {\vec \kappa}_1^2} + \sqrt{m^2_2 + {\vec \kappa}_2^2})^2
  - {\vec \kappa}_{+}^2})}}.\nonumber \\
  &&{}
 \label{II20}
 \eea

Then after some straightforward algebra we get (note that in
Ref.\cite{iten1} we used the notation $M_{(free)} = M_{sys}$ and
${\cal M}^2_{(free)} = \Pi$)\medskip

\bea
 && \sqrt{m^2_i + {\vec \kappa}_i^2} = {1\over 2}\, \sqrt{{\cal M}_{(free)}^2 +
  {\vec \kappa}^2_{+}}\, \Big( 1 + (-)^{i+1}\, {{m^2_1 -
  m^2_2}\over {{\cal M}^2_{(free)}}}\Big) + (-)^{i+1}\, {{{\vec \pi}_q \cdot
  {\vec \kappa}_{+}}\over {{\cal M}_{(free)}}},\nonumber \\
  &&{}\nonumber \\
  &&M_{(free)} = \sqrt{m^2_1 + {\vec \kappa}_1^2} + \sqrt{m^2_2 + {\vec \kappa}_2^2}
 = \sqrt{{\cal M}^2_{(free)} + {\vec \kappa}^2_{+}} \approx {\cal
 M}_{(free)}\, {\buildrel {def}\over =}\, \sqrt{m^2_1 + {\vec
 \pi}_q^2} + \sqrt{m^2_2 + {\vec \pi}_q^2},\nonumber \\
 &&{}\nonumber \\
 &&\sqrt{m^2_1 + {\vec \kappa}_1^2} - \sqrt{m^2_2 + {\vec
 \kappa}_2^2} = {{2\, {\vec \pi}_q \cdot {\vec \kappa}_{+}}\over
 {{\cal M}_{(free)}}} + {{m^2_1 - m^2_2}\over {{\cal
 M}^2_{(free)}}}\, \sqrt{{\cal M}^2_{(free)} + {\vec
 \kappa}_{+}^2}\, {\buildrel {def}\over =}\, E,\nonumber \\
 &&{}\nonumber \\
 &&{}\nonumber \\
 &&\vec \pi = {\vec \pi}_q + {{{\vec \kappa}_{+}}\over {{\cal M}_{(free)}\,
 \sqrt{{\cal M}^2_{(free)} + {\vec \kappa}_{+}^2}}}\, \Big[ {\vec
 \pi}_q \cdot {\vec \kappa}_{+}\, \Big(1 - (\sqrt{{\cal M}^2_{(free)} +
 {\vec \kappa}_{+}^2} - {\cal M}_{(free)})\, {{{\cal M}_{(free)}}\over
 {{\vec \kappa}_{+}^2}}\Big) +\nonumber \\
 &&+ (m^2_1 - m^2_2)\, \sqrt{{\cal M}^2_{(free)} + {\vec
 \kappa}_{+}^2}\Big]\, {\buildrel {def}\over =}\, {\vec \pi}_q +
 F\, {\vec \kappa}_{+},\nonumber \\
 &&\vec \rho = {\vec \rho}_q - {A\over { B}}\, {{{\vec \kappa}_{+}
  \cdot {\vec \rho}_q\, {\vec \pi}_q}\over {{\cal M}_{(free)}\, \sqrt{{\cal M}^2_{(free)}}
  + {\vec \kappa}_{+}^2}}\, {\buildrel {def}\over =}\, {\vec
  \rho}_q + C\, {\vec \pi}_q,\nonumber \\
  &&\quad A = {{\sqrt{m^2_1 + {\vec \kappa}_1^2}}\over {\sqrt{m^2_2 + {\vec \pi}_q^2}}}
  + {{\sqrt{m^2_2 + {\vec \kappa}_2^2}}\over {\sqrt{m^2_1 + {\vec
  \pi}_q^2}}},\quad B = 1 + {{A\, {\vec \pi}_q \cdot {\vec \kappa}_{+}}\over
  {{\cal M}_{(free)}\, \sqrt{{\cal M}^2_{(free)} + {\vec \kappa}_{+}^2}}},\nonumber \\
 &&{}\nonumber \\
  &&{\vec \pi}_q = \vec \pi - {{{\vec \kappa}_{+}}\over {\sqrt{M^2_{(free)} -
  {\vec \kappa}_{+}^2}}}\, \Big( {1\over 2}\, (\sqrt{m^2_1 + {\vec
  \kappa}_1^2}-  \sqrt{m^2_2 + {\vec \kappa}_2^2}) -\nonumber \\
  &&- {{{\vec \kappa}_{+} \cdot \vec \pi}\over {{\vec \kappa}_{+}^2}}\,
  [M_{(free)} - \sqrt{M^2_{(free)} - {\vec \kappa}_{+}^2}]\Big)\,
  {\buildrel {def}\over =}\, \vec \pi - D\, {\vec
  \kappa}_{+},\nonumber \\
  &&{\vec \rho}_q = \vec \rho + {{A\, {\vec \kappa}_{+} \cdot \vec \rho}\over
  { M_{(free)}\, \sqrt{M^2_{(free)} - {\vec \kappa}^2_{+}}}}\, {\vec
  \pi}_q, \nonumber \\
  &&{}\nonumber \\
  && {\vec S}_q = \vec S - D\, \vec \rho \times {\vec
  \kappa}_{+},\nonumber \\
  &&{}\nonumber \\
  &&{\vec q}_{+} = {\vec R}_{+} + G\, {\vec S}_q \times {\vec
  \kappa}_{+},\nonumber \\
  &&\quad G = {1\over {M_{(free)}\, (M_{(free)} + \sqrt{M^2_{(free)}
   - {\vec \kappa}_{+}^2})}} = {1\over {\sqrt{{\cal M}^2_{(free)} +
   {\vec \kappa}_{+}^2}\, (\sqrt{{\cal M}^2_{(free)} + {\vec \kappa}_{+}^2}
    + {\cal M}_{(free)})}},\nonumber \\
 &&{\vec \eta}_{+} = {\vec q}_{+} - {E\over {2\, \sqrt{{\cal M}^2_{(free)}
  + {\vec \kappa}_{+}^2}}}\, \Big[ {\vec \rho}_q + C\, {\vec
  \pi}_q \Big] - G\, {\vec S}_q \times {\vec \kappa}_{+},\nonumber \\
  &&{}\nonumber \\
  &&{}\nonumber \\
  &&{\vec \eta}_i = {\vec q}_{+} + {1\over 2}\, \Big( (-)^{i+1} -
  {E\over {\sqrt{{\cal M}^2_{(free)} + {\vec
  \kappa}_{+}^2}}}\Big)\, \Big( {\vec \rho}_q + C\, {\vec
  \pi}_q\Big) - G\, {\vec S}_q \times {\vec \kappa}_{+},\nonumber \\
  &&{\vec \kappa}_i = \Big({1\over 2} + (-)^{i+1}\, F\Big)\, {\vec
  \kappa}_{+} + (-)^{i+1}\, {\vec \pi}_q.
 \label{II21}
 \eea

\bigskip

2) For $N > 2$ the results concerning the coordinates (and also
${\vec q}_{+}$) are much more involved due to the complexity of
Eqs. (\ref{c15}) so that we give only the following results for
the momenta

\bea
 && M_{(free)} = \sum_i\, \sqrt{m^2_i + {\vec \kappa}_i^2} =
 \sqrt{{\cal M}^2_{(free)} + {\vec \kappa}_{+}^2} \approx {\cal
 M}_{(free)} = \sum_i\, \sqrt{m^2_i + N\, \sum_{ab}\, \gamma_{ai}\,
  \gamma_{bi}\, {\vec \pi}_{qa} \cdot {\vec \pi}_{qb}},\nonumber \\
 &&{}\nonumber \\
 &&\sqrt{m^2_i + {\vec \kappa}_i^2} = {1\over {{\cal
 M}_{(free)}}}\, \Big(  \sqrt{N}\, \sum_a\, \gamma_{ai}\, {\vec \kappa}_{+} \cdot {\vec
  \pi}_{qa} + \nonumber \\
 &&\quad + \sqrt{m^2_i + N\, \sum_{ab}\, \gamma_{ai}\,
  \gamma_{bi}\, {\vec \pi}_{qa} \cdot {\vec \pi}_{qb}}\,
  \sqrt{{\cal M}^2_{(free)} + {\vec \kappa}_{+}^2}\Big),\nonumber  \\
 &&{}\nonumber \\
 &&{}\nonumber \\
 &&{\vec \kappa}_i = {{{\vec \kappa}_{+}}\over N} + \sqrt{N}\,
 \sum_a\, \gamma_{ai}\, \Big[ {\vec \pi}_{qa} + ({{ M_{(free)} }\over
  { {\cal M}_{(free)} }} - 1) {{{\vec \kappa}_{+} \cdot {\vec \pi}_{qa}}
  \over {{\vec \kappa}_{+}^2}}\, {\vec \kappa}_{+} +\nonumber \\
  &&\quad + {{{\vec \kappa}_{+}}\over
  {{\cal M}_{(free)}\, \sqrt{N}}}\, \sum_i\, \gamma_{ai}\,
  \sqrt{m^2_i + N\, \sum_{ab}\, \gamma_{ai}\, \gamma_{bi}\, {\vec \pi}_{qa}
   \cdot {\vec \pi}_{qb}}\Big].
 \label{II22}
 \eea

\vfill\eject

\section{The Energy-Momentum Tensor on the Wigner Hyperplane.}

\subsection{The Euler-Lagrange Equations and the Energy-Momentum Tensor
of Parametrized Minkowski Theories.}

The Euler-Lagrange equations associated with the Lagrangian
(\ref{II1}) are  (the symbol '${\buildrel \circ \over =}$' means
evaluated on the solutions of the equations of motion)

\begin{eqnarray}
&&\Big( {{\partial {\cal L}}\over {\partial z^{\mu}}}-\partial_A
{{\partial {\cal L}}\over {\partial z^{\mu}_A}}\Big) (\tau ,\vec
\sigma )=\eta_{\mu\nu}\partial_A[\sqrt{g} T^{AB} z_B^{\nu}](\tau
,\vec \sigma )\, {\buildrel \circ \over =}\, 0,\nonumber \\
 &&{{\partial L}\over
{\partial {\vec \eta}_i}}-\partial_{\tau} {{\partial L}\over
{\partial {\dot {\vec \eta}}_i}}=\nonumber \\
 &&-[{1\over 2} {{T^{AB}}\over {\sqrt{g}}}]{|}_{\vec \sigma ={\vec \eta}_i}
 {{\partial g_{AB}}\over {\partial{\vec \eta}_i}} -\partial_{\tau}
 {{ g_{\tau r}+g_{rs}{\dot \eta}_i^s}\over {\sqrt{g_{\tau\tau}+2g_{\tau u}{\dot \eta}^u_i
 +g_{uv}{\dot \eta}^u_i{\dot \eta}^v_i}}}{|}_{\vec \sigma ={\vec \eta}_i}
\, {\buildrel \circ \over =}\, 0, \label{III1}
\end{eqnarray}

\noindent where we have introduced the energy-momentum tensor
[here ${\dot \eta}^A_i(\tau )=(1; {\dot {\vec \eta}}_i(\tau ))$]

\begin{equation}
T^{AB}(\tau ,\vec \sigma )=-[{2\over {\sqrt{g}}}{{\delta S}\over
{\delta g_{AB}}}](\tau ,\vec \sigma )=-\sum_{i=1}^N\delta^3(\vec
\sigma -{\vec \eta}_i(\tau )) {{m_i{\dot \eta}^A_i(\tau ){\dot
\eta}^B_i(\tau )}\over {\sqrt{g_{\tau\tau}+2g_{\tau u}{\dot
\eta}^u_i
 +g_{uv}{\dot \eta}^u_i{\dot \eta}^v_i}}}(\tau ,\vec \sigma ).
\label{III2}
\end{equation}
\medskip

Because of the delta functions, the Euler-Lagrange equations for
the fields $z^{\mu}(\tau ,\vec \sigma )$ are trivial ($0\,
{\buildrel \circ \over =}\, 0$) everywhere except at the positions
of the particles. They may be rewritten in a form valid for every
isolated system

\begin{equation}
 \partial_AT^{AB} z^{\mu}_B\, {\buildrel \circ \over =}\, -{1\over {\sqrt{g}}}
 \partial_A[\sqrt{g} z^{\mu}_B] T^{AB}.
\label{III3}
\end{equation}

\noindent When $\partial_A[\sqrt{g} z^{\mu}_B]=0$, as it happens
on the Wigner hyper-planes in the gauge ${\vec q}_{+}\approx 0$
and $T_s-\tau \approx 0$, we get the conservation of the
energy-momentum tensor $T^{AB}$, i.e. $\partial_AT^{AB}\,
{\buildrel \circ \over =}\, 0$. Otherwise, there is a compensation
coming from the dynamics of the surface.
\bigskip

On the Wigner hyperplane, where we have

\begin{eqnarray}
x^{\mu}_i(\tau )&=&z^{\mu}(\tau ,{\vec \eta}_i(\tau
))=x^{\mu}_s(\tau )+\epsilon^{\mu}_r(u(p_s))\eta^r_i(\tau
),\nonumber \\
 &&{\dot x}^{\mu}_i(\tau )=z^{\mu}_{\tau}(\tau ,{\vec \eta}_i(\tau
))+z^{\mu}_r(\tau ,{\vec \eta}_i(\tau)) {\dot \eta}^r_i(\tau
)={\dot x}_s^{\mu}(\tau )+\epsilon^{\mu}_r(u(p_s)) {\dot
\eta}^r_i(\tau ),\nonumber \\
 &&{\dot x}^2_i(\tau )=g_{\tau\tau}(\tau ,{\vec
\eta}_i(\tau ))+2g_{\tau r}(\tau ,{\vec \eta}_i(\tau )){\dot
\eta}^r _i(\tau )+g_{rs}(\tau ,{\vec \eta}_i(\tau )){\dot
\eta}_i^r(\tau ){\dot \eta} ^s_i(\tau )=\nonumber \\
 &&{\dot x}_s^2(\tau )+2{\dot x}_{s\mu}(\tau
)\epsilon^{\mu}_r(u(p_s)){\dot \eta}^r_i(\tau )- {\dot {\vec
\eta}}_i^2(\tau ),\nonumber \\
 &&{}\nonumber \\
p^{\mu}_i(\tau )&=&\sqrt{m^2_i-\gamma^{rs}(\tau ,{\vec \eta}
_i(\tau ))\kappa_{ir}(\tau )\kappa_{is}(\tau )} l^{\mu}(\tau
,{\vec \eta} _i(\tau ))-\kappa_{ir}(\tau )\gamma^{rs}(\tau ,{\vec
\eta}_i(\tau )) z^{\mu}_s (\tau ,{\vec \eta}_i(\tau ))=\nonumber
\\
 &=&\sqrt{m_i^2+{\vec \kappa}_i^2(\tau )}u^{\mu}(p_s)+
 \epsilon^{\mu}_r(u(p_s))\kappa_i^r(\tau )\,\, \Rightarrow p^2_i=m^2_i, \nonumber \\
 p^{\mu}_s&=&\int d^3\sigma \rho_{\mu}(\tau ,\vec \sigma )\approx
\sum_{i=1}^N p^{\mu}_i(\tau ), \label{III4}
\end{eqnarray}

\noindent the energy-momentum tensor $T^{AB}(\tau ,\vec \sigma )$
takes the form

\begin{eqnarray}
T^{\tau\tau}(\tau ,\vec \sigma )&=&-\sum_{i=1}^N \delta^3(\vec
\sigma -{\vec \eta}_i(\tau )) {{m_i}\over {\sqrt{{\dot x}_s^2(\tau
)+ 2{\dot x}_{s\mu}(\tau )\epsilon^{\mu}_r(u(p_s))-{\dot {\vec
\eta}}_i^2(\tau )}}},\nonumber \\ T^{\tau r}(\tau ,\vec \sigma
)&=&-\sum_{i=1}^N \delta^3(\vec \sigma -{\vec \eta}_i(\tau ))
{{m_i{\dot \eta}^r_i(\tau )}\over {\sqrt{{\dot x}_s^2(\tau )+
2{\dot x}_{s\mu}(\tau )\epsilon^{\mu}_r(u(p_s))-{\dot {\vec
\eta}}_i^2(\tau )}}},\nonumber \\ T^{rs}(\tau ,\vec \sigma
)&=&-\sum_{i=1}^N \delta^3(\vec \sigma -{\vec \eta}_i(\tau ))
{{m_i{\dot \eta}_i^r(\tau ) {\dot \eta}_i^s(\tau )}\over
{\sqrt{{\dot x}_s^2(\tau )+ 2{\dot x}_{s\mu}(\tau
)\epsilon^{\mu}_r(u(p_s))-{\dot {\vec \eta}}_i^2(\tau )}}}.
\label{III5}
\end{eqnarray}

\subsection{The Energy-Momentum Tensor in the Standard Lorentz-Covariant Theory.}

With the position $x^{\mu} = z^{\mu}(\tau ,\vec \sigma )$, the
same form is obtained from the energy momentum tensor of the
standard manifestly Lorentz covariant theory with Lagrangian
$S_S=\int d\tau L_S(\tau )=-\sum_{i=1}^Nm_i \int d\tau \sqrt{{\dot
x}^2_i(\tau )}$, restricted to positive energies\cite{c21}

\begin{eqnarray}
T^{\mu\nu}(z(\tau ,\vec \sigma ))&=&- \Big( {2\over {\sqrt{g}}}\,
{{\delta S_S}\over {\delta g_{\mu\nu}}}\Big) {|}_{x=z(\tau ,\vec
\sigma )}=\nonumber \\
 &=&\sum_{i=1}^N m_i
\int d\tau_1 {{ {\dot x}_i^{\mu}(\tau_1){\dot
x}_i^{\nu}(\tau_1)}\over {\sqrt{{\dot x}^2_i(\tau_1)}}}
\delta^4\Big(x_i(\tau_1)-z(\tau ,\vec \sigma )\Big)=\nonumber \\
 &=&\epsilon^{\mu}_A(u(p_s))\epsilon^{\nu}_B(u(p_s)) T^{AB}(\tau ,\vec
\sigma ). \label{III6}
\end{eqnarray}

\bigskip

1) On arbitrary {\it space-like hyper-surfaces} we have
\medskip

\begin{eqnarray}
T^{\mu\nu}(z(\tau ,\vec \sigma ))&=&\sum_{i=1}^N m_i \int d\tau_1
{{ {\dot x}_i^{\mu}(\tau_1){\dot x}_i^{\nu}(\tau_1)}\over
{\sqrt{{\dot x}^2_i(\tau_1)}}} \delta^4\Big(x_i(\tau_1)-z(\tau
,\vec \sigma )\Big)=\nonumber \\ &=&\sum_{i=1}^N m_i
 \int {{d\tau_1 }\over {\sqrt{{\dot x}^2_i(\tau_1)}}} \delta^4\Big(z(\tau_1,
{\vec \eta}_i(\tau_1) )-z(\tau ,\vec \sigma )\Big)\nonumber \\
&&[z^{\mu}_{\tau}(\tau_1,{\vec \eta}
_i(\tau_1))+z^{\mu}_r(\tau_1,{\vec \eta}_i(\tau_1)){\dot \eta}
^r_i(\tau_1)]\nonumber \\ && [z^{\nu}_{\tau}(\tau_1,{\vec
\eta}_i(\tau_1))+z^{\nu}_r(\tau_1 ,{\vec \eta}_i(\tau_1)){\dot
\eta}^r_i(\tau_1)] =\nonumber \\ &=&\sum_{i=1}^N m_i \int
{{d\tau_1}\over {\sqrt{{\dot x}^2_i(\tau_1)}}} \delta^4\Big(
z(\tau_1,{\vec \eta}_i(\tau_1) )-z(\tau ,\vec \sigma )\Big)
\nonumber \\ && \Big[ z^{\mu}_{\tau}(\tau_1,{\vec \eta}
_i(\tau_1))\, z^{\nu}_{\tau}(\tau_1,{\vec
\eta}_i(\tau_1))+\nonumber \\ &+&\Big( z^{\mu}_{\tau}(\tau_1,{\vec
\eta}_i(\tau_1))z^{\nu}_r(\tau_1,{\vec \eta}
_i(\tau_1))+z^{\nu}_{\tau}(\tau_1,{\vec
\eta}_i(\tau_1))z^{\mu}_r(\tau_1,{\vec \eta}_i(\tau_1))\Big) {\dot
\eta}_i^r(\tau_1)+\nonumber \\ &+&z^{\mu}_r(\tau_1,{\vec
\eta}_i(\tau_1)) z^{\nu}_s(\tau_1,{\vec \eta}_i(\tau_1)) {\dot
\eta}^r_i(\tau_1) {\dot \eta}^s_i(\tau_1) \Big] =\nonumber \\
&=&\sum_{i=1}^N {{m_i}\over {\sqrt{{\dot x}^2_i(\tau )}}}
\delta^3\Big( \vec \sigma -{\vec \eta}_i(\tau )\Big) (det\,
|z^{\mu}_A(\tau ,\vec \sigma )|)^{-1}\nonumber \\ &&\Big[
z^{\mu}_{\tau}(\tau,{\vec \eta}_i(\tau_1))\, z^{\nu}_{\tau}(\tau
,{\vec \eta}_i(\tau ))+\Big( z^{\mu}_{\tau}(\tau,{\vec
\eta}_i(\tau )) z^{\nu}_r(\tau ,{\vec \eta}_i(\tau ))+\nonumber \\
&+&z^{\nu}_{\tau}(\tau ,{\vec \eta}_i(\tau ))z^{\mu}_r(\tau ,{\vec
\eta}_i(\tau ))\Big) {\dot \eta}_i^r(\tau )+ z^{\mu}_r(\tau ,{\vec
\eta}_i(\tau )) z^{\nu}_s(\tau ,{\vec \eta}_i(\tau )) {\dot
\eta}^r_i(\tau ) {\dot \eta}^s_i(\tau ) \Big] =\nonumber \\
&=&\sum_{i=1}^N {{m_i}\over {\sqrt{{\dot x}^2_i(\tau
)}\sqrt{g(\tau ,\vec \sigma )}}} \delta^3\Big( \vec \sigma -{\vec
\eta}_i(\tau )\Big)\nonumber \\ &&\Big[ z^{\mu}_{\tau}(\tau,{\vec
\eta}_i(\tau_1))\, z^{\nu}_{\tau}(\tau ,{\vec \eta}_i(\tau
))+\Big( z^{\mu}_{\tau}(\tau,{\vec \eta}_i(\tau )) z^{\nu}_r(\tau
,{\vec \eta}_i(\tau ))+\nonumber \\ &+&z^{\nu}_{\tau}(\tau ,{\vec
\eta}_i(\tau ))z^{\mu}_r(\tau ,{\vec \eta}_i(\tau ))\Big) {\dot
\eta}_i^r(\tau )+ z^{\mu}_r(\tau ,{\vec \eta}_i(\tau ))
z^{\nu}_s(\tau ,{\vec \eta}_i(\tau )) {\dot \eta}^r_i(\tau ) {\dot
\eta}^s_i(\tau ) \Big],
 \label{III7}
 \end{eqnarray}

\noindent since $det\, |z^{\mu}_A| =\sqrt{g}=\sqrt{\gamma
(g_{\tau\tau}-\gamma^{rs}g _{\tau r}g_{\tau s})}$, $\gamma =|det\,
g_{rs}|$.\bigskip

2) On arbitrary {\it space-like hyper-planes}, where it
holds\cite{lus}\medskip

\begin{eqnarray}
z^{\mu}(\tau ,\vec \sigma )&=&x^{\mu}_s(\tau )+b^{\mu}_u(\tau
)\sigma^u,\quad\quad x^{\mu}_i(\tau )=x^{\mu}_s(\tau )+
b^{\mu}_u(\tau )\eta^u_i(\tau ),\nonumber \\
 &&{}\nonumber \\
 z^{\mu}_r(\tau ,\vec \sigma
)&=&b^{\mu}_r(\tau ),\quad\quad z^{\mu}_{\tau}(\tau ,\vec \sigma
)={\dot x}_s^{\mu}(\tau )+{\dot b}^{\mu}_u (\tau
)\sigma^u=l^{\mu}/\sqrt{g^{\tau\tau}}-g_{\tau
r}z^{\mu}_r,\nonumber \\
 &&{}\nonumber \\
g_{\tau \tau}&=& [{\dot x}_s^{\mu}+ {\dot
b}^{\mu}_r\sigma^r]^2,\quad\quad g_{\tau r}=b_{r\mu}[{\dot
x}_s^{\mu}+ {\dot b}^{\mu}_s\sigma^s], \nonumber \\
g_{rs}&=&-\delta_{rs},\quad\quad
\gamma^{rs}=-\delta^{rs},\quad\quad \gamma =1,\nonumber \\
g&=&g_{\tau\tau}+\sum_rg^2_{\tau r}, \nonumber \\
  g^{\tau\tau}&=&1/[l_{\mu}({\dot x}_s^{\mu}+{\dot b}^{\mu}_u\sigma^u)]^2,
\quad\quad g^{\tau r} =g^{\tau\tau}g_{\tau r}=b_{r\mu}({\dot
x}_s^{\mu}+{\dot b}^{\mu}_u\sigma^u)/ [l_{\mu}({\dot
x}_s^{\mu}+{\dot b}^{\mu}_u\sigma^u)]^2,\nonumber \\
 &&{}\nonumber \\
g^{rs}&=&-\delta^{rs} +g^{\tau\tau}g_{\tau r}g_{\tau
s}=-\delta^{rs}+ b_{r\mu}({\dot x}_s^{\mu}+{\dot
b}^{\mu}_u\sigma^u) b_{r\nu}({\dot x}_s^{\nu}+{\dot
b}^{\nu}_v\sigma^v) /[l_{\mu}({\dot x}_s^{\mu}+{\dot
b}^{\mu}_u\sigma^u)]^2,
 \label{III8}
 \end{eqnarray}

\noindent we have

\begin{eqnarray}
T^{\mu\nu}[x^{\beta}_s(\tau )+b^{\beta}_r(\tau )\sigma^r]&=&
\sum_{i=1}^N {{m_i \delta^3\Big( \vec \sigma -{\vec \eta}_i(\tau
)\Big) } \over {\sqrt{g(\tau ,\vec \sigma )}
\sqrt{g_{\tau\tau}(\tau ,\vec \sigma )+2g_{\tau r}(\tau ,\vec
\sigma ){\dot \eta}_i(\tau )-{\dot {\vec \eta}}^2_i(\tau )}
}}\nonumber \\ &&\Big[ ({\dot x}^{\mu}_s(\tau )+{\dot
b}^{\mu}_r(\tau )\eta^r_i(\tau )) ({\dot x}^{\nu}_s(\tau )+{\dot
b}^{\nu}_s(\tau )\eta^s_i(\tau ))+\nonumber \\ &+&\Big( ({\dot
x}^{\mu}_s(\tau )+{\dot b}^{\mu}_s(\tau )\eta^s_i(\tau ))
b^{\nu}_r(\tau )+({\dot x}^{\nu}_s(\tau )+{\dot b}^{\nu}_s(\tau
)\eta^s _i(\tau )) b^{\mu}_r(\tau )\Big) {\dot \eta}^r_i(\tau
)+\nonumber \\ &+&b^{\mu}_r(\tau ) b^{\nu}_s(\tau ) {\dot
\eta}^r_i(\tau ){\dot \eta}^s_i(\tau ) \Big] .
 \label{III9}
 \end{eqnarray}

\bigskip

3) Finally, on {\it Wigner hyper-planes}, where it holds\cite{lus}
\medskip

\begin{eqnarray}
z^{\mu}(\tau ,\vec \sigma )&=&x^{\mu}_s(\tau
)+\epsilon^{\mu}_u(u(p_s))\sigma^u,\quad\quad x^{\mu}_i(\tau
)=x^{\mu}_s(\tau )+\epsilon^{\mu}_u(u(p_s)) \eta^u_i(\tau
),\nonumber \\
 &&{}\nonumber \\
z^{\mu}_r&=&\epsilon^{\mu}_r(u(p_s),\quad\quad
l^{\mu}=u^{\mu}(p_s),\quad\quad z^{\mu}_{\tau}={\dot
x}^{\mu}_s(\tau ),\nonumber \\
 g&=&[{\dot x}_s(\tau )\cdot u(p_s)]^2,\quad\quad g_{\tau\tau}={\dot x}^2_s,
 \quad\quad g_{\tau r}={\dot x}_{s\mu} \epsilon^{\mu}_r(u(p_s)),\quad\quad
 g_{rs}=-\delta_{rs},\nonumber \\
g^{\tau\tau}&=&1/[{\dot x}_{s\mu}u^{\mu}(p_s)]^2,\quad\quad
g^{\tau r}={\dot x}_{s\mu}\epsilon^{\mu} _r(u(p_s))/[{\dot
x}_{s\mu}u^{\mu}(p_s)]^2,\nonumber \\ g^{rs}&=&-\delta^{rs}+{\dot
x} _{s\mu}\epsilon^{\mu}_r(u(p_s)) {\dot
x}_{s\nu}\epsilon^{\nu}_s(u(p_s)) /[{\dot
x}_{s\mu}u^{\mu}(p_s)]^2,\nonumber \\
 &&{}\nonumber \\
ds^2&=& {\dot x}_s^2(\tau )d\tau^2+2{\dot x}_s(\tau )\cdot
\epsilon_r(u(p_s))d\tau d\sigma^r-d{\vec \sigma}^2,
 \label{III10}
 \end{eqnarray}

\noindent we have

\begin{eqnarray}
T^{\mu\nu}[x^{\beta}_s(\tau
)+\epsilon^{\beta}_r(u(p_s))\sigma^r]&=& {1\over {{{\dot x}_s(\tau
)\cdot u(p_s)} }} \sum_{i=1}^N {{m_i}\over {\sqrt{{\dot
x}_s^2(\tau )+2{\dot x}_{s\beta}(\tau )
\epsilon^{\beta}_r(u(p_s)){\dot \eta}^r_i(\tau )-{\dot {\vec
\eta}}_i^2(\tau )} }}\nonumber \\ &&\delta^3\Big( \vec \sigma
-{\vec \eta} _i(\tau )\Big) \Big[ {\dot x}^{\mu}_s(\tau ){\dot
x}^{\nu}_s(\tau )+ \nonumber \\ &+& \Big( {\dot x}^{\mu}_s(\tau
)\epsilon^{\nu}_r(u(p_s))+{\dot x}^{\nu}_s(\tau
)\epsilon^{\mu}_r(u(p_s))\Big) {\dot \eta}^r_i(\tau )+\nonumber \\
&+&\epsilon^{\mu}_r(u(p_s)) \epsilon^{\nu}_s(u(p_s)) {\dot
\eta}^r_i(\tau ){\dot \eta}^s_i(\tau ) \Big] = \nonumber \\
&=&{1\over {{{\dot x}_s(\tau )\cdot u(p_s)}}} \sum_{i=1}^N
{{m_i}\over {\sqrt{{\dot x}_s^2(\tau )+2{\dot x}_{s\beta}(\tau )
\epsilon^{\beta}_r(u(p_s)){\dot \eta}^r_i(\tau )-{\dot {\vec
\eta}}_i^2(\tau )} }}\nonumber \\ &&\Big[{\dot x}^{\mu}_s(\tau
){\dot x}^{\nu} _s(\tau )  \delta^3\Big( \vec \sigma -{\vec \eta}
_i(\tau )\Big)+\nonumber \\ &+&\Big( {\dot x}^{\mu}_s(\tau
)\epsilon^{\nu}_r(u(p_s))+{\dot x}^{\nu}_s(\tau
)\epsilon^{\mu}_r(u(p_s))\Big)  \delta^3\Big( \vec \sigma - {\vec
\eta}_i(\tau )\Big) {\dot \eta}^r_i(\tau )+\nonumber \\
&+&\epsilon^{\mu}_r(u(p_s))\epsilon^{\nu}_s(u(p_s))
 \delta^3\Big( \vec \sigma -{\vec \eta}
 _i(\tau )\Big) {\dot \eta}^r_i(\tau ){\dot \eta}^s_i(\tau ) \Big].
 \label{III11}
 \end{eqnarray}
\bigskip

Since the volume element on the Wigner hyperplane is $u^{\mu}(p_s)
d^3\sigma$, we obtain the following total 4-momentum and total
mass of the N free particle system   (Eqs.(\ref{II7}) are used)
\medskip

\begin{eqnarray}
P^{\mu}_T&=&\int d^3\sigma T^{\mu\nu}[x^{\beta}_s(\tau
)+\epsilon^{\beta} _u(u(p_s))\sigma^u] u_{\nu}(p_s)=\nonumber \\
&=& \sum_{i=1}^N {{m_i}\over {\sqrt{\lambda^2(\tau )- [{\dot {\vec
\eta}}_i(\tau )+\vec \lambda (\tau )]^2}}}\nonumber \\ &&\Big[
-\lambda (\tau )u^{\mu}(p_s)+ [{\dot \eta}^r_i(\tau
)+\lambda^r(\tau )] \epsilon^{\mu}_r(u(p_s))\Big]  \nonumber \\
&{\buildrel \circ \over =}\,& {|}_{\lambda (\tau )=-1}\,\,
\sum_{i=1}^N\Big [ \sqrt{m^2_ic^2+{\vec \kappa}_i^2(\tau
)}u^{\mu}(p_s)+\kappa^r_i(\tau ) \epsilon^{\mu}_r(u(p_s))\Big]
=\sum_{i=1}^Np^{\mu}_i(\tau )=p^{\mu}_s, \nonumber \\
&&{}\nonumber \\
 M_{sys}&=&P^{\mu}_T u_{\mu}(p_s)=-\lambda(\tau )
\sum_{i=1}^N {{m_i}\over {\sqrt{\lambda^2(\tau )- [{\dot {\vec
\eta}}_i(\tau )+\vec \lambda (\tau )]^2}}} \nonumber \\
&{\buildrel \circ \over =}\,& {|}_{\lambda (\tau )=-1}\,\,
\sum_{i=1}^N \sqrt{m^2_i+{\vec \kappa}^2_i(\tau )},
 \label{III12}
\end{eqnarray}

\noindent which turn out to be in the correct form only if
$\lambda (\tau )=-1$. This shows that the agreement with
parametrized Minkowski theories on arbitrary space-like
hyper-surfaces is obtained {\it only on Wigner hyper-planes in the
gauge $T_s-\tau \approx 0$, which indeed implies $\lambda (\tau
)=-1$}.

\subsection{The Phase-Space Version of the Standard Energy-Momentum Tensor.}

The same result may be obtained by first rewriting the standard
energy-momentum tensor in phase space and then imposing the
restriction $p^{\mu}_i(\tau )=$\hfill\break
$\sqrt{m^2_i-\gamma^{rs}(\tau ,{\vec \eta} _i(\tau
))\kappa_{ir}(\tau )\kappa_{is}(\tau )} l^{\mu}(\tau ,{\vec \eta}
_i(\tau ))-\kappa_{ir}(\tau )\gamma^{rs}(\tau ,{\vec \eta}_i(\tau
)) z^{\mu}_s (\tau ,{\vec \eta}_i(\tau ))\, \rightarrow \,
\sqrt{m^2_i+{\vec \kappa}^2_i (\tau
)}u^{\mu}(p_s)+\kappa^r_i\epsilon^{\mu}_r(u(p_s))$ \cite{c22}\,\,
[the last equality refers to the Wigner hyperplane, see
Eq.(\ref{II14})]. We have:\medskip

\begin{eqnarray}
T^{\mu\nu}(z(\tau ,\vec \sigma ))&=&\sum_{i=1}^N{1\over {m_i}}
\int d\tau_1 \sqrt{{\dot
x}^2_i(\tau_1)}p^{\mu}_i(\tau_1)p^{\nu}_i(\tau_1)
\delta^4(x_i(\tau_1)-z(\tau ,\vec \sigma ))=\nonumber \\
&=&\sum_{i=1}^N {{\sqrt{{\dot x}_i^2(\tau )}}\over {m_i
\sqrt{g(\tau ,\vec \sigma )}}} p^{\mu}_i(\tau )p^{\nu}_i(\tau )
\delta^3(\vec \sigma -{\vec \eta}_i(\tau )), \nonumber \\
 &&{}\nonumber \\
\Downarrow && on\, Wigner's\, hyper-planes \nonumber \\
 &&{}\nonumber \\
 T^{\mu\nu}[x^{\beta}_s(\tau
)+\epsilon^{\beta}_u(u(p_s))\sigma^u]&=&\sum_{i=1} ^N {{
\sqrt{{\dot x}^2_s+2{\dot x}_{s\beta}\epsilon^{\beta}_u(u(p_s))
{\dot \eta}^u_i-{\dot {\vec \eta}}^2_i}}\over {m_i \sqrt{{\dot
x}_s\cdot u(p_s)}}} (\tau ) p^{\mu}_i(\tau )p^{\nu}_i (\tau
)\delta^3(\vec \sigma -{\vec \eta}_i(\tau ))=\nonumber \\
&=&\sum_{i=1}^N \delta^3(\vec \sigma -{\vec \eta}_i(\tau )) {{
\sqrt{{\dot x}^2_s+2{\dot x}_{s\beta}\epsilon^{\beta}_u(u(p_s))
{\dot \eta}^u_i-{\dot {\vec \eta}}_i^2}}\over {m_i \sqrt{{\dot
x}_s\cdot u(p_s)}}} (\tau )\nonumber \\ &&\Big[(m_i^2+{\vec
\kappa}^2_i(\tau ))u^{\mu}(p_s)u^{\nu}(p_s)+\nonumber \\
&+&k^r_i(\tau ) \sqrt{m_i^2+{\vec \kappa}^2_i(\tau
)}\Big(u^{\mu}(p_s)\epsilon^{\nu}_r(u(p_s))
+u^{\nu}(p_s)\epsilon^{\mu}_r(u(p_s))\Big) +\nonumber \\
&+&\kappa^r_i(\tau )\kappa^s_i(\tau )
\epsilon^{\mu}_r(u(p_s))\epsilon^{\nu}_s(u(p_s))\Big] =\nonumber
\\ &=&\sum_{i=1}^N \delta^3(\vec \sigma -{\vec \eta}_i(\tau )) {{
\sqrt{\lambda^2(\tau )-[{\dot {\vec \eta}}_i(\tau )+\vec \lambda
(\tau )]^2 }}\over { \sqrt{-\lambda (\tau )}}} \nonumber \\
&&{{\sqrt{m_i^2+{\vec \kappa}^2_i(\tau )}}\over {m_i}}
\Big[\sqrt{m_i^2+{\vec \kappa}^2_i(\tau
)}u^{\mu}(p_s)u^{\nu}(p_s)+\nonumber \\ &+&k^r_i(\tau
)\Big(u^{\mu}(p_s)\epsilon^{\nu}_r(u(p_s))
+u^{\nu}(p_s)\epsilon^{\mu}_r(u(p_s))\Big) +\nonumber \\
&+&{{\kappa^r_i(\tau )\kappa^s_i(\tau )}\over {\sqrt{m_i^2+{\vec
\kappa}^2 _i(\tau
)}}}\epsilon^{\mu}_r(u(p_s))\epsilon^{\nu}_s(u(p_s))\Big] .
 \label{III13}
\end{eqnarray}

\medskip

The total 4-momentum and total mass are then
\medskip

\begin{eqnarray}
P^{\mu}_T&=&\int d^3\sigma T^{\mu\nu}[x^{\beta}_s(\tau
)+\epsilon^{\beta} _u(u(p_s))\sigma^u] u_{\nu}(p_s)=\nonumber \\
&=&\sum_{i=1}^N {{ {\dot x}^2_s+2{\dot
x}_{s\beta}\epsilon^{\beta}_u(u(p_s)){\dot \eta}^u_i- {\dot {\vec
\eta}}_i^2}\over { {\dot x}_s\cdot u(p_s)}} (\tau ) {{\sqrt{m^2_i+
{\vec \kappa}^2_i(\tau )} }\over {m_i}}\nonumber \\
&&[\sqrt{m^2_i+{\vec \kappa}^2_i(\tau )}
u^{\mu}(p_s)+\kappa^r_i(\tau ) \epsilon^{\mu}_r(u(p_s))]=\nonumber
\\ &=&\sum_{i=1}^N {{ \sqrt{\lambda^2(\tau )-[{\dot {\vec
\eta}}_i(\tau )+\vec \lambda (\tau )]^2 }}\over { \sqrt{-\lambda
(\tau )}}}{{\sqrt{m^2_i+ {\vec \kappa}^2_i(\tau )} }\over {m_i}}
p^{\mu}_i(\tau ),\nonumber \\
 M_{sys}&=&P^{\mu}_T u_{\mu}(p_s)=\sum_{i=1}^N
{{ \sqrt{\lambda^2(\tau )-[{\dot {\vec \eta}}_i(\tau )+\vec
\lambda (\tau )]^2 }}\over { {-\lambda (\tau )}}} {{\sqrt{m^2_i+
{\vec \kappa}^2_i(\tau )} }\over {m_i}} \sqrt{m^2_i+{\vec
\kappa}^2_i(\tau )}.
 \label{III14}
\end{eqnarray}
\medskip

Since in this case we have $m_i/\sqrt{1- [{\dot {\vec
\eta}}_i(\tau )+\vec \lambda (\tau )]^2}\, {\buildrel \circ \over
=}\, \sqrt{m^2_ic^2+{\vec \kappa}^2_i(\tau )}$, the equations
above show that the total 4-momentum evaluated from the
energy-momentum tensor of the standard theory restricted to
positive energy particles is consistent \cite{c23} with the
description on the Wigner hyper-planes with its gauge freedom
$\lambda (\tau )$, $\vec \lambda (\tau )$, {\it provided} one
works with Dirac brackets of the gauge $T_s\equiv \tau$, where one
has $\lambda (\tau )=-1$ and

\begin{eqnarray}
{\dot x}^{\mu}_s(\tau )&=&u^{\mu}(p_s)+\epsilon^{\mu}_r(u(p_s))
\lambda_r(\tau ),\nonumber \\ x^{\mu}_s(\tau )&=&x^{\mu}_s(0)+\tau
u^{\mu}(p_s)+\epsilon^{\mu}_r (u(p_s)) \int^{\tau}_0d\tau_1
\lambda_r(\tau_1),
 \label{III15}
\end{eqnarray}

\bigskip

Therefore, for every $\vec \lambda (\tau )$, we have
\medskip

\begin{eqnarray}
T^{\mu\nu}[x^{\beta}_s(T_s)+\epsilon^{\beta}_u(u(p_s))\sigma^u]&=&
\epsilon^{\mu}_A(u(p_s)) \epsilon^{\nu}_B(u(p_s)) T^{AB}(T_s,\vec
\sigma ) =\nonumber \\
 &=&\sum_{i=1}^N \delta^3(\vec \sigma -{\vec \eta}_i(T_s))
\Big[\sqrt{m_i^2+{\vec \kappa}^2_i(T_s)}
u^{\mu}(p_s)u^{\nu}(p_s)+\nonumber \\
&+&k^r_i(T_s)\Big(u^{\mu}(p_s)\epsilon^{\nu}_r(u(p_s))
+u^{\nu}(p_s)\epsilon^{\mu}_r(u(p_s))\Big) +\nonumber \\
&+&{{\kappa^r_i(T_s)\kappa^s_i(T_s)}\over {\sqrt{m_i^2+{\vec
\kappa}^2 _i(T_s)} }}
\epsilon^{\mu}_r(u(p_s))\epsilon^{\nu}_s(u(p_s))\Big] , \nonumber
\\ T^{\tau\tau}(T_s,\vec \sigma )&=&\sum_{i=1}^N\delta^3(\vec
\sigma -{\vec \eta}_i(T_s)) \sqrt{m_i^2+{\vec
\kappa}_i^2(T_s))},\nonumber \\ T^{\tau r}(T_s,\vec \sigma
)&=&\sum_{i=1}^N\delta^3(\vec \sigma -{\vec \eta}_i(T_s))
\kappa_i^r(T_s),\nonumber \\ T^{rs}(T_s,\vec \sigma
)&=&\sum_{i=1}^N\delta^3(\vec \sigma -{\vec \eta}_i(T_s)) {{
\kappa_i^r(T_s) \kappa_i^s(T_s)}\over { \sqrt{m_i^2+{\vec
\kappa}_i^2(T_s)}}},\nonumber \\
 &&{}\nonumber \\
P^{\mu}_T&=&p^{\mu}_s=Mu^{\mu}(p_s)+\epsilon^{\mu}_r(u(p_s))\kappa^r_{+}
\approx M u^{\mu}(p_s),\nonumber \\ M&=&\sum_{i=1}^N
\sqrt{m^2_ic^2+{\vec \kappa}_i^2(T_s)},\nonumber \\
 &&{}\nonumber \\
T^{\mu\nu}[x^{\beta}_s(T_s)+\epsilon^{\beta}_u(u(p_s))\sigma^u]
&&u_{\nu}(p_s)=\epsilon^{\mu}_A(u(p_s)) T^{A\tau}(T_s,\vec \sigma
)= \nonumber \\
 &=&\sum_{i=1}^N \delta^3(\vec \sigma -{\vec \eta}_i(T_s))
\nonumber \\ &&[\sqrt{m^2_ic^2+{\vec
\kappa}_i^2(T_s)}u^{\mu}(p_s)+\kappa^r_i(T_s)\epsilon
^{\mu}_r(u(p_s))],\nonumber \\ T^{\mu}{}_{\mu}[x^{\beta}_s(T_s)
+\epsilon^{\beta}_u(u(p_s))\sigma^u]&=& T^A{}_A(T_s,\vec \sigma
)=\nonumber \\
 &=&\sum_{i=1}^N \delta^3(\vec \sigma -{\vec \eta}_i(T_s)) {{m^2_i}\over
{\sqrt{m^2_i+{\vec \kappa}^2_i(T_s)}}}.
 \label{III16}
 \end{eqnarray}

\vfill\eject

\section{Dixon's Multipoles for Free Particles on the Wigner Hyperplane.}

In this Section we  shall define the special relativistic Dixon
multipoles on the Wigner hyperplane  with $T_s - \tau \equiv 0$
for the N-body problem \cite{c24} [see Eqs.(\ref{II13}) with
$x^{\mu}_s(\tau ) = x^{\mu}_o + u^{\mu}(p_s)\, T_s +
\epsilon^{\mu}_r(u(p_s))\, \int_o^{\tau} d\tau_1\,
\lambda_r(\tau_1) = x_s^{({\vec q}_{+})\mu}(\tau ) + \int_o^{\tau}
d\tau_1\, \lambda_r(\tau_1)$]. By comparison, a list of the {\it
non-relativistic multipoles} for N free particles is given in
Appendix A .
\bigskip

Consider an arbitrary time-like world-line $w^{\mu}(\tau
)=z^{\mu}(\tau ,\vec \eta (\tau ))= x^{\mu}_s(\tau ) +
\epsilon^{\mu}_r(u(p_s))\, \eta^r(\tau ) = x_s^{({\vec
q}_{+})\mu}(\tau ) + \epsilon^{\mu}_r(u(p_s))\, {\tilde
\eta}^r(\tau )$ [${\tilde \eta}^r(\tau ) = \eta^r(\tau ) +
\int_o^{\tau} d\tau_1\, \lambda_r(\tau_1)$] and evaluate the Dixon
multipoles \cite{dixon} \cite{c25} on the Wigner hyper-planes in
the natural gauge with respect to the given world-line. A generic
point will be parametrized by

\begin{eqnarray}
z^{\mu}(\tau ,\vec \sigma ) &=& x^{\mu}_s(\tau ) +
\epsilon^{\mu}_r(u(p_s))\, \sigma^r =\nonumber \\
 &=& x_s^{({\vec q}_{+})\mu}(\tau ) +
\epsilon^{\mu}_r(u(p_s))\, \Big[\sigma^r + \int_o^{\tau} d\tau_1\,
\lambda_r(\tau_1)\Big] =\nonumber \\
 &=& w^{\mu}(\tau ) + \epsilon^{\mu}_r(u(p_s)) [\sigma^r-\eta^r(\tau
)]\, {\buildrel {def} \over =}\, w^{\mu}(\tau )+\delta
z^{\mu}(\tau ,\vec \sigma ),
 \label{IV1}
\end{eqnarray}

\noindent so that $\delta z_{\mu}(\tau ,\vec \sigma
)u^{\mu}(p_s)=0$.\medskip

While for ${\vec {\tilde \eta}} (\tau )=0$  [$\vec \eta (\tau ) =
\int_o^{\tau} d\tau_1\, \lambda_r(\tau_1)$] we get the multipoles
relative to the centroid $x^{\mu}_s(\tau )$, for $\vec \eta (\tau
) = 0$ we get those relative to the centroid $x_s^{({\vec
q}_{+})\mu}(\tau )$. In the gauge ${\vec R}_{+} \approx {\vec
q}_{+} \approx {\vec y}_{+} \approx 0$, where $\vec \lambda (\tau
) = 0$, it follows that $\vec \eta (\tau ) = {\vec {\tilde
\eta}}(\tau ) = 0$ identifies the {\it barycentric} multipoles
with respect to the centroid $x_s^{({\vec q}_{+})\mu}(\tau )$,
which now carries the internal 3-center of mass.

\subsection{Dixon's Multipoles.}

Lorentz covariant {\it Dixon's multipoles} and their Wigner
covariant counterparts on the Wigner hyper-planes are then defined
as
\medskip

\begin{eqnarray}
t_T^{\mu_1...\mu_n\mu\nu}(T_s,\vec \eta
)&=&t_T^{(\mu_1...\mu_n)(\mu\nu)}(T_s,\vec \eta )= \nonumber \\
 &&{}\nonumber \\
 &=&\epsilon^{\mu_1}_{r_1}(u(p_s))...\epsilon^{\mu_n}_{r_n}(u(p_s))\,
 \epsilon^{\mu}_A(u(p_s))\epsilon^{\nu}_B(u(p_s)) q_T^{r_1..r_nAB}(T_s,\vec \eta )
 =\nonumber \\
 &&{}\nonumber \\
&=&\int d^3\sigma \delta z^{\mu_1}(T_s,\vec \sigma )...\delta
z^{\mu_n}(T_s,\vec \sigma ) T^{\mu\nu}[x^{({\vec
q}_{+})\beta}_s(T_s)+\epsilon^{\beta}_u(u(p_s))
\sigma^u]=\nonumber \\
&=&\epsilon^{\mu}_A(u(p_s))\epsilon^{\nu}_B(u(p_s)) \int d^3\sigma
\delta z^{\mu_1}(T_s,\vec \sigma )....\delta z^{\mu_n}(T_s,\vec
\sigma ) T^{AB}(T_s,\vec \sigma )=\nonumber \\
 &&{}\nonumber \\
&=&\epsilon^{\mu_1}_{r_1}(u(p_s))...\epsilon^{\mu_n}_{r_n}(u(p_s))
\nonumber \\
 &&\Big[ u^{\mu}(p_s) u^{\nu}(p_s)
\sum_{i=1}^N[\eta^{r_1}_i(T_s)-\eta^{r_1}(T_s)]...[\eta^{r_n}
_i(T_s)-\eta^{r_n}(T_s)]\sqrt{m^2_i+{\vec
\kappa}^2_i(T_s)}+\nonumber \\
 &+&\epsilon^{\mu}_r(u(p_s))\epsilon^{\nu}_s(u(p_s))\nonumber \\
 &&\sum_{i=1}^N[\eta^{r_1}
_i(T_s)-\eta^{r_1}(T_s)]...[\eta^{r_n}_i(T_s)-\eta^{r_n}(T_s)]
{{\kappa^r_i(T_s)\kappa^s_i(T_s)}\over {\sqrt{m_i^2+{\vec
\kappa}^2_i(T_s)}}}+\nonumber \\
&+&[u^{\mu}(p_s)\epsilon^{\nu}_r(u(p_s))+u^{\nu}(p_s)\epsilon^{\mu}_r(u(p_s))]
  \nonumber \\
&&\sum_{i=1}^N[\eta^{r_1}_i(T_s)-\eta^{r_1}(T_s)]...[\eta^{r_n}_i(T_s)-
\eta^{r_n}(T_s)]\kappa^r_i(T_s)\Big] ,\nonumber \\
 &&{}\nonumber \\
 &&{}\nonumber \\
 q_T^{r_1...r_nAB}(T_s,\vec \eta ) &=& \int d^3\sigma\,
 [\sigma^{r_1} - \eta^{r_1}(T_s)] ... [\sigma^{r_n} -
 \eta^{r_n}]\, T^{AB}(T_s, \vec \sigma ) =\nonumber \\
 &=&\delta^A_{\tau}\delta^B_{\tau}
 \sum_{i=1}^N[\eta^{r_1}_i(T_s)-\eta^{r_1}(T_s)]...[\eta^{r_n}
_i(T_s)-\eta^{r_n}(T_s)]\sqrt{m^2_i+{\vec
\kappa}^2_i(T_s)}+\nonumber \\ &+&\delta^A_u\delta^B_v
\sum_{i=1}^N[\eta^{r_1}
_i(T_s)-\eta^{r_1}(T_s)]...[\eta^{r_n}_i(T_s)-\eta^{r_n}(T_s)]
{{\kappa^u_i(T_s)\kappa^v_i(T_s)}\over {\sqrt{m_i^2+{\vec
\kappa}^2_i(T_s)}}}+\nonumber \\
&+&(\delta^A_{\tau}\delta^B_u+\delta^A_u\delta^B_{\tau})
\sum_{i=1}^N[\eta^{r_1}_i(T_s)-\eta^{r_1}(T_s)]...[\eta^{r_n}_i(T_s)-
\eta^{r_n}(T_s)]\kappa^r_i(T_s),\nonumber \\
 &&{}\nonumber \\
u_{\mu_1}(p_s)&& t_T^{\mu_1...\mu_n\mu\nu}(T_s,\vec \eta )=0,
\nonumber \\
 &&{}\nonumber \\
 t_T^{\mu_1...\mu_n\mu}{}_{\mu}(T_s,\vec \eta )
 &{\buildrel {def}
\over =}&\epsilon^{\mu_1}_{r_1}(u(p_s))...\epsilon^{\mu_n}
_{r_n}(u(p_s))q_T^{r_1...r_nA}{}_A(T_s,\vec \eta )=\nonumber \\
 &=&\int d^3\sigma
\delta z^{\mu_1}(\tau ,\vec \sigma )...\delta z^{\mu_n}(\tau ,\vec
\sigma ) T^{\mu}{}_{\mu} [x^{({\vec
q}_{+})\mu}_s(T_s)+\epsilon^{\mu}_u(u(p_s))\sigma^u] =\nonumber \\
&=&\epsilon^{\mu_1}_{r_1}(u(p_s))...\epsilon^{\mu_n}_{r_n}(u(p_s))
\nonumber \\
 &&\sum_{i=1}^N[\eta^{r_1}_i(T_s)-\eta^{r_1}(T_s)]...
[\eta^{r_n}_i(T_s)-\eta^{r_n}(T_s)] {{m^2_i}\over {\sqrt{m^2_i+
{\vec \kappa}_i^2(T_s)}}}=\nonumber \\
 &&{}\nonumber \\
 {\tilde t}_T^{\mu_1...\mu_n}(T_s,\vec \eta )&=&
 t_T^{\mu_1...\mu_n\mu\nu}(T_s,\vec \eta )u_{\mu}(p_s)
u_{\nu}(p_s)=\nonumber \\
&=&\epsilon^{\mu_1}_{r_1}(u(p_s))...\epsilon^{\mu_n}_{r_n}(u(p_s))
q_T^{r_1...r_n\tau \tau}(T_s,\vec \eta )=\nonumber \\
&=&\epsilon^{\mu_1}_{r_1}(u(p_s))...\epsilon^{\mu_n}_{r_n}(u(p_s))
\nonumber \\
 &&\sum_{i=1}^N[\eta^{r_1}_i(T_s)-\eta^{r_1}(T_s)]...[\eta^{r_n}
_i(T_s)-\eta^{r_n}(T_s)]\sqrt{m^2_i+{\vec \kappa}^2_i(T_s)}.
\label{IV2}
\end{eqnarray}

\medskip

\noindent  Related multipoles are
\medskip

\begin{eqnarray}
p_T^{\mu_1..\mu_n\mu}(T_s,\vec \eta )&=&
 t_T^{\mu_1...\mu_n\mu\nu}(T_s,\vec \eta ) u_{\nu}(p_s)= \nonumber \\
 &&{}\nonumber \\
 &=&\epsilon^{\mu_1}_{r_1}(u(p_s))...\epsilon^{\mu_n}_{r_n}(u(p_s))
 \epsilon^{\mu}_A(u(p_s)) q_T^{r_1...r_nA\tau}(T_s,\vec \eta )=\nonumber \\
 &&{}\nonumber \\
 &=&\epsilon^{\mu_1}_{r_1}(u(p_s)).... \epsilon^{\mu_n}_{r_n}(u(p_s))
   \nonumber \\
 &&\sum_{i=1}^N [\eta^{r_1}_i(T_s)-\eta^{r_1}(T_s)]...[\eta^{r_n}_i(T_s)-
 \eta^{r_n}(T_s)]\nonumber \\
&&\Big[\sqrt{m_i^2+{\vec \kappa}^2_i(\tau )}u^{\mu}(p_s)+
k^r_i(T_s) \epsilon^{\mu}_r(u(p_s))\Big] ,\nonumber \\
 &&{}\nonumber \\
&&u_{\mu_1}(p_s) p_T^{\mu_1...\mu_n\mu}(T_s,\vec \eta
)=0,\nonumber \\
 &&{}\nonumber \\
 &&p_T^{\mu_1...\mu_n\mu}(T_s,\vec \eta
)u_{\mu}(p_s)={\tilde t}_T^{\mu_1...\mu_n}(T_s,\vec \eta
),\nonumber \\
 &&{}\nonumber \\
n=0&&\Rightarrow p^{\mu}_T(T_s,\vec \eta
)=\epsilon^{\mu}_A(u(p_s)) q_T^{A\tau}(T_s)=P^{\mu}_T\approx
p^{\mu}_s.
 \label{IV3}
 \end{eqnarray}

\medskip

The inverse formulas, giving the {\it multipolar expansion},  are
\medskip

\begin{eqnarray}
T^{\mu\nu}[w^{\beta}(T_s)+\delta z^{\beta}(T_s,\vec \sigma )]&=&
T^{\mu\nu}[x^{({\vec q}_{+})\beta}_s(T_s) +
\epsilon^{\beta}_r(u(p_s))\, \sigma^r]=\nonumber \\
 &&{}\nonumber \\
 &=&\epsilon^{\mu}_A(u(p_s))\epsilon^{\nu}_B(u(p_s)) T^{AB}(T_s,\vec \sigma )
 =\nonumber \\
  &&{}\nonumber \\
  &=&\epsilon^{\mu}_A(u(p_s))\epsilon^{\nu}_B(u(p_s)) \sum_{n=0}^{\infty}
  (-1)^n\, {{ q_T^{r_1...r_nAB}(T_s,\vec \eta )}\over {n!}} \nonumber \\
  &&{{\partial^n}\over
  {\partial \sigma^{r_1}...\partial \sigma^{r_n}}} \delta^3(\vec \sigma -
  \vec \eta (T_s))=\nonumber \\
 &&{}\nonumber \\
  &=&\sum_{n=0}^{\infty} (-1)^n\, {{t_T^{\mu_1...\mu_n\mu\nu}(T_s,\vec \eta )}\over
  {n!}} \epsilon_{r_1\mu_1}(u(p_s))...\epsilon_{r_n\mu_n}(u(p_s))
  \nonumber \\
  &&{{\partial^n}\over
  {\partial \sigma^{r_1}...\partial \sigma^{r_n}}} \delta^3(\vec \sigma -
  \vec \eta (T_s)).
 \label{IV4}
\end{eqnarray}
\medskip

Note however that, as pointed out by Dixon \cite{dixon}, the
distributional equation (\ref{IV4}) is valid only if analytic test
functions are used, defined on the support of the energy-momentum
tensor.

\bigskip

The quantities $q_T^{r_1...r_n\tau\tau}(T_s,\vec \eta )$,
$q_T^{r_1...r_n r\tau}(T_s,\vec \eta )=q_T^{r_1...r_n \tau
r}(T_s,\vec \eta )$, $q_T^{r_1...r_n uv}(T_s,\vec \eta )$ are the
{\it mass density, momentum density and stress tensor multipoles}
with respect to the world-line $w^{\mu}(T_s)$  (barycentric for
$\vec \eta = {\vec {\tilde \eta}} = 0$).

\subsection{Monopoles.}

The {\it monopoles} correspond to $n=0$ \cite{c26} and have the
following expression \cite{c27} (see Appendix C for the definition
of $\rightarrow_{\alpha\rightarrow\infty}$)
\medskip

\begin{eqnarray}
q^{AB}_T(T_s,\vec \eta )&=&\delta^A_{\tau}\delta^B_{\tau} M
+\delta^A_u\delta^B_v\sum_{i=1}^N{{\kappa^u_i\kappa^v_i}\over
{\sqrt{m_i^2+{\vec \kappa}_i^2}}}+
(\delta^A_{\tau}\delta^B_u+\delta^A_u\delta^B_{\tau}) \kappa^u_{+}
\approx \nonumber \\ &{\rightarrow}_{\alpha \rightarrow \infty}\,&
\delta^A_{\tau}\delta^B_{\tau}  \sum_{i=1}^N
 \sqrt{m_i^2+N\sum_{de}\gamma_{di}\gamma_{ei}{\vec \pi}_{qd}\cdot {\vec \pi}_{qe}}
+\nonumber \\
 &+&\delta^A_u\delta^B_v N \sum_{i=1}^N {{ \sum_{ab}^{1..N-1}\gamma_{ai}
 \gamma_{bi} {\vec \pi}_{qa}\cdot{\vec \pi}_{qb}}\over
 {\sqrt{m_i^2+N\sum_{de}\gamma_{di}\gamma_{ei}{\vec \pi}_{qd}\cdot {\vec \pi}_{qe}}}},
 \nonumber \\
 &&{}\nonumber \\
 &&{}\nonumber \\
 q^{\tau\tau}_T(T_s,\vec \eta )\, &{\rightarrow}_{c \rightarrow \infty}\,& \sum_{i=1}^N
 m_ic^2 +{1\over 2}\sum_{ab}^{1..N-1}\sum_{i=1}^N{{N\gamma_{ai}\gamma_{bi}}\over {m_i}}
 {\vec \pi}_{qa}\cdot {\vec \pi}_{qb} +O(1/c)=\nonumber \\
 &=&\sum_{i=1}^Nm_ic^2+ H_{rel,nr} +O(1/c),\nonumber \\
 &&{}\nonumber \\
 q_T^{r\tau}(T_s,\vec \eta ) &=&\kappa^r_{+} \approx 0,\quad rest-frame\, condition\,
 (also\, at\, the \, non-relativistic\, level),\nonumber \\
 &&{}\nonumber \\
 q^{uv}_T(T_s,\vec \eta )\, &{\rightarrow}_{c \rightarrow \infty}\,& \sum_{ab}^{1..N-1}
 \sum_{i=1}^N {{N\gamma_{ai}\gamma_{bi}}\over {m_i}} \pi^u_{qa}\pi^v_{qb} +O(1/c)
 =\nonumber \\
 &=& \sum_{ab}^{1..N-1} k^{-1}_{ab} \pi^u_{qa}\pi^v_{qb} +O(1/c) =\sum_{ab}^{1..N-1}
 k_{ab} {\dot \rho}^u_a{\dot \rho}^v_b +O(1/c), \nonumber \\
 &&{}\nonumber \\
 q^A_{T A}(T_s,\vec \eta )&=& t^{\mu}_{T \mu}(T_s,\vec \eta )=
 \sum_{i=1}^N {{m^2_i}\over {\sqrt{m_i^2+
 {\vec \kappa}_i^2}}}\nonumber \\
 &{\rightarrow}_{\alpha \rightarrow \infty}\,&
 \sum_{i=1}^N{{m_i^2}\over
 {\sqrt{m_i^2+N\sum_{de}\gamma_{di}\gamma_{ei}{\vec \pi}_{qd}\cdot {\vec \pi}_{qe}}}}
 \nonumber \\
 &{\rightarrow}_{c \rightarrow \infty}\,& \sum_{i=1}^N m_ic^2 -{1\over 2}\sum_{ab}^{1..N-1}
 \sum_{i=1}^N {{N\gamma_{ai} \gamma_{bi}}\over {m_i}} {\vec \pi}_{qa}\cdot {\vec \pi}_{qb} +O(1/c)
 =\nonumber \\
 &=&\sum_{i=1}^Nm_ic^2 -H_{rel,nr} +O(1/c).
 \label{IV5}
\end{eqnarray}

\noindent where we have exploited Eqs. (5.10), (5.11) of
Ref.\cite{iten1} to obtain the expression in terms of the internal
relative variables.
\medskip

Therefore, independently of the choice of the world-line
$w^{\mu}(\tau )$, in the rest-frame instant form the {\it mass
monopole} $q^{\tau\tau}_T$ is the invariant mass
$M=\sum_{i=1}^N\sqrt{m_i^2+{\vec \kappa}_i^2}$, while the {\it
momentum monopole} $q^{r\tau}_T$ vanishes and $q^{uv}_T$ is the
{\it stress tensor monopole}.

\subsection{Dipoles.}

The mass, momentum and stress tensor {\it dipoles} correspond to
$n=1$ \cite{c28}\medskip

\begin{eqnarray}
q_T^{rAB}(T_s,\vec \eta )&=&\delta^A_{\tau}\delta^B_{\tau} M
[R^r_{+}(T_s)-\eta^r(T_s)] +\delta^A_u\delta^B_v\Big[
\sum_{i=1}^N{{\eta_i^r\kappa^u_i\kappa^v_i}\over {\sqrt{m_i^2+
{\vec \kappa}_i^2}}}(T_s)-\eta^r(T_s)q_T^{uv}(T_s,\vec \eta )\Big]
+ \nonumber \\
&+&(\delta^A_{\tau}\delta^B_u+\delta^A_u\delta^B_{\tau}) \Big[
\sum_{i=1}^N[\eta^r_i\kappa_i^u](T_s)-\eta^r(T_s)\kappa^u_{+}\Big]
, \nonumber \\
 &&{}\nonumber \\
 q_T^{rA}{}_A(T_s,\vec \eta )&=&\epsilon^{r_1}_{\mu_1}(u(p_s))
 t_T^{\mu_1\mu}{}_{\mu}(T_s,\vec \eta )=\sum_{i=1}^N {{[\eta^r_i-\eta^r]
 m_i^2}\over {\sqrt{m_i^2+{\vec \kappa}_i^2}}}(T_s).
 \label{IV6}
\end{eqnarray}
\medskip

The vanishing of the {\it mass dipole} $q^{r\tau\tau}_T$ implies
$\vec \eta (\tau ) = {\vec {\tilde \eta}}(\tau ) -
\int_o^{\tau}d\tau_1\, \vec \lambda (\tau_1) = {\vec R}_{+}$ and
identifies the world-line $w^{\mu}(\tau ) = x^{({\vec q}_{+})
\mu}_s(\tau ) + \epsilon^{\mu}_r(u(p_s))\, \Big[ R^r_{+} +
\int_o^{\tau}d\tau_1\, \lambda_r(\tau_1)\Big]$. In the gauge
${\vec R}_{+} \approx {\vec q}_{+} \approx {\vec y}_{+} \approx
0$, where $\vec \lambda (\tau ) = 0$, this is the world-line
$w^{\mu}(\tau ) = x^{({\vec q}_{+}) \mu}_s(\tau )$ of the centroid
associated with the {\it internal M$\o$ller 3-center of energy}
and, as a consequence of the rest frame condition, also with the
{\it rest-frame internal 3-center of mass} ${\vec q}_{+}$.
Therefore we have the implications following from the vanishing of
the barycentric (i.e. $\vec \lambda (\tau ) = 0$) mass dipole
\medskip

\begin{eqnarray}
q_T^{r\tau\tau}(T_s,\vec \eta )&=&\epsilon^{r_1}_{\mu_1}(u(p_s))
{\tilde t}_T^{\mu_1}(T_s,\vec \eta )= M\, \Big[ R^r_{+}(T_s) -
\eta^r(T_s)\Big] = 0,\quad and\,\, \vec \lambda (\tau ) =
0,\nonumber \\
 &&{}\nonumber \\
 &&\Rightarrow \vec \eta (T_s)= {\vec {\tilde \eta}}(T_s) = {\vec R}_{+}
 \approx {\vec q}_{+} \approx {\vec y}_{+}.
 \label{IV7}
 \end{eqnarray}
\medskip

In the gauge ${\vec R}_{+} \approx {\vec q}_{+} \approx {\vec
y}_{+} \approx 0$,  Eq.(\ref{IV7}) with $\vec \eta = {\vec {\tilde
\eta}} = 0$ implies the vanishing of the time derivative of the
barycentric mass dipole: this identifies the {\it center-of-mass
momentum-velocity relation} (or constitutive equation) for the
system

\begin{equation}
{{d q_T^{r\tau\tau}(T_s,\vec \eta )}\over {dT_s}}\, {\buildrel
\circ \over =}\, \kappa^r_{+}-M {\dot R}^r_{+}\, = 0.
 \label{IV8}
\end{equation}

\medskip

The expression of the barycentric dipoles in terms of the internal
relative variables, when $\vec \eta = {\vec {\tilde \eta}} = {\vec
R}_{+} \approx {\vec q}_{+} \approx 0$ and $\vec{\kappa}_+\approx
0$, is obtained by using the results of Appendix C.
\medskip

\begin{eqnarray}
q_T^{r\tau\tau}(T_s,{\vec R}_{+})&=&0,\nonumber \\
 &&{}\nonumber \\
 q_T^{ru\tau}(T_s,{\vec R}_{+})&=&\sum_{i=1}^N\eta_i^r\kappa_i^u-R_{+}^r\kappa_{+}^u=
 \sum_{a=1}^{N-1} \rho_a^r\pi^u_a+(\eta^r_{+}-R^r_{+})\kappa^u_{+}\nonumber \\
 &{\rightarrow}_{\alpha \rightarrow \infty}\,& \sum_{a=1}^{N-1} \rho^r_{qa}
 \pi^u_{qa}\nonumber \\
 &{\rightarrow}_{c \rightarrow \infty}\,& \sum_{a=1}^{N-1} \rho^r_a\pi^u_{qa} =
 \sum_{ab}^{1..N-1} k_{ab} \rho^r_a{\dot \rho}^u_b,\nonumber \\
 &&{}\nonumber \\
 q_T^{ruv}(T_s,{\vec R}_{+})&=& \sum_{i=1}^N \eta^r_i {{\kappa_i^u\kappa_i^v}\over {H_i}}-
 R^r_{+} \sum_{i=1}^N {{\kappa_i^u\kappa_i^v}\over {H_i}}=\nonumber \\
 &=&{1\over {\sqrt{N}}} \sum_{i=1}^N \sum_{a=1}^{N-1} \gamma_{ai} \rho^r_a
 {{\kappa^u_i\kappa_i^v}\over {H_i}}+(\eta^r_{+}-R^r_{+})\sum_{i=1}^N
 {{\kappa_i^u\kappa_i^v}\over {H_i}}\nonumber \\
 &{\rightarrow}_{\alpha \rightarrow \infty}\,& \sum_{a=1}^{N-1}
 \Big( c\sqrt{N}\sum_{ij}^{1..N}(\gamma_{ai}-\gamma_{aj})
{{ \sqrt{m_j^2+N\sum_{de}\gamma_{dj}\gamma_{ej}{\vec
\pi}_{qd}\cdot {\vec \pi}_{qe}} }\over
{\sqrt{m_i^2+N\sum_{de}\gamma_{di}\gamma_{ei}{\vec \pi}_{qd}\cdot
{\vec \pi}_{qe}}}}  \times \nonumber \\
 &&{{ \sum_{bc}^{1..N-1} \gamma_{bi}\gamma_{ci} \pi_{qb}^u \pi^v_{qc}}\over
 {\sum_{k=1}^N\sqrt{m_k^2+N\sum_{de}\gamma_{dk}\gamma_{ek}
 {\vec \pi}_{qd}\cdot {\vec \pi}_{qe}} }} \Big) \rho^r_{qa}\nonumber \\
&{\rightarrow}_{c \rightarrow
\infty}\,&\sum_{ij}^{1..N}\sum_{a=1}^{N-1}
{{\gamma_{ai}-\gamma_{aj}}\over {\sqrt{N}}} \rho^r_a {{m_jN}\over
{m_im}}
\sum_{bc}^{1..N-1}\gamma_{bi}\gamma_{ci}\pi^u_{qb}\pi^v_{qc}+O(1/c)
=\nonumber \\ &=&{1\over {\sqrt{N}}}\sum_{abc}^{1..N-1}\Big[
N\sum_{i=1}^N{{\gamma_{ai}\gamma_{bi}\gamma_{ci}}\over
{m_i}}-{{\sum_{j=1}^Nm_j\gamma_{aj}}\over m}\Big]
\rho^r_a\pi^u_{qb}\pi^v_{qc}+O(1/c),\nonumber \\
 &&{}\nonumber \\
 q_T^{rA}{}_A(T_s,{\vec R}_{+})&=& \sum_{i=1}^N (\eta_i^r-R^r_{+}){{m_i^2}\over
 {H_i}}=\nonumber \\
 &{\rightarrow}_{\alpha \rightarrow \infty}\,& \sum_{a=1}^{N-1}\Big( {1\over
 {\sqrt{N}}} \sum_{ij}^{1..N}(\gamma_{ai}-\gamma_{aj})
 {{\sqrt{m_j^2+N\sum_{de}\gamma_{dj}\gamma_{ej}{\vec \pi}_{qd}\cdot {\vec \pi}_{qe}}}\over
 {\sqrt{m_i^2+N\sum_{de}\gamma_{di}\gamma_{ei}{\vec \pi}_{qd}\cdot {\vec \pi}_{qe}}}}
 \times \nonumber \\
 &&{{m_i}\over {\sum_{k=1}^N
 \sqrt{m_k^2+N\sum_{de}\gamma_{dk}\gamma_{ek}{\vec \pi}_{qd}\cdot {\vec \pi}_{qe}}}}
 \Big) \rho^r_a\nonumber \\
 &{\rightarrow}_{c \rightarrow \infty}\,&\sum_{ij}^{1..N}\sum_{a=1}^{N-1}\sqrt{N}
 (\gamma_{ai}-\gamma_{aj}) \rho^r_a {{m_jN}\over {m_im}}\sum_{bc}^{1..N-1}\gamma_{bi}\gamma_{ci}
 {\vec \pi}_{qb}\cdot {\vec \pi}_{qc}+O(1/c)=\nonumber \\
 &=&\sqrt{N}\sum_{abc}^{1..N-1}\Big[
N\sum_{i=1}^N{{\gamma_{ai}\gamma_{bi}\gamma_{ci}}\over
{m_i}}-{{\sum_{j=1}^Nm_j\gamma_{aj}}\over m}\Big] \rho^r_a {\vec
\pi}_{qb}\cdot {\vec \pi}_{qc} +O(1/c).
 \label{IV9}
\end{eqnarray}

\medskip

The antisymmetric part of the related dipole
$p_T^{\mu_1\mu}(T_s,\vec \eta )$ identifies the {\it spin tensor}.
Indeed, the {\it spin dipole} \cite{c29} is
\medskip

\begin{eqnarray}
S^{\mu\nu}_T(T_s)[\vec \eta]&=&2 p_T^{[\mu\nu ]}(T_s,\vec \eta ) =
2 \epsilon^{[\mu}_r(u(p_s))\, \epsilon^{\nu]}_A(u(p_s))\,
q_T^{rA\tau}(T_s, \vec \eta ) = \nonumber \\
 &=&M [R^r_{+}(T_s)-\eta^r(T_s)]
\Big[
\epsilon^{\mu}_r(u(p_s))u^{\nu}(p_s)-\epsilon^{\nu}_r(u(p_s))u^{\mu}(p_s)
\Big] +\nonumber \\
&+&\sum_{i=1}^N[\eta^r_i(T_s)-\eta^r(T_s)]\kappa^s_i(T_s)
\Big[\epsilon^{\mu}_r(u(p_s))
\epsilon^{\nu}_s(u(p_s))-\epsilon^{\nu}_r(u(p_s))\epsilon^{\mu}_s(u(p_s))
\Big],\nonumber \\
 &&{}\nonumber \\
 m^{\nu}_{u(p_s)}(T_s,\vec \eta )&=& u_{\mu}(p_s)
S^{\mu\nu}_T(T_s)[\vec \eta ]=-\epsilon^{\nu}_r(u(p_s)) [{\bar
S}^{\tau r}_s-M\eta^r(T_s)]= \nonumber \\
&=&-\epsilon^{\nu}_r(u(p_s)) M[R^r_{+}(T_s)-\eta^r(T_s)]=-
 \epsilon^{\nu}_r(u(p_s)) q_T^{r\tau\tau}(T_s,\vec \eta ),\nonumber \\
 &&{}\nonumber \\
 \Rightarrow&& u_{\mu}(p_s) S^{\mu\nu}_T(T_s)[\vec \eta ]=0,\quad\quad
 \Rightarrow \vec \eta =\vec{R}_+,\nonumber \\
 &&{}\nonumber \\
 &&\Downarrow \quad {\it barycentric\, spin}\,\,
 for\,\, \vec \eta = {\vec {\tilde \eta}} = 0,\,\, see\, Eq(2.9),\nonumber \\
 &&{}\nonumber \\
S^{\mu\nu}_T(T_s)[\vec \eta =0]&=&S^{\mu\nu}_s\, {\buildrel \circ
\over =}\nonumber \\
 &{\buildrel \circ \over =}\,&
\sum_{i=1}^N{{m_i\eta^r_i(T_s)}\over {\sqrt{1-{\dot {\vec \eta}}
_i^2(T_s)}}}\Big[
\epsilon^{\mu}_r(u(p_s))u^{\nu}(p_s)-\epsilon^{\nu}
_r(u(p_s))u^{\mu}(p_s)\Big] +\nonumber \\
 &+&\sum_{i=1}^N{{m_i\eta^r_i(T_s){\dot \eta}^s_i(T_s)}\over
{\sqrt{1-{\dot {\vec
\eta}}_i^2(T_s)}}}\Big[\epsilon^{\mu}_r(u(p_s))
\epsilon^{\nu}_s(u(p_s))-\epsilon^{\nu}_r(u(p_s))\epsilon^{\mu}_s(u(p_s))
\Big] \cir\nonumber \\
 &\cir& \sum_i\, \eta^r_i(T_s)\, \sqrt{m^2_i + {\vec \kappa}_i^2}\,
\Big[ \epsilon^{\mu}_r(u(p_s))u^{\nu}(p_s)-\epsilon^{\nu}
_r(u(p_s))u^{\mu}(p_s)\Big] +\nonumber \\
 &+& \epsilon^{rsu}\, {\bar S}^u_s\, \epsilon^{\mu}_r(u(p_s))
\epsilon^{\nu}_s(u(p_s)).
 \label{IV10}
\end{eqnarray}
\medskip

This explains why $m^{\mu}_{u(p_s)}(T_s,\vec \eta )$ is also
called the {\it mass dipole moment}.
\medskip

We find, therefore, that in the gauge ${\vec R}_{+} \approx {\vec
q}_{+} \approx {\vec y}_{+} \approx 0$ with  $ P^{\mu}_T = M\,
u^{\mu}(p_s) = M\, {\dot x}_s^{({\vec q} _{+})\mu}(T_s)$ the
M$\o$ller and barycentric centroid $x^{({\vec q}_{+})\mu}_s(T_s)$
is simultaneously the {\it Tulczyjew
centroid}\cite{mul4,ehlers,mul8,c30} (defined by $S^{\mu\nu}\,
P_{\nu} =0$) and also the {\it Pirani centroid}\cite{pirani,c31}
(defined by $S^{\mu\nu}\, {\dot x}^{({\vec q} _{+})}_{s \nu} =0$).
In general, lacking a relation between 4-momentum and 4-velocity,
they are different centroids \cite{c32}.
\medskip

Note that non-covariant centroids could also be connected with the
non-covariant external center of mass ${\tilde x}^{\mu}_s$ and the
non-covariant external M\o ller center of energy.

\subsection{Quadrupoles and the Barycentric Tensor of Inertia.}

The {\it quadrupoles} correspond to $n=2$ \cite{c33}\medskip

\begin{eqnarray}
q_T^{r_1r_2AB}(T_s,\vec \eta )&=&\delta^A_{\tau}\delta^B_{\tau}
\sum_{i=1}^N[\eta_i^{r_1}(T_s)-\eta^{r_1}(T_s)][\eta_i^{r_2}(T_s)-\eta^{r_2}(T_s)]
\sqrt{m_i^2+{\vec \kappa}_i^2(T_s)}+\nonumber \\
&+&\delta^A_u\delta^B_v
\sum_{i=1}^N[\eta_i^{r_1}(T_s)-\eta^{r_1}(T_s)][\eta_i^{r_2}(T_s)-\eta^{r_2}(T_s)]
{{\kappa_i^u\kappa_i^v}\over {\sqrt{m_i^2+{\vec
\kappa}_i^2}}}(T_s)+\nonumber \\
&+&(\delta^A_{\tau}\delta^B_u+\delta^A_u\delta^B_{\tau})
\sum_{i=1}^N[\eta_i^{r_1}(T_s)-\eta^{r_1}(T_s)][\eta_i^{r_2}(T_s)-\eta^{r_2}(T_s)]
\kappa_i^u(T_s),\nonumber
\end{eqnarray}

\noindent and when the mass dipole vanishes, i.e.
$\vec{\eta}=\vec{R}_+ = \sum_i\, {\vec \eta}_i\, \sqrt{m_i^2 +
{\vec \kappa}_i^2} / M$, we get

\begin{eqnarray}
 &&{}\nonumber \\
 q_T^{r_1r_2\tau\tau}(T_s,{\vec R}_{+})&=&\sum_{i=1}^N(\eta_i^{r_1}-R^{r_1}_{+})(\eta_i^{r_2}-
 R^{r_2}_{+}) \sqrt{m_i^2+{\vec \kappa}_i^2(T_s)},\nonumber \\
 &&{}\nonumber \\
  q_T^{r_1r_2u\tau}(T_s,{\vec R}_{+})&=&\sum_{i=1}^N (\eta_i^{r_1}-R^{r_1}_{+})
 (\eta_i^{r_2}-R^{r_2}_{+}) \kappa_i^u,\nonumber \\
 &&{}\nonumber \\
 q_T^{r_1r_2uv}(T_s,{\vec R}_{+})&=& \sum_{i=1}^N(\eta_i^{r_1}-R^{r_1}_{+})(\eta_i^{r_2}-
 R^{r_2}_{+}) {{\kappa_i^u\kappa_i^v}\over {\sqrt{m_i^2+{\vec \kappa}_i^2(T_s)}}}\nonumber \\
 &=&{1\over N} \sum_{ijk}^{1..N}\sum_{ab}^{1..N-1}(\gamma_{ai}-\gamma_{aj}).
 \label{IV11}
 \end{eqnarray}
\medskip

Following the non-relativistic pattern, Dixon starts from the {\it
mass quadrupole}
\medskip

\begin{equation}
q_T^{r_1r_2\tau\tau}(T_s,{\vec R}_{+} )=
\sum_{i=1}^N[\eta_i^{r_1}\eta_i^{r_2}\sqrt{m_i^2+{\vec
\kappa}_i^2}](T_s) - M\, R_{+}^{r_1}\, R^{r_2}_{+},
 \label{IV12}
\end{equation}

\noindent and defines the following {\it barycentric tensor of
inertia}

\begin{eqnarray}
I_{dixon}^{r_1r_2}(T_s)&=&\delta^{r_1r_2} \sum_u
q_T^{uu\tau\tau}(T_s,\vec{R}_+)-
q_T^{r_1r_2\tau\tau}(T_s,\vec{R}_+)=\nonumber \\
&=&\sum_{i=1}^N[(\delta^{r_1r_2} ({\vec
\eta}_i-\vec{R}_+)^2-(\eta_i^{r_1}-R_+^{r_1})
(\eta_i^{r_2}-R_+^{r_2}))\sqrt{m_i^2+{\vec \kappa}_i^2}](T_s)
\nonumber \\ &{\rightarrow}_{\alpha \rightarrow
\infty}\,&\sum_{ab}^{1..N-1}\Big( {1\over N}
\sum_{ijk}^{1..N}(\gamma_{ai}-\gamma_{aj})(\gamma_{bi}-\gamma_{bk})\nonumber
\\ &&{{ \sqrt{m_i^2+N\sum_{de}\gamma_{di}\gamma_{ei}{\vec
\pi}_{qd}\cdot {\vec \pi}_{qe}}
 }\over {(\sum_{h=1}^N
\sqrt{m_h^2+N\sum_{de}\gamma_{dh}\gamma_{eh}{\vec \pi}_{qd}\cdot
{\vec \pi}_{qe}})^2}}
 \nonumber \\
&&\sqrt{m_j^2+N\sum_{de}\gamma_{dj}\gamma_{ej}{\vec \pi}_{qd}\cdot
{\vec \pi}_{qe}} \sqrt{m_k^2+N\sum_{de}\gamma_{dk}\gamma_{ek}{\vec
\pi}_{qd}\cdot {\vec \pi}_{qe}}   \Big) \nonumber \\
 &&[{\vec \rho}_{qa}\cdot {\vec \rho}_{qb}
\delta^{r_1r_2}-\rho_{qa}^{r_1}\rho_{qb}^{r_2}]\nonumber \\
&{\rightarrow}_{c \rightarrow \infty}\,&
\sum_{ab}^{1..N-1}\sum_{ijk}^{1..N}{{m_im_jm_k}\over {Nm^2}}
(\gamma_{ai}-\gamma_{aj})(\gamma_{bi}-\gamma_{bk}) {\vec
\rho}_{qa}\cdot {\vec \rho}_{qb}
 \delta^{r_1r_2}-\rho_{qa}^{r_1}\rho_{qb}^{r_2}]\times \nonumber \\
 &&\times \Big[ 1+{1\over c} \Big( {{N\sum_{cd}^{1..N-1}\gamma_{ci}\gamma_{di}
 {\vec \pi}_{qc}\cdot {\vec \pi}_{qd}}\over {2m_i^2}}+{{N\sum_{cd}^{1..N-1}
 \gamma_{cj}\gamma_{dj}{\vec \pi}_{qc}\cdot {\vec \pi}_{qd}}\over {2m_j^2}}+
 \nonumber \\
 &+&{{N\sum_{cd}^{1..N-1}\gamma_{ck}\gamma_{dk}{\vec \pi}_{qc}\cdot
 {\vec \pi}_{qd}}\over {2m_k^2}} -{1\over m}\sum_{h=1}^N{{N\sum_{cd}^{1..N-1}
 \gamma_{ch}\gamma_{dh}{\vec \pi}_{qc}\cdot {\vec \pi}_{qd}}\over {m_h}}\Big) +O(1/c^2)\Big]
 =\nonumber \\
 &=&\sum_{ab}^{1..N-1} k_{ab} [ {\vec \rho}_{qa}\cdot {\vec \rho}_{qb}
 \delta^{r_1r_2}-\rho_{qa}^{r_1}\rho_{qb}^{r_2}] +O(1/c)=\nonumber \\
 &=& I^{r_1r_2}[{\vec q}_{nr}] + O(1/c).
\label{IV13}
\end{eqnarray}

Note that in the non-relativistic limit we recover the {\it tensor
of inertia} of Eqs.(\ref{a11}).

\bigskip

On the other hand, Thorne's definition of {\it barycentric tensor
of inertia}\cite{thorne} is

\begin{eqnarray}
I_{thorne}^{r_1r_2}(T_s)&=&\delta^{r_1r_2} \sum_u
q_T^{uuA}{}_A(T_s,\vec{R}_+)-
q_T^{r_1r_2A}{}_A(T_s,\vec{R}_+)=\nonumber \\
 &=& \sum_{i=1}^N {{m_i^2 (\delta^{r_1r_2}{(\vec \eta}_i-\vec{R}_+)^2-(\eta_i^{r_1}-R_+^{r_1})
(\eta_i^{r_2}-R_+^{r_2}))}\over {\sqrt{m_i^2+{\vec
\kappa}_i^2}}}(T_s)\nonumber \\
 &{\rightarrow}_{\alpha \rightarrow \infty}\,&\sum_{ab}^{1..N-1}
 \Big( {c\over N} \sum_{ijk}^{1..N}(\gamma_{ai}-\gamma_{aj})
 (\gamma_{bi}-\gamma_{bk}) \nonumber \\
 &&{{ m_i^2
 \sqrt{m_j^2+N\sum_{de}\gamma_{dj}\gamma_{ej}{\vec \pi}_{qd}\cdot {\vec \pi}_{qe}}
 \sqrt{m_k^2+N\sum_{de}\gamma_{dk}\gamma_{ek}{\vec \pi}_{qd}\cdot {\vec \pi}_{qe}} }\over
 { \sqrt{m_i^2+N\sum_{de}\gamma_{di}\gamma_{ei}{\vec \pi}_{qd}\cdot {\vec \pi}_{qe}}
 (\sum_{h=1}^N\sqrt{m_h^2+N\sum_{de}\gamma_{dh}\gamma_{eh}
 {\vec \pi}_{qd}\cdot {\vec \pi}_{qe}})^2}} \Big) \nonumber \\
 &&[{\vec \rho}_{qa}\cdot {\vec \rho}_{qb}
 \delta^{r_1r_2}-\rho_{qa}^{r_1}\rho_{qb}^{r_2}]\nonumber \\
 &{\rightarrow}_{c \rightarrow \infty}\,&
\sum_{ab}^{1..N-1}\sum_{ijk}^{1..N}{{m_im_jm_k}\over {Nm^2}}
(\gamma_{ai}-\gamma_{aj})(\gamma_{bi}-\gamma_{bk}) {\vec
\rho}_{qa}\cdot {\vec \rho}_{qb}
 \delta^{r_1r_2}-\rho_{qa}^{r_1}\rho_{qb}^{r_2}]\times \nonumber \\
 &&\times \Big[ 1+{1\over c} \Big( -{{N\sum_{cd}^{1..N-1}\gamma_{ci}\gamma_{di}
 {\vec \pi}_{qc}\cdot {\vec \pi}_{qd}}\over {2m_i^2}}+{{N\sum_{cd}^{1..N-1}
 \gamma_{cj}\gamma_{dj}{\vec \pi}_{qc}\cdot {\vec \pi}_{qd}}\over {2m_j^2}}+
 \nonumber \\
 &+&{{N\sum_{cd}^{1..N-1}\gamma_{ck}\gamma_{dk}{\vec \pi}_{qc}\cdot
 {\vec \pi}_{qd}}\over {2m_k^2}} -{1\over m}\sum_{h=1}^N{{N\sum_{cd}^{1..N-1}
 \gamma_{ch}\gamma_{dh}{\vec \pi}_{qc}\cdot {\vec \pi}_{qd}}\over {m_h}}\Big) +O(1/c^2)\Big]
 =\nonumber \\
 &=&\sum_{ab}^{1..N-1} k_{ab} [ {\vec \rho}_{qa}\cdot {\vec \rho}_{qb}
 \delta^{r_1r_2}-\rho_{qa}^{r_1}\rho_{qb}^{r_2}] +O(1/c)=\nonumber \\
 &=& I^{r_1r_2}[{\vec q}_{nr}] + O(1/c).
\label{IV14}
\end{eqnarray}

In this case too we recover the {\it tensor of inertia} of
Eq.(\ref{a11}).
\medskip

Note that the Dixon and Thorne barycentric tensors of inertia
differ at the post-Newtonian level\medskip

\begin{eqnarray*}
I^{r_1r_2}_{dixon}(T_s)-I^{r_1r_2}_{thorne}(T_s)&=& {1\over c}
\sum_{ab}^{1..N-1}\sum_{ijk}^{1..N}{{m_jm_k}\over {Nm^2}}
(\gamma_{ai}-\gamma_{aj})(\gamma_{bi}-\gamma_{bk}) \nonumber \\
 &&\Big[{\vec \rho}_{qa}\cdot {\vec \rho}_{qb}
 \delta^{r_1r_2}-\rho_{qa}^{r_1}\rho_{qb}^{r_2}\Big] {{N\sum_{cd}^{1..N-1}
 \gamma_{ci}\gamma_{di}{\vec \pi}_{qc}\cdot {\vec \pi}_{qd}}\over {m_i}}+O(1/c^2).
 \end{eqnarray*}

\subsection{The Multipolar Expansion.}

By further using the types of Dixon's multipoles analyzed in
Appendix D as well as the consequences of Hamilton equations for
an isolated system (equivalent to $\partial_{\mu}T^{\mu\nu}\,
{\buildrel \circ \over =}\, 0$), it turns out that the {\it
multipolar expansion} (\ref{IV4}) can be rearranged [see
Eqs.(\ref{d11})] in the following form
\medskip

\begin{eqnarray}
&&T^{\mu\nu}[ x_s^{({\vec q}_{+})
\beta}(T_s)+\epsilon^{\beta}_r(u(p_s))\sigma^r ]= T^{\mu\nu}[
w^{\beta}(T_s)+\epsilon^{\beta}_r(u(p_s)) (\sigma^r-\eta^r(T_s))
]=\nonumber \\
 &&{}\nonumber \\
 &=& u^{(\mu}(p_s) \epsilon^{\nu )}_A(u(p_s)) [\delta^A_{\tau}
  M+\delta^A_u \kappa^u_{+}] \delta^3(\vec \sigma -\vec \eta
  (T_s))+\nonumber \\
  &&{}\nonumber \\
  &+&{1\over 2} S_T^{\rho (\mu}(T_s)[\vec \eta ] u^{\nu )}(p_s)
   \epsilon^r_{\rho}(u(p_s)) {{\partial}\over {\partial \sigma^r}}
   \delta^3(\vec \sigma -\vec \eta (T_s))+\nonumber \\
 &&{}\nonumber \\
   &+&\sum_{n=2}^{\infty} {{(-1)^n}\over {n!}} I_T^{\mu_1..\mu_n\mu\nu}(T_s,\vec \eta )
   \epsilon^{r_1}_{\mu_1}(u(p_s))..\epsilon^{r_n}_{\mu_n}(u(p_s))
   {{\partial^n}\over {\partial \sigma^{r_1}..\partial \sigma^{r_n}}}
   \delta^3(\vec \sigma -\vec \eta (T_s)),
\label{IV15}
\end{eqnarray}

\noindent where for $n\geq 2$ and $\vec \eta =0$ $\quad
I_T^{\mu_1..\mu_n\mu\nu}(T_s)={{4(n-1)}\over {n+1}}
J_T^{(\mu_1..\mu_{n-1} | \mu | \mu_n)\nu}(T_s)$, with
$J_T^{\mu_1..\mu_n\mu\nu\rho\sigma}(T_s)$ being the Dixon {\it
$2^{2+n}$-pole inertial moment tensors} given in Eqs.(\ref{d10}).
With this form of the multipolar expansion, the quadrupole term
($n = 2$) has the form [see Eq.(\ref{d11})]

\begin{eqnarray*}
 &&{1\over 2}\, \Big( {5\over 3}\, u^{\mu}(p_s)\, u^{\nu}(p_s)\,
 q_F^{r_1r_2\tau\tau}(T_s, \vec \eta ) + {1\over 2}\,
 [u^{\mu}(p_s)\, \epsilon^{\nu}_u(u(p_s)) + u^{\nu}(p_s)\,
 \epsilon^{\mu}_u(u(p_s))]\, q_T^{r_1r_2u\tau}(T_s, \vec \eta )
 +\nonumber \\
 &+& \epsilon^{\mu}_{u_1}(u(p_s))\, \epsilon^{\nu}_{u_2}(u(p_s))\,
 \Big[ q_T^{r_1r_2u_1u_2}(T_s, \vec \eta ) - {3\over 2}\, \Big(
 q_T^{(r_1r_2u_1)u_2}(T_s, \vec \eta ) + q_T^{(r_1r_2u_2)u_1}(T_s,
 \vec \eta ) \Big) +\nonumber \\
 &+& q_T^{(r_1r_2u_1u_2)}(T_s, \vec \eta )\Big]
 \Big).
 \end{eqnarray*}
 \medskip

 Note that, as said in Appendix D, Eq.(\ref{IV15}) holds
 only if the multipoles are evaluated with respect to world-lines
 $w^{\mu}(\tau ) = z^{\mu}(\tau , {\vec \eta}(\tau ))$ with $\vec
 \eta (\tau ) = \vec \eta = const.$, namely with respect to one of
 the integral lines of the vector field $z^{\mu}_{\tau}(\tau ,\vec
 \sigma )\, \partial_{\mu}$.

\bigskip

For an isolated system  described by the multipoles appearing in
Eq.(\ref{IV15}) [this is not true for those in Eq.(\ref{IV4})] the
equations $\partial_{\mu}T^{\mu\nu}\, {\buildrel \circ \over =}\,
0$  [see Eqs.(\ref{d4}) and (\ref{d7})] imply no more than the
following {\it Papapetrou-Dixon-Souriau equations of motion}
\cite{mul5,dixon1,souriau,mul14} for the total momentum
$P^{\mu}_T(T_s)=\epsilon^{\mu}_A(u(p_s)) q_T^{A\tau}(T_s)\approx
p^{\mu}_s$ and the spin tensor $S^{\mu\nu}_T(T_s)[\vec \eta =0]$

\medskip

\begin{eqnarray}
{{d P^{\mu}_T(T_s)}\over {dT_s}}\, &{\buildrel \circ \over =}\,&
0,\nonumber \\
 {{d S^{\mu\nu}_T(T_s)[\vec \eta =0]}\over {dT_s}}\, &{\buildrel
 \circ \over =}\,& 2 P^{[\mu}_T(T_s) u^{\nu ]}(p_s)=2 \kappa^u_{+} \epsilon^{[\mu}_u(u(p_s))
 u^{\nu ]}(p_s) \approx 0,\nonumber \\
 &&{}\nonumber \\
 &&{}\nonumber \\
 or&& {{d M}\over {dT_s}} \cir 0,\qquad {{d {\vec
 \kappa}_{+}}\over {dT_s}} \cir 0,\qquad {{d S_s^{\mu\nu}}\over
 {dT_s}} \cir 0.
\label{IV16}
\end{eqnarray}

\vfill\eject

\section{Dixon's Multipoles and Relevant Centroids for Closed and Open Systems of Interacting Relativistic Particles.}

In this Section we present new applications of the multipolar
expansion to interacting systems of particles and fields. We first
deal with the case of an isolated system of positive-energy
relativistic particles with mutual action-at-a-distance
interaction (see Section VIII of Ref.\cite{iten1} and Section VI
of Ref.\cite{crater}); then we deal with the case of an {\it open
particle sub-system} of an isolated system consisting of $N$
charged positive-energy relativistic particles (with
Grassmann-valued electric charges to regularize the Coulomb
self-energies) plus the electro-magnetic field \cite{crater}.
\bigskip

\subsection{An isolated System of Positive-Energy Particles with
Action-at-a-Distance Interactions.}

As said in Section VIII of Ref.\cite{iten1}, in the rest-frame
instant form the most general expression of the internal energy
for an isolated system of $N$ positive-energy particles with
mutual action-at-a-distance interactions is

\beq
 M = \sum_i\, \sqrt{m_i^2 + U_i + ({\vec \kappa}_i - {\vec V}_i)^2} + V,
 \label{V1}
 \eeq

\noindent where all the potentials $U_i$, ${\vec V}_i$, $V$ are
functions of ${\vec \kappa}_i \cdot {\vec \kappa}_j$, $|{\vec
\eta}_i - {\vec \eta}_j|$, ${\vec \kappa}_k \cdot ({\vec \eta}_i -
{\vec \eta}_j)$. On the other hand, as shown at the end of Section
II, in the free case we have

\bea
 M_{(free)} &=& \sum_i\, \sqrt{m^2_i + {\vec \kappa}_i^2} =
 \sqrt{{\cal M}^2_{(free)} + {\vec \kappa}_{+}^2} \approx\nonumber
 \\
 &\approx& {\cal M}_{(free)} = \sum_i\, \sqrt{m^2_i + N\, \sum_{ab}\,
  \gamma_{ai}\, \gamma_{bi}\, {\vec \pi}_{qa} \cdot {\vec
  \pi}_{qb}}.
  \label{V2}
  \eea

Since the 3-centers ${\vec R}_{+}$ and ${\vec q}_{+}$ became
interaction dependent, in the interacting case we do not know the
final canonical basis ${\vec q}_{+}$, ${\vec \kappa}_{+}$, ${\vec
\rho}_{qa}$, ${\vec \pi}_{qa}$ explicitly. For an isolated system,
however, we have $M = \sqrt{{\cal M}^2 + {\vec \kappa}^2_{+}}
\approx {\cal M}$ with ${\cal M}$ independent of ${\vec q}_{+}$
($\{ M, {\vec \kappa}_{+} \} = 0$ in the internal Poincare'
algebra). This suggests that also in the interacting case the same
result should hold true. Indeed, by its definition, the
Gartenhaus-Schwartz transformation gives ${\vec \rho}_{qa} \approx
{\vec \rho}_a$, ${\vec \pi}_{qa} \approx {\vec \pi}_a$ also in
presence of interactions, so that we get

\bea
 M{|}_{{\vec \kappa}_{+} =0} &=& \Big( \sum_i\, \sqrt{m_i^2 + U_i +
  ({\vec \kappa}_i - {\vec V}_i)^2} + V \Big){|}_{{\vec \kappa}_{+} =0}
= \sqrt{{\cal M}^2 + {\vec \kappa}^2_{+}} {|}_{{\vec \kappa}_{+}}
=\nonumber
\\
 &=& {\cal M}{|}_{{\vec \kappa}_{+} =0} = \sum_i\, \sqrt{m_i^2 +
  {\tilde U}_i + ({\vec \kappa}_i - {\vec {\tilde V}}_i)^2} + \tilde
  V,
  \label{V3}
  \eea

 \noindent where the potentials ${\tilde U}_i$, ${\vec {\tilde
 V}}_i$, $\tilde V$ are now functions of ${\vec \pi}_{qa} \cdot {\vec
 \pi}_{qb}$, ${\vec \pi}_{qa} \cdot {\vec \rho}_{qb}$, ${\vec
 \rho}_{qa} \cdot {\vec \rho}_{qb}$.

A relevant example of this type of isolated system has been
studied in Ref.\cite{crater} starting from the isolated system of
$N$ charged positive-energy particles (with Grassmann-valued
electric charges $Q_i = \theta^*\, \theta$, $Q^2_i =0$, $Q_i\, Q_j
= Q_j\, Q_i \not= 0$ for $i \not= j$) plus the electro-magnetic
field. After a Shanmugadhasan canonical transformation, this
system can be expressed only in terms of transverse Dirac
observables corresponding to a radiation gauge for the
electro-magnetic field.  The expression of the energy-momentum
tensor in this gauge will be shown in the next Subsection. In the
semi-classical approximation of Ref.\cite{crater}, the
electro-magnetic degrees of freedom are re-expressed in terms of
the particle variables by means of the Lienard-Wiechert solution
in the framework of the rest-frame instant form. In this way it
has been possible to derive the exact semi-classical relativistic
form of the action-at-a-distance Darwin potential in the reduced
phase space of the particles. Note that this form is independent
of the choice of the Green function in the Lienard-Wiechert
solution. In Ref.\cite{crater} the associated energy-momentum
tensor for the case $N = 2$ [Eqs.(6.48)] is also given. The
internal energy is $M = \sqrt{{\cal M}^2 + {\vec \kappa}_{+}^2}
\approx {\cal M} = \sum_{i=1}^2\, \sqrt{m^2_i + {\vec \pi}^2} +
{{Q_1\, Q_2}\over {4\pi\, \rho}}\, [1 + \tilde V({\vec \pi}^2,
{\vec \pi} \cdot {{\vec \rho}\over {\rho}})]$ where $\tilde V$ is
given in Eqs.(6.34), (6.35) [in Eqs. (6.36), (6.37) for $m_1 =
m_2$]. The internal boost ${\vec {\cal K}}$ [Eq.(6.46)] allows the
determination of the 3-center of energy ${\vec R}_{+} = - {{{\vec
{\cal K}}}\over M} \approx {\vec q}_{+} \approx {\vec y}_{+}$ in
the present interacting case.

The knowledge of the energy-momentum tensor $T^{AB}(\tau ,\vec
\sigma )$ and of ${\vec R}_{+} \approx {\vec q}_{+}$ allows to
apply our formalism to find the barycentric multipoles of this
interacting case. It turns out that, in the gauge ${\vec R}_{+}
\approx {\vec q}_{+} \approx {\vec y}_{+} \approx 0$,  all the
formal properties  studied in the previous Section (like the
coincidence of all the relevant centroids) are reproduced in
presence of mutual action-at-a-distance interactions.
\bigskip

\subsection{Open Subsystem of the Isolated System of N Positive-Energy
Particles with Grassmann-valued Electric Charge plus the
Electromagnetic Field.}

Let us now consider an open sub-system of the isolated system of
$N$ charged positive-energy particles plus the electro-magnetic
field in the radiation gauge. The energy-momentum tensor and the
Hamilton equations on the Wigner hyper-plane of the isolated
system are, respectively, [${\vec \kappa}_{+} = \sum_i\, {\vec
\kappa}_i$; $\stackrel{\circ }{=}$ means evaluated on the
equations of motion; to avoid degenerations we assume that all the
masses $m_i$ are different]

\begin{eqnarray}
T^{\tau\tau}(\tau ,\vec \sigma )&=&\sum_{i=1}^N \delta^3(\vec
\sigma - {\vec \eta}_i(\tau )) \sqrt{m^2_i+[{ {\vec
\kappa}}_i(\tau ) -Q_i{{ \vec A}}_{\perp}(\tau ,{\vec \eta}_i(\tau
))]^2} + \nonumber \\
 &+&{\frac{1}{2}}[\Big( {{\vec \pi}}_{\perp}+\sum_{i=1}^NQ_i{\frac{{
\vec \partial}}{{\triangle}}}\delta^3(\vec \sigma -{\vec
\eta}_i(\tau ))\Big) ^2 +{\vec B}^2](\tau ,\vec \sigma )=
\nonumber \\
 &=& \sum_{i=1}^N \delta^3(\vec
\sigma - {\vec \eta}_i(\tau )) \sqrt{m^2_i+[{ {\vec
\kappa}}_i(\tau ) -Q_i{{ \vec A}}_{\perp}(\tau ,{\vec \eta}_i(\tau
))]^2} + \nonumber \\
 &+& \sum_{i=1}^N\, Q_i\, {\vec \pi}_{\perp}(\tau ,\vec \sigma )\,
 \times {{\vec \partial}\over {\triangle}}\, \delta^3(\vec \sigma - {\vec
 \eta}_i(\tau )) + {1\over 2}\, [{\vec \pi}^2_{\perp} + {\vec
 B}^2](\tau ,\vec \sigma ) +\nonumber \\
 &+& {1\over 2}\, \sum_{i,k,i\not= k}^{1..N}\, Q_i\, Q_k\,
 {{\vec \partial}\over {\triangle}}\, \delta^3(\vec \sigma - {\vec
 \eta}_i(\tau )) \cdot  {{\vec \partial}\over {\triangle}}\, \delta^3(\vec \sigma - {\vec
 \eta}_k(\tau )),\nonumber \\
 &&{}\nonumber \\
 T^{r\tau}(\tau ,\vec \sigma )&=&\sum_{i=1}^N\delta^3(\vec \sigma -{\vec \eta}
_i(\tau )) [{\kappa}_i^r(\tau )-Q_i {A}_{\perp}^r(\tau ,{\vec
\eta}_i(\tau ))] +  \nonumber \\
 &+&[\Big( {{\vec \pi}}_{\perp}+\sum_{i=1}^NQ_i{\frac{{\vec \partial}}{
{\triangle}}}\delta^3(\vec \sigma -{\vec \eta}_i(\tau
))\Big)\times \vec B ](\tau ,\vec \sigma ),  \nonumber \\
 &&{}\nonumber \\
 T^{rs}(\tau ,\vec \sigma )&=& \sum_{i=1}^N \delta^3(\vec \sigma -{\vec \eta}
_i(\tau )) {\frac{{\ [{\kappa}_i^r(\tau )-Q_i{A} _{\perp}^r(\tau
,{\vec \eta}_i(\tau ))] [{\kappa}_i^s(\tau )-Q_i{
A}_{\perp}^s(\tau ,{\vec \eta}_i(\tau ))]}}{{\sqrt{m_i^2+[{ { \vec
\kappa}}_i(\tau )-Q_i{
 {\vec A}}_{\perp}(\tau ,{\vec \eta}_i(\tau ))]^2} }}} -  \nonumber \\
 &-&\Big[{\frac{1}{2}}\delta^{rs} [\Big( {{\vec \pi}}
_{\perp}+\sum_{i=1}^NQ_i{\frac{{\vec
\partial}}{{\triangle}}}\delta^3(\vec \sigma -{\vec \eta}_i(\tau
))\Big)^2+{\vec B}^2] -  \nonumber \\
 &-&[\Big( {{\vec \pi}}_{\perp}+\sum_{i=1}^NQ_i{\frac{{\vec
\partial}}{ {\triangle}}}\delta^3(\vec \sigma -{\vec \eta}_i(\tau
))\Big)^r \Big( { {\vec \pi}}_{\perp}+\sum_{i=1}^NQ_i{\frac{{\vec
\partial}}{{\triangle} }}\delta^3(\vec \sigma -{\vec \eta}_i(\tau
))\Big)^s +  \nonumber \\
 &+&B^rB^s]\Big] (\tau ,\vec \sigma ).
 \label{V4}
\end{eqnarray}
\medskip

\begin{eqnarray}
\dot{\vec{\eta}}_{i}(\tau )\, &\,\stackrel{\circ
}{=}&\,\frac{{\vec{ \kappa}}_{i}(\tau )-Q_{i}{\vec{A}}_{\perp
}(\tau ,\vec{\eta}_{i}(\tau
))}{\sqrt{m_{i}^{2}+({\vec{\kappa}}_{i}(\tau )-Q_{i}{\vec{A}}
_{\perp }(\tau ,\vec{\eta}_{i}(\tau )))^{2}}}, \nonumber \\
 \dot{{\vec{\kappa}}}_{i}(\tau )\,\stackrel{\circ
}{=} &&\, \sum_{k\neq i}\frac{Q_{i}Q_{k}({\vec{\eta}}_{i}(\tau
)-{\vec{\eta}}_{k}(\tau ))}{4\pi \mid \vec{\eta}_{i}(\tau
)-\vec{\eta}_{k}(\tau )\mid ^{3}}+
 Q_{i}\, \dot{\eta}_{i}^{u}(\tau )\, {\frac{{\partial }}{
\partial {{\vec{\eta}}_{i}}}}{{A}}_{\perp }^{u}(\tau ,\vec{\eta}
_{i}(\tau ))],  \nonumber \\
  &&{}\nonumber \\
  &&{}\nonumber \\
 &&\dot{{A}}_{\perp r}(\tau ,\vec{\sigma})\,\stackrel{\circ }{=}-{
\pi} _{\perp r}(\tau ,\vec{\sigma}),  \nonumber \\
 &&{}\nonumber \\
 \dot{{\pi}}_{\perp }^{r}(\tau ,\vec{\sigma})\, &\stackrel{\circ }{=}
&\,\Delta {A}_{\perp }^{r}(\tau ,\vec{\sigma}) -
\sum_{i}Q_{i}P_{\perp }^{rs}(\vec{\sigma})\dot{\eta}_{i}^{s}(\tau
)\delta ^{3}(\vec{\sigma}-\vec{\eta}_{i}(\tau )),\nonumber \\
 &&{}\nonumber \\
 &&{}\nonumber \\
 {{\vec{\kappa}}}_{+}(\tau )
&+&\int d^{3}\sigma \lbrack {{\vec{ \pi}}}_{\perp }\times
{{\vec{B}}}](\tau ,\vec{\sigma})\approx 0\,\, (rest-frame\,
condition).
 \label{V5}
\end{eqnarray}
\medskip

Let us note that in this reduced phase space there are only either
particle-field interactions or action-at-a-distance 2-body
interactions.

\bigskip

The particle world-lines are $x^{\mu}_i(\tau ) = x^{\mu}_o +
u^{\mu}(p_s)\, \tau + \epsilon^{\mu}_r(u(p_s))\, \eta^r_i(\tau )$,
while their 4-momenta are $p^{\mu}_i(\tau ) = \sqrt{m^2_i + [{\vec
\kappa}_i - Q_i\, {\vec A}_{\perp}(\tau ,{\vec \eta}_i) ]^2}\,
u^{\mu}(p_s) + \epsilon^{\mu}_r(u(p_s))\, [\kappa^r_i - Q_i\,
A^r_{\perp}(\tau ,{\vec \eta}_i)]$.
\bigskip

The generators of the internal Poincar\'{e} group are

\begin{eqnarray}
 {\cal P}_{(int)}^{\tau } &=&M=\sum_{i=1}^{N}\sqrt{m_{i}^{2}+({\vec{
\kappa}}_{i}(\tau )-Q_{i}{\vec{A}}_{\perp }(\tau ,\vec{\eta}
_{i}(\tau )))^{2}}+  \nonumber \\
 &+&{1\over 2}\, \sum_{i\neq j}\frac{Q_{i}Q_{j}}{4\pi \mid \vec{\eta}_{i}(\tau )-\vec{\eta}
_{j}(\tau )\mid }+\int d^{3}\sigma
{\frac{1}{2}}[{\vec{\pi}}_{\perp }^{2}+{\vec{B}}^{2}](\tau
,\vec{\sigma}),  \nonumber \\
 &&{}\nonumber \\
 {\cal \vec{P}}_{(int)} &=& {{\vec{\kappa}}}
_{+}(\tau )+\int d^{3}\sigma \lbrack {{\vec{\pi}}}_{\perp }\times
{ {\vec{B}}}](\tau ,\vec{\sigma})\approx 0, \nonumber \\
 &&{} \nonumber \\
 {\cal J}_{(int)}^{r} &=& \sum_{i=1}^{N}( \vec{\eta}_{i}(\tau )\times
{{\vec{\kappa}}}_{i}(\tau ))^{r}+\int d^{3}\sigma
\,(\vec{\sigma}\times \,{[{{\vec{\pi}}}}_{\perp }{\times {
{\vec{B}}]}}^{r}{(\tau ,\vec{\sigma})},  \nonumber \\
 {\cal K}_{(int)}^{r} &=& -\sum_{i=1}^{N}\vec{\eta}_{i}(\tau
)\sqrt{m_{i}^{2}+[{{{\vec{\kappa}} }}_{i}(\tau
)-Q_{i}{{\vec{A}}}_{\perp }(\tau ,{\ \vec{\eta}}_{i}(\tau
))]{}^{2}}+  \nonumber \\
 &+&{1\over 2}\, \Big[ Q_i\, \sum_{i=1}^{N} \sum_{j\not=i}^{1..N}\,
 Q_j\, \int d^3\sigma\, \sigma^r\, \vec c(\vec \sigma -
 {\vec \eta}_i(\tau )) \cdot \vec c(\vec \sigma - {\vec
 \eta}_j(\tau )) + \nonumber \\
 &+& Q_{i}\, \int d^{3}\sigma {{\pi}}_{\perp }^{r}(\tau ,\vec{\sigma})c(
\vec{\sigma} - {\ \vec{\eta}}_{i}(\tau )) \Big]-{\frac{1}{2}}\int
d^{3}\sigma \sigma ^{r}\,({{{\vec{\pi}}}}_{\perp }^{2}+{{{\vec{B}
}}}^{2})(\tau ,\vec{\sigma}),
 \label{V6}
\end{eqnarray}

\noindent with $c({\vec{\eta}}_{i}-{\vec{\eta}}_{j})= 1/(4\pi
|\vec{\eta}_{j}-\vec{ \eta}_{i}|)$ [$\triangle\, c(\vec \sigma ) =
\delta^3(\vec \sigma )$, $\triangle = - {\vec \partial}^2$, $\vec
c(\vec \sigma ) = \vec \partial\, c(\vec \sigma ) = \vec \sigma /
(4\pi\, |\vec \sigma |^3)$]. \medskip

Note that ${\cal P}^{\tau}_{(int)} = q^{\tau\tau}$ and ${\cal
P}^r_{(int)} = q^{r\tau}$ are the mass and momentum monopoles,
respectively.

\bigskip

For the sake of simplicity let us consider the sub-system formed
by the two particles of mass $m_1$ and $m_2$. Our considerations
may be extended to any cluster of particles both in this case and
in the case discussed in the previous Subsection. This sub-system
is {\it open}: besides their mutual interaction the two particles
have Coulomb interaction with the other $N - 2$ particles and feel
the transverse electric and magnetic fields.
\medskip

By using the multipoles we will select a set of {\it effective
parameters} (mass, 3-center of motion, 3-momentum, spin)
describing the two-particle cluster as a global entity subject to
external forces in the global rest-frame instant form. This was
the original motivation of the multipolar expansion in general
relativity: actually replacing an extended object (an open system
due to the presence of the gravitational field) with a set of
multipoles concentrated on a center of motion. In the rest-frame
instant form it is possible to show that there is no preferred
centroid for an open system, namely different centers of motion
may be selected according to different conventions unlike the case
of isolated systems where, in the rest frame ${\vec \kappa}_{+}
\approx 0$, all these conventions identify the same centroid. We
will see, however, that there is a choice which seems preferable
due to its properties.
\medskip

Given the energy-momentum tensor $T^{AB}(\tau ,\vec \sigma )$
(\ref{V4}) of the isolated system,  it would seem natural to
define {\it the energy-momentum tensor $T^{AB}_{c(n)}(\tau ,\vec
\sigma )$ of an open sub-system composed by a cluster of $n \leq
N$ particles} as the sum of all the terms in Eq.(\ref{V4})
containing a dependence on the variables ${\vec \eta}_i$, ${\vec
\kappa}_i$, of the particles of the cluster. Besides kinetic
terms, this tensor would contain  internal mutual interactions as
well as external interactions of the cluster particles with the
environment composed by the other $N-n$ particles and by the
transverse electro-magnetic field. There is an ambiguity, however:
why attributing just to the cluster {\it all the external
interactions with the other $N-n$ particles} (no such ambiguity
exists for the interaction with the electro-magnetic field)? Since
we have 2-body interactions, it seems more reasonable to attribute
only {\it half} of these external interactions to the cluster and
consider the other half as a property of the remaining $N-n$
particles. Let us remark that according to the first choice, if we
consider two clusters composed by two non-overlapping sets of
$n_1$ and $n_2$ particles respectively, we would get
$T^{AB}_{c(n_1+n_2)} \not= T^{AB}_{c(n_1)} + T^{AB}_{c(n_2)}$,
since the mutual Coulomb interactions between the two clusters are
present in both $T^{AB}_{c(n_1)}$ and $T^{AB}_{c(n_2)}$. Instead
according to the second choice we would get $T^{AB}_{c(n_1+n_2)} =
T^{AB}_{c(n_1)} + T^{AB}_{c(n_2)}$. Since this property is
important for studying the mutual relative motion of two clusters
in actual cases, we will adopt {\it the convention that the
energy-momentum tensor of a $n$ particle cluster contains only
half of the external interaction with the other $N-n$ particles}.

\bigskip

Let us remark that, in the case of $k$-body forces, this
convention should be replaced by the following rule: i) for each
particle $m_i$ of the cluster and each $k$-body term in the
energy-momentum tensor involving this particle, we write $k = h_i
+ (k - h_i)$, where $h_i$ is the number of particles of the
cluster participating to this particular $k$-body interaction; ii)
then only the fraction $h_i/k$ of this particular $k$-body
interaction term containing $m_i$ has to be attributed to the
cluster.

\bigskip

Let us consider the cluster composed by the two particles with
mass $m_1$ and $m_2$. The knowledge of $T^{AB}_c\, {\buildrel
{def}\over =}\, T^{AB}_{c(2)}$ on the Wigner hyper-plane of the
global rest-frame instant form allows to find the following 10
{\it non conserved} charges [due to $Q^2_i =0$ we have
$\sqrt{m^2_i + [{\vec \kappa}_i - Q_i\, {\vec A}_{\perp}(\tau
,{\vec \eta}_i)]^2} = \sqrt{m^2_i + {\vec \kappa}_1^2} - Q_i\,
{{{\vec \kappa}_i \cdot {\vec A}_{\perp}(\tau ,{\vec
\eta}_i)}\over {\sqrt{m^2_i + {\vec \kappa}_i^2 }}} $] \bigskip

\bea
 M_c &=& \int d^3\sigma\, T^{\tau\tau}_c(\tau ,\vec \sigma ) =
 \sum_{i=1}^2\, \sqrt{m^2_i + [{\vec \kappa}_i(\tau ) - Q_i\,
  {\vec A}_{\perp}(\tau ,{\vec \eta}_i(\tau ))]^2} +\nonumber \\
  &+& {{Q_1\, Q_2}\over {4\pi\, |{\vec \eta}_1(\tau ) - {\vec \eta}_2(\tau )|^2}}
  + {1\over 2}\, \sum_{i=1}^2\, \sum_{k\not= 1,2}\, {{Q_i\, Q_k}\over {4\pi\,
  |{\vec \eta}_i(\tau ) - {\vec \eta}_k(\tau )|^2}} =\nonumber \\
  &=& M_{c(int)} + M_{c(ext)},\nonumber \\
  &&M_{c(int)} = \sum_{i=1}^2\, \sqrt{m_i^2 + {\vec \kappa}_i^2} -
  {{Q_1\, Q_2}\over {4\pi\, |{\vec \eta}_1(\tau ) - {\vec \eta}_2(\tau
  )|^2}},\nonumber \\
  &&{}\nonumber \\
  {\vec {\cal P}}_c &=& \{ \int d^3\sigma\, T^{r\tau}_c(\tau ,\vec \sigma ) \} =
  {\vec \kappa}_1(\tau ) + {\vec \kappa}_2(\tau
  ),\nonumber \\
  &&{}\nonumber \\
  {\vec {\cal J}}_c &=& \{ \epsilon^{ruv}\, \int d^3\sigma\, [\sigma^u\, T^{v\tau}_c
  - \sigma^v\, T^{u\tau}_c](\tau ,\vec \sigma ) \} =\nonumber \\
  &=& {\vec \eta}_i(\tau ) \times {\vec
  \kappa}_1(\tau ) + {\vec \eta}_2(\tau ) \times {\vec
  \kappa}_2(\tau ),\nonumber \\
  &&{}\nonumber \\
  {\vec {\cal K}}_c &=& - \int d^3\sigma\, \vec \sigma\,
  T^{\tau\tau}_c(\tau ,\vec \sigma ) =\nonumber \\
  &=& - \sum_{i=1}^2\, {\vec \eta}_i(\tau )\, \sqrt{m^2_i + [{\vec \kappa}_i(\tau ) - Q_i\,
  {\vec A}_{\perp}(\tau ,{\vec \eta}_i(\tau ))]^2} -\nonumber \\
  &-&  \sum_{i=1}^2\, Q_i\, \int d^3\sigma\, {\vec
  \pi}_{\perp}(\tau ,\vec \sigma )\, c(\vec \sigma - {\vec
  \eta}_i(\tau )) -\nonumber \\
 &-&  Q_1\, Q_2\, \int d^3\sigma\, \vec \sigma\, \vec c(\vec
 \sigma - {\vec \eta}_1(\tau )) \cdot \vec c(\vec \sigma - {\vec
 \eta}_2(\tau )) -\nonumber \\
 &-& {1\over 2}\, \sum_{i=1}^2\, Q_i\, \sum_{k\not= 1,2}\, Q_k\,
\int d^3\sigma\, \vec \sigma\, \vec c(\vec
 \sigma - {\vec \eta}_i(\tau )) \cdot \vec c(\vec \sigma - {\vec
 \eta}_k(\tau )) =\nonumber \\
  &=& {\vec {\cal K}}_{c(int)} + {\vec {\cal
  K}}_{c(ext)},\nonumber \\
  &&{\vec {\cal K}}_{c(int)} = - \sum_{i=1}^2\, {\vec \eta}_i(\tau
  )\, \sqrt{m^2_i + {\vec \kappa}_i^2} - Q_1\, Q_2\,
 \int d^3\sigma\, \vec \sigma\, \vec c(\vec
 \sigma - {\vec \eta}_1(\tau )) \cdot \vec c(\vec \sigma - {\vec
 \eta}_2(\tau )),
 \label{V7}
 \eea

\noindent which do not satisfy the algebra of an internal
Poincare' group due to the openess of the system. Since we work in
an instant form of dynamics only the cluster internal energy and
boosts depend on the (internal and external) interactions. Again
$M_c = q_c^{\tau\tau}$ and ${\cal P}^r_c = q_c^{r\tau}$ are the
mass and momentum monopoles of the cluster.
\bigskip

Another needed quantity  is the momentum dipole

\bea
 p^{ru}_c &=& \int d^3\sigma\, \sigma^r\, T^{u\tau}_c(\tau ,\vec
 \sigma ) =\nonumber \\
 &=& \sum_{i=1}^2\, \eta^r_i(\tau )\, \kappa^u_i(\tau ) -
  \sum_{i=1}^2\, Q_i\, \int d^3\sigma\, c(\vec
 \sigma - {\vec \eta}_i(\tau ))\, [\partial^r\, A^s_{\perp} + \partial^s\, A^r_{\perp}]
 (\tau ,\vec \sigma),\nonumber \\
 &&{}\nonumber \\
 &&p^{ru}_c + p^{ur}_c = \sum_{i=1}^2\, [\eta^r_i(\tau )\,
 \kappa^u_i(\tau ) + \eta^u_i(\tau )\, \kappa^r_i(\tau
 )] -\nonumber \\
 &-& 2\, \sum_{i=1}^2\, Q_i\, \int d^3\sigma\, c(\vec
 \sigma - {\vec \eta}_i(\tau ))\, [\partial^r\, A^s_{\perp} + \partial^s\, A^r_{\perp}]
 (\tau ,\vec \sigma ),\nonumber \\
 &&{}\nonumber \\
 &&p^{ru}_c - p^{ur}_c = \epsilon^{ruv}\, {\cal J}^v_c.
 \label{V8}
 \eea
\medskip

The time variation of the 10 charges (\ref{V7}) can be evaluated
by using the equations of motion (\ref{V5}) \bigskip

\bea
 {{d M_c}\over {d\tau}} &=& \sum_{i=1}^2\, Q_i\, \Big( {{{\vec
 \kappa}_i(\tau ) \cdot {\vec \pi}_{\perp}(\tau ,{\vec
 \eta}_i(\tau ))}\over {\sqrt{m^2_i + {\vec \kappa}_i^2}}}
 +\nonumber \\
 &+&  {1\over 2}\, \sum_{k\not= 1,2}\, Q_k\, \Big[{{{\vec \kappa}_i(\tau
)}\over {\sqrt{m^2_i + {\vec \kappa}_i^2}}} + {{{\vec
\kappa}_k(\tau )}\over {\sqrt{m^2_k + {\vec \kappa}_k^2}}}\Big]
\cdot {{{\vec \eta}_i(\tau ) - {\vec \eta}_k(\tau ) }\over {4\pi\,
 |{\vec \eta}_i(\tau ) - {\vec \eta}_k(\tau ) |^3}}\Big),\nonumber \\
 &&{}\nonumber \\
 {{d {\cal P}^r_c}\over {d\tau }} &=& \sum_{i=1}^2\, Q_i\, \Big( {{{\vec \kappa}_i(\tau )}\over
{\sqrt{m^2_i + {\vec \kappa}_i^2}}} \cdot {{\partial
A^r_{\perp}(\tau ,{\vec \eta}_i(\tau ))}\over {\partial {\vec
\eta}_i}} + \sum_{k\not= 1,2}\, Q_k\, {{\eta^r_i(\tau ) -
\eta^r_k(\tau ) }\over {4\pi\, |{\vec \eta}_i(\tau ) - {\vec
\eta}_k(\tau ) |^3}} \Big),\nonumber \\
 &&{}\nonumber \\
 {{d {\vec {\cal J}}_c}\over {d\tau }} &=&
 \sum_{i=1}^2\, Q_i\, \Big( {{{\vec \kappa}_i(\tau )}\over
{\sqrt{m^2_i + {\vec \kappa}_i^2}}} \times {\vec A}_{\perp}(\tau
,{\vec \eta}_i(\tau )) + {\vec \eta}_i(\tau ) \times \Big[
 {{{\vec \kappa}_i(\tau )}\over
{\sqrt{m^2_i + {\vec \kappa}_i^2}}} \cdot {{\partial}\over
{\partial {\vec \eta}_i}}\Big]\, {\vec A}_{\perp}(\tau ,{\vec
\eta}_i(\tau )) -\nonumber \\
 &-& \sum_{k \not= i}\, Q_k\, {{{\vec \eta}_i(\tau ) \times {\vec \eta}_k(\tau )}
 \over {4\pi\, |{\vec \eta}_i(\tau ) - {\vec \eta}_k(\tau )|^3}}\Big),\nonumber \\
 &&{}\nonumber \\
 {{d {\cal K}^r_c}\over {d\tau}} &=& - {\cal P}^r_c - \nonumber \\
 &-& \sum_{i=1}^2\, Q_i\, {\vec \eta}_i(\tau )\, {{{\vec
 \kappa}_i(\tau )}\over {\sqrt{m^2_i + {\vec \kappa}_i^2(\tau )}}}
 \cdot \Big[ {\vec \pi}_{\perp}(\tau ,{\vec \eta}_i(\tau )) +
 \sum_{k\not= i}\, Q_k\, \vec c({\vec \eta}_i(\tau ) - {\vec
 \eta}_k(\tau ))\Big] +\nonumber \\
 &+& \sum_{i=1}^2\, Q_i\,  \Big[ \sum_{k\not= i}\, Q_k\, {{{\vec
 \kappa}_k(\tau )}\over {\sqrt{m^2_k + {\vec \kappa}_k^2(\tau )}}}
 \, c({\vec \eta}_i(\tau ) - {\vec \eta}_k(\tau )) - \int
 d^3\sigma\, {\vec \pi}_{\perp}(\tau ,\vec \sigma )\, {{{\vec
 \kappa}_i(\tau ) \cdot \vec c(\vec \sigma - {\vec \eta}_i(\tau ))}
 \over {\sqrt{m^2_i + {\vec \kappa}_i^2(\tau )}}}\Big] +\nonumber \\
 &+& \sum_{i=1}^2\, Q_i\, \sum_{k\not= i}\, Q_k\, \int d^3\sigma\,
 c(\vec \sigma - {\vec \eta}_i(\tau ))\, \Big({{{\vec
 \kappa}_k(\tau )}\over {\sqrt{m^2_k + {\vec \kappa}_k^2(\tau )}}}
 \cdot \vec \partial \Big)\, \vec c(\vec \sigma - {\vec
 \eta}_k(\tau )) -\nonumber \\
 &=& Q_1Q_2\, \int d^3\sigma\, \vec \sigma\, \Big( \Big[ \Big( {{{\vec
 \kappa}_1(\tau )}\over {\sqrt{m^2_1 + {\vec \kappa}_1^2(\tau )}}}
\cdot \vec \partial\Big)\, \vec c(\vec \sigma - {\vec \eta}_1(\tau
))\Big] \cdot \vec c(\vec \sigma - {\vec \eta}_2(\tau ))
+\nonumber \\
 &+& \vec c(\vec \sigma - {\vec \eta}_1(\tau )) \cdot \Big[ \Big( {{{\vec
 \kappa}_2(\tau )}\over {\sqrt{m^2_2 + {\vec \kappa}_2^2(\tau )}}}
 \cdot \vec \partial \Big)\, \vec c(\vec \sigma - {\vec
 \eta}_2(\tau ))\Big] \Big) -\nonumber \\
 &-& {1\over 2}\, \sum_{i=1}^2\, Q_i\, \sum_{k\not= 1,2}\, Q_k\,
  \int d^3\sigma\, \vec \sigma\, \Big( \Big[ \Big( {{{\vec
 \kappa}_i(\tau )}\over {\sqrt{m^2_i + {\vec \kappa}_i^2(\tau )}}}
\cdot \vec \partial\Big)\, \vec c(\vec \sigma - {\vec \eta}_i(\tau
))\Big] \cdot \vec c(\vec \sigma - {\vec \eta}_k(\tau ))
+\nonumber \\
 &+& \vec c(\vec \sigma - {\vec \eta}_i(\tau )) \cdot \Big[ \Big( {{{\vec
 \kappa}_k(\tau )}\over {\sqrt{m^2_k + {\vec \kappa}_k^2(\tau )}}}
 \cdot \vec \partial \Big)\, \vec c(\vec \sigma - {\vec
 \eta}_k(\tau ))\Big] \Big).
 \label{V9}
 \eea

\medskip

Let us remark that, if we have two clusters of $n_1$ and $n_2$
particles respectively, our definition of cluster energy-momentum
tensor implies

\bea
 M_{c(n_1+n_2)} &=& M_{c(n_1)} + M_{c(n_2)},\nonumber \\
 {\vec {\cal P}}_{c(n_1+n_2)} &=& {\vec {\cal P}}_{c(n_1)} + {\vec
 {\cal P}}_{c(n_2)},\nonumber\\
 {\vec {\cal J}}_{c(n_1+n_2)} &=& {\vec {\cal J}}_{c(n_1)} + {\vec
 {\cal J}}_{c(n_2)},\nonumber \\
 {\vec {\cal K}}_{c(n_1+n_2)} &=& {\vec {\cal K}}_{c(n_1)} + {\vec
 {\cal K}}_{c(n_2)}.
  \label{V10}
 \eea
\bigskip

The main problem is the determination of an effective center of
motion $\zeta^r_c(\tau )$ with world-line $w^{\mu}_c(\tau ) =
x^{\mu}_o + u^{\mu}(p_s)\, \tau + \epsilon^{\mu}_r(u(p_s))\,
\zeta^r_c(\tau )$ in the gauge $T_s \equiv \tau$, ${\vec q}_{+} =
{\vec R}_{+} = {\vec y}_{+} \equiv 0$ of the isolated system. The
unit 4-velocity of this center of motion is $u^{\mu}_c(\tau ) =
{\dot w}_c^{\mu}(\tau ) / \sqrt{1 - {\dot {\vec \zeta}}^2_c(\tau
)}$ with ${\dot w}_c^{\mu}(\tau ) = u^{\mu}(p_s) +
\epsilon^{\mu}_r(u(p_s))\, {\dot \zeta}^r_c(\tau )$. By using
$\delta\, z^{\mu}(\tau ,\vec \sigma ) = \epsilon^{\mu}_r(u(p_s))\,
(\sigma^r - \zeta^r(\tau ))$ we can define the multipoles of the
cluster with respect to the world-line $w^{\mu}_c(\tau )$

\beq
 q_c^{r_1..r_nAB}(\tau ) = \int d^3\sigma \, [\sigma^{r_1} -
 \zeta_c^{r_1}(\tau )] .. [\sigma^{r_n} - \zeta_c^{r_n}(\tau )]\,
 T^{AB}_c(\tau ,\vec \sigma ).
 \label{V11}
 \eeq

\medskip

The mass and momentum monopoles and the mass, momentum and spin
dipoles are respectively
\medskip

\bea
 q^{\tau\tau}_c &=& M_c,\qquad q_c^{r\tau} = {\cal
 P}_c^r,\nonumber \\
 &&{}\nonumber \\
 q_c^{r\tau\tau} &=& - {\cal K}^r_c - M_c\, \zeta^r_c(\tau ) = M_c\,
 (R^r_c(\tau ) - \zeta^r_c(\tau )),\nonumber \\
 q_c^{ru\tau} &=& p^{ru}_c(\tau ) - \zeta^r_c(\tau ) \, {\cal
 P}^u_c,\nonumber \\
 &&{}\nonumber \\
 S^{\mu\nu}_c &=& [\epsilon^{\mu}_r(u(p_s))\, u^{\nu}(p_s) -
 \epsilon^{\nu}_r(u(p_s))\, u^{\mu}(p_s)]\, q_c^{r\tau\tau} +
 \epsilon^{\mu}_r(u(p_s))\, \epsilon^{\nu}_u(u(p_s))\,
 (q^{ru\tau}_c - q_c^{ur\tau}) =\nonumber \\
 &=& [\epsilon^{\mu}_r(u(p_s))\, u^{\nu}(p_s) -
 \epsilon^{\nu}_r(u(p_s))\, u^{\mu}(p_s)]\, M_c\, (R^r_c -
 \zeta^r_c) +\nonumber \\
 &+&  \epsilon^{\mu}_r(u(p_s))\, \epsilon^{\nu}_u(u(p_s))\,
 \Big[ \epsilon^{ruv}\, {\cal J}^v_c - (\zeta^r_c\, {\cal P}_c^u -
 \zeta^u_c\, {\cal P}_c^r)\Big],\nonumber \\
 &&{}\nonumber \\
 &&\Rightarrow \, m^{\mu}_{c(p_s)} = - S^{\mu\nu}_c\, u_{\nu}(p_s)
 = - \epsilon^{\mu}_r(u(p_s))\, q_c^{r\tau\tau}.
 \label{V12}
 \eea
\medskip

Let us now consider the following possible definitions of
effective centers of motion (many other possibilities
exist)\bigskip

1) {\it Center of energy as center of motion}, ${\vec
\zeta}_{c(E)}(\tau ) = {\vec R}_c(\tau )$, where ${\vec R}_c(\tau
)$ is {\it a 3-center of energy} for the cluster built by means of
the standard definition

\beq
 {\vec R}_c = - {{ {\vec {\cal K}}_c}\over {M_c}}.
 \label{V13}
 \eeq

\medskip

It is determined by the requirement that either the mass dipole
vanishes, $q_c^{r\tau\tau} = 0$ or the mass dipole moment with
respect to $u^{\mu}(p_s)$ vanishes, $m^{\mu}_{c(p_s)} = 0$.
\medskip

The center of energy seems to be the only center of motion
enjoying the simple composition rule

\beq
 {\vec R}_{c(n_1+n_2)} = {{M_{c(n_1)}\, {\vec R}_{c(n_1)} +
 M_{c(n_2)}\, {\vec R}_{c(n_2)}}\over {M_{c(n_1+n_2)}}}.
 \label{V14}
 \eeq

\medskip

The constitutive relation between ${\vec {\cal P}}_c$ and ${\dot
{\vec R}}_c(\tau )$, see Eq.(\ref{IV8}), is

\bea
 0 &=& {{d q^{r\tau\tau}_c}\over {d\tau}} = - {\dot {\cal K}}^r_c -
{\dot M}_c\, R^r_c - M_c\, {\dot R}^r_c,\nonumber \\
 &&{}\nonumber \\
 &&\Downarrow\nonumber \\
 &&{}\nonumber \\
 {\vec {\cal P}}_c &=& M_c\, {\dot {\vec R}}_c + {\dot M}_c\,
 {\vec R}_c -\nonumber \\
 &-& \sum_{i=1}^2\, Q_i\, {\vec \eta}_i(\tau )\, {{{\vec
 \kappa}_i(\tau )}\over {\sqrt{m^2_i + {\vec \kappa}_i^2(\tau )}}}
 \cdot \Big[ {\vec \pi}_{\perp}(\tau ,{\vec \eta}_i(\tau )) +
 \sum_{k\not= i}\, Q_k\, \vec c({\vec \eta}_i(\tau ) - {\vec
 \eta}_k(\tau ))\Big] +\nonumber \\
 &+& \sum_{i=1}^2\, Q_i\,  \Big[ \sum_{k\not= i}\, Q_k\, {{{\vec
 \kappa}_k(\tau )}\over {\sqrt{m^2_k + {\vec \kappa}_k^2(\tau )}}}
 \, c({\vec \eta}_i(\tau ) - {\vec \eta}_k(\tau )) - \int
 d^3\sigma\, {\vec \pi}_{\perp}(\tau ,\vec \sigma )\, {{{\vec
 \kappa}_i(\tau ) \cdot \vec c(\vec \sigma - {\vec \eta}_i(\tau ))}
 \over {\sqrt{m^2_i + {\vec \kappa}_i^2(\tau )}}}\Big] +\nonumber \\
 &+& \sum_{i=1}^2\, Q_i\, \sum_{k\not= i}\, Q_k\, \int d^3\sigma\,
 c(\vec \sigma - {\vec \eta}_i(\tau ))\, \Big({{{\vec
 \kappa}_k(\tau )}\over {\sqrt{m^2_k + {\vec \kappa}_k^2(\tau )}}}
 \cdot \vec \partial \Big)\, \vec c(\vec \sigma - {\vec
 \eta}_k(\tau )) -\nonumber \\
 &=& Q_1Q_2\, \int d^3\sigma\, \vec \sigma\, \Big( \Big[ \Big( {{{\vec
 \kappa}_1(\tau )}\over {\sqrt{m^2_1 + {\vec \kappa}_1^2(\tau )}}}
\cdot \vec \partial\Big)\, \vec c(\vec \sigma - {\vec \eta}_1(\tau
))\Big] \cdot \vec c(\vec \sigma - {\vec \eta}_2(\tau ))
+\nonumber \\
 &+& \vec c(\vec \sigma - {\vec \eta}_1(\tau )) \cdot \Big[ \Big( {{{\vec
 \kappa}_2(\tau )}\over {\sqrt{m^2_2 + {\vec \kappa}_2^2(\tau )}}}
 \cdot \vec \partial \Big)\, \vec c(\vec \sigma - {\vec
 \eta}_2(\tau ))\Big] \Big) -\nonumber \\
 &-& {1\over 2}\, \sum_{i=1}^2\, Q_i\, \sum_{k\not= 1,2}\, Q_k\,
  \int d^3\sigma\, \vec \sigma\, \Big( \Big[ \Big( {{{\vec
 \kappa}_i(\tau )}\over {\sqrt{m^2_i + {\vec \kappa}_i^2(\tau )}}}
\cdot \vec \partial\Big)\, \vec c(\vec \sigma - {\vec \eta}_i(\tau
))\Big] \cdot \vec c(\vec \sigma - {\vec \eta}_k(\tau ))
+\nonumber \\
 &+& \vec c(\vec \sigma - {\vec \eta}_i(\tau )) \cdot \Big[ \Big( {{{\vec
 \kappa}_k(\tau )}\over {\sqrt{m^2_k + {\vec \kappa}_k^2(\tau )}}}
 \cdot \vec \partial \Big)\, \vec c(\vec \sigma - {\vec
 \eta}_k(\tau ))\Big] \Big).
 \label{V15}
 \eea

\medskip

From Eq.(\ref{IV10}) the associated cluster spin tensor is

\bea
  S^{\mu\nu}_c &=& \epsilon^{\mu}_r(u(p_s))\, \epsilon^{\nu}_u(u(p_s))\,
  [q_c^{ru\tau} - q_c^{ur\tau}] =\nonumber \\
  &=& \epsilon^{\mu}_r(u(p_s))\, \epsilon^{\nu}_u(u(p_s))\,
 \epsilon^{ruv}\, \Big[ {\cal J}^v_c - ({\vec R}_c \times {\vec {\cal P}}_c)^v\Big].
 \label{V16}
 \eea

\bigskip

2)  {\it Pirani centroid ${\vec \zeta}_{c(P)}(\tau )$ as center of
motion}. It is determined by the requirement that the mass dipole
moment with respect to 4-velocity ${\dot w}^{\mu}_c(\tau )$
vanishes (it involves the anti-symmetric part of $p^{ur}_c$)

\bea
  m^{\mu}_{c({\dot w}_c)} &=& - S^{\mu\nu}_c\, {\dot w}_{c\nu} = 0,\quad
 \Rightarrow {\dot {\vec \zeta}}_{c(P)} \cdot {\vec \zeta}_{c(P)} = {\dot {\vec
 \zeta}}_{c(P)} \cdot {\vec R}_c,\nonumber \\
 &&\Downarrow \nonumber \\
 &&{}\nonumber \\
 {\vec \zeta}_{c(P)}(\tau ) &=& {1\over {M_c - {\vec {\cal P}}_c
 \cdot {\dot {\vec \zeta}}_{c(P)}(\tau )}}\, \Big[ M_c\, {\vec R}_c -
 {\vec R}_c \cdot {\dot {\vec \zeta}}_{c(P)}(\tau )\, {\vec {\cal P}}_c
 - {\dot {\vec \zeta}}_{c(P)}(\tau ) \times  {\vec {\cal J}}_c\Big].
 \label{V17}
 \eea

Therefore this centroid is implicitly defined as the solution of
these three coupled first order ordinary differential equations.

\bigskip

3) {\it Tulczyjew centroid ${\vec \zeta}_{c(T)}(\tau )$ as center
of motion}. If we define the cluster 4-momentum  $P^{\mu}_c =
M_c\, u^{\mu}(p_s) + {\cal P}^s_c\, \epsilon^{\mu}_s(u(p_s))$
[$P^2_c = M^2_c - {\vec {\cal P}}_c^2 \, {\buildrel {def}\over
=}\, {\cal M}_c^2$], its definition is the requirement that the
mass dipole moment with respect to $P^{\mu}_c$ vanishes (it
involves the anti-symmetric part of $p^{ur}_c$)

\bea
 m^{\mu}_{c(P_c)} &=& - S^{\mu\nu}_c\, P_{c\nu} = 0,\quad
 \Rightarrow {\vec {\cal P}}_c \cdot {\vec \zeta}_{c(T)} = {\vec
 {\cal P}}_c \cdot {\vec R}_c,\nonumber \\
 &&\Downarrow \nonumber \\
 &&{}\nonumber \\
 {\vec \zeta}_{c(T)}(\tau ) &=& {1\over {M^2_c - {\vec {\cal
 P}}_c^2}}\, \Big[ M^2_c\, {\vec R}_c - {\vec {\cal P}}_c \cdot
 {\vec R}_c\, {\vec {\cal P}}_c -
 {\vec {\cal P}}_c \times  {\vec {\cal J}}_c\Big].
 \label{V18}
 \eea
\medskip

Let us show that this centroid satisfies the free particle
relation as constitutive relation

\bea
 {\vec {\cal P}}_c &=& M_c\, {\dot {\vec \zeta}}_{c(T)},\nonumber
 \\
 &&{}\nonumber \\
 &&\Downarrow \nonumber \\
 &&{}\nonumber \\
 &&P^{\mu}_c = M_c\, \Big[ u^{\mu}(p_s) + {\dot \zeta}^s_{c(T)}\,
  \epsilon^{\mu}_s(u(p_s))\Big],\nonumber \\
 &&q^{r\tau\tau}_{c(T)} = {{M_c}\over {M^2_c - {\vec {\cal
 P}}_c^2}}\, \Big[ {\vec {\cal P}}^2_c\, {\vec R}_c + {\vec {\cal
 P}}_c \cdot {\vec R}_c\, {\vec {\cal P}}_c + {\vec {\cal P}}_c
 \times {\vec {\cal J}}_c\Big],\nonumber \\
 &&S^{\mu\nu}_c = [\epsilon^{\mu}_r(u(p_s))\, u^{\nu}(p_s) -
 \epsilon^{\nu}_r(u(p_s))\, u^{\mu}(p_s)]\, q_{c(T)}^{r\tau\tau}
 +\nonumber \\
 &&\quad +  \epsilon^{\mu}_r(u(p_s))\, \epsilon^{\nu}_u(u(p_s))\,
 \epsilon^{ruv}\, \Big[ {\cal J}^v_c - ({\vec \zeta}_{c(T)} \times {\vec {\cal P}}_c)^v\Big].
 \label{V19}
 \eea

If we use Eq.(\ref{V17}) to find a Pirani centroid such that
${\dot {\vec \zeta}}_c = {\vec {\cal P}}_c / M_c$, it turns out
that the condition (\ref{V17}) becomes Eq.(\ref{V18}) and this
implies Eq.(\ref{V19}).

\medskip

The equations of motion

\beq
 M_c(\tau )\, {\ddot {\vec \zeta}}_{c(T)}(\tau ) =
{\dot {\vec {\cal P}}}_c(\tau ) - {\dot M}_c(\tau )\, {\dot {\vec
\zeta}}_{c(T)}(\tau ),
 \label{V20}
  \eeq
\medskip

\noindent  contain both internal and external forces.
Notwithstanding the nice properties (\ref{V19}) and (\ref{V20}) of
the Tulczyjew centroid, this effective center of motion suffers
the drawback of not satisfying a simple composition property. The
relation among the Tulczyjew centroids of clusters with $n_1$,
$n_2$ and $n_1+n_2$ particles respectively is much more
complicated of the composition (\ref{V14}) of  the centers of
energy.

\medskip

All the previous centroids coincide for an isolated system in the
rest-frame instant form with ${\vec {\cal P}}_c = {\vec
\kappa}_{+} \approx 0$ in the gauge ${\vec q}_{+} \approx {\vec
R}_{+} \approx {\vec y}_{+} \approx 0$.
\medskip

4) The {\it Corinaldesi-Papapetrou centroid with respect to a
time-like observer with 4-velocity $v^{\mu}(\tau )$, ${\vec
\zeta}^{(v)}_{c(CP)}(\tau )$ as center of motion}.

\beq
 m^{\mu}_{c(v)} = - S^{\mu\nu}_c\, v_{\nu} = 0.
 \label{V21}
 \eeq

Clearly these centroids are unrelated to the previous ones being
dependent on the choice of an arbitrary observer.

\medskip

5) The {\it Pryce center of spin or classical canonical
Newton-Wigner centroid} ${\vec \zeta}_{c(NW)}$. It defined as the
solution of the differential equations implied by the requirement
$\{ \zeta^r_{c(NW)}, \zeta^s_{c(NW)} \} = 0$, $\{ \zeta^r_{c(NW)},
{\cal P}_c^s \} = \delta^{rs}$. Let us remark that, being in an
instant form of dynamics, we have $\{ {\cal P}_c^r, {\cal P}_c^s
\} = 0$ also for an open system.

\bigskip

The two effective centers of motion which look more useful for
applications seems to be the center of energy ${\vec
\zeta}_{c(E)}(\tau )$ and Tulczyjew's centroid ${\vec
\zeta}_{c(T)}(\tau )$, with ${\vec \zeta}_{c(E)}(\tau )$ preferred
for the study of the mutual motion of clusters due to
Eq.(\ref{V14}).

\bigskip

Therefore, in the spirit of the multipolar expansion, our two-body
cluster may be described by an effective non-conserved internal
energy (or mass) $M_c(\tau ) $, by the world-line $w^{\mu}_c =
x^{\mu}_o + u^{\mu}(p_s)\, \tau + \epsilon^{\mu}_r(u(p_s))\,
\zeta_{c(E\, or\, T)}^r(\tau )$ associated with the effective
center of motion ${\vec \zeta}_{c(E\, or\, T)}(\tau )$ and by the
effective 3-momentum ${\vec {\cal P}}_c(\tau )$, with ${\vec
\zeta}_{c(E\, or\, T)}(\tau )$ and ${\vec {\cal P}}_c(\tau )$
forming a non-canonical basis for the collective variables of the
cluster. A non-canonical effective spin for the cluster in the 1)
and 3) cases is defined by\medskip

\bea
 &&a)\, case\, of\, the\, center\, of\, energy,\nonumber \\
 &&{}\nonumber \\
 {\vec {\cal S}}_{c(E)}(\tau ) &=& {\vec {\cal J}}_c(\tau ) - {\vec R}_c(\tau )
 \times {\vec {\cal P}}_c(\tau ),\nonumber \\
 &&{}\nonumber \\
 {{d {\vec \zeta}_{c(E)}(\tau )}\over {d\tau}} &=& {{d {\vec {\cal
 J}}_c(\tau )}\over {d\tau }} - {{d {\vec R}_c(\tau )}\over {d\tau}}\, \times {\vec
 {\cal P}}_c(\tau ) - {\vec R}_c(\tau ) \times
 {{d {\vec {\cal P}}_c(\tau )}\over {d\tau}}, \nonumber \\
 &&{}\nonumber \\
 &&{}\nonumber \\
 &&b)\, case\, of\, the\, Tulczyjew\, centroid,\nonumber \\
 &&{}\nonumber \\
 {\vec {\cal S}}_{c(T)}(\tau ) &=& {\vec {\cal J}}_c(\tau ) - {\vec \zeta}_{c(T)}(\tau )
 \times {\vec {\cal P}}_c(\tau )= \nonumber \\
 &=& {{M^2_c(\tau )\, {\vec {\cal S}}_{c(E)}(\tau ) - {\vec {\cal
 P}}_c(\tau ) \cdot {\vec {\cal J}}_c(\tau )\, {\vec {\cal
 P}}_c(\tau )}\over {M^2_c(\tau ) - {\vec {\cal P}}_c^2(\tau
 )}},\nonumber \\
 &&{}\nonumber \\
 {{d {\vec \zeta}_{c(T)}(\tau )}\over {d\tau}} &=& {{d {\vec {\cal
 J}}_c(\tau )}\over {d\tau }} - {\vec \zeta}_{c(T)}(\tau ) \times
 {{d {\vec {\cal P}}_c(\tau )}\over {d\tau}}.
 \label{V22}
 \eea

Since our cluster contains only two particles, this pole-dipole
description concentrated on the world-line $w^{\mu}_c(\tau )$ is
equivalent to the original description in terms of the canonical
variables ${\vec \eta}_i(\tau )$, ${\vec \kappa}_i(\tau )$ (all
the higher multipoles are not independent quantities in this
case).
\medskip

Let us see whether it is possible to replace the description of
the two body system as an effective pole-dipole system with a
description as an effective extended two-body system by
introducing two non-canonical relative variables ${\vec
\rho}_{c(E\, or\, T)}(\tau )$, ${\vec \pi}_{c(E\, or\, T)}(\tau )$
with the following definitions
\medskip

\bea
 {\vec \eta}_1 &{\buildrel {def}\over =}& {\vec \zeta}_{c(E\, or\, T)} + {1\over
 2}\, {\vec \rho}_{c(E\, or\, T)},\qquad {\vec \zeta}_{c(E\, or\, T)} = {1\over 2}\, ({\vec
 \eta}_1 + {\vec \eta}_2),\nonumber \\
 {\vec \eta}_2  &{\buildrel {def}\over =}& {\vec \zeta}_{c(E\, or\, T)} - {1\over
 2}\, {\vec \rho}_{c(E\, or\, T)},\qquad {\vec \rho}_{c(E\, or\, T)} = {\vec \eta}_1 - {\vec
 \eta}_2,\nonumber \\
 &&{}\nonumber \\
 {\vec \kappa}_1  &{\buildrel {def}\over =}& {1\over 2}\, {\vec
 {\cal P}}_c + {\vec \pi}_{c(E\, or\, T)},\qquad {\vec {\cal P}}_c = {\vec
 \kappa}_1 + {\vec \kappa}_2,\nonumber \\
 {\vec \kappa}_2  &{\buildrel {def}\over =}& {1\over 2}\, {\vec
 {\cal P}}_c - {\vec \pi}_{c(E\, or\, T)},\qquad {\vec \pi}_{c(E\, or\, T)} = {1\over 2}\,
 ({\vec \kappa}_1 - {\vec \kappa}_2),\nonumber \\
 &&{}\nonumber \\
 {\vec {\cal J}}_c &=& {\vec \eta}_1 \times {\vec \kappa}_1 +
 {\vec \eta}_2 \times {\vec \kappa}_2 = {\vec \zeta}_{c(E\, or\, T)} \times {\vec
 {\cal P}}_c + {\vec \rho}_{c(E\, or\, T)} \times {\vec \pi}_{c(E\, or\, T)},\nonumber \\
 &&{}\nonumber \\
 &&\Rightarrow\,\, {\vec {\cal S}}_{c(E\, or\, T)} = {\vec \rho}_{c(E\, or\, T)} \times {\vec
 \pi}_{c(E\, or\, T)}.
 \label{V23}
 \eea
\medskip

\noindent Even if suggested by a canonical transformation, it is
{\it not} a canonical transformation and it only exists because we
are working in an instant form of dynamics in which both ${\vec
{\cal P}}_c$ and ${\vec {\cal J}}_c$ do not depend on the
interactions.
\medskip

Note that we know everything about this new basis except for the
unit vector ${\vec \rho}_{c(E\, or\, T)}/|{\vec \rho}_{c(E\, or\,
T)}|$ and the momentum ${\vec \pi}_{c(E\, or\, T)}$. The relevant
lacking information can be extracted from the symmetrized momentum
dipole $p^{ru}_c + p^{ur}_c$, which  is a known effective quantity
due to Eq.(\ref{V9}) having the following expression in terms of
the variables (\ref{V23})

\bea
  &&p^{ru}_c + p^{ur}_c +  2\, \sum_{i=1}^2\, Q_i\, \int d^3\sigma\, c(\vec
 \sigma - {\vec \eta}_i(\tau ))\, [\partial^r\, A^s_{\perp} + \partial^s\, A^r_{\perp}]
 (\tau ,\vec \sigma ) =\nonumber \\
  &&\quad = \sum_{i=1}^2\,
 (\eta^r_i\, \kappa_i^u + \eta^u_i\, \kappa^r_i) = \zeta^r_{c(E\, or\, T)}\,
 {\cal P}_c^u + \zeta^u_{c(E\, or\, T)}\, {\cal P}^r_c +\nonumber \\
 &+& \rho^r_{c(E\, or\, T)}\, \pi^u_{c(E\, or\, T)} +
 \rho^u_{c(E\, or\, T)}\, \pi^r_{c(E\, or\, T)}.\nonumber \\
 &&{}
 \label{V24}
 \eea

\medskip

A strategy for getting this information is to construct a {\it
spin frame}, which, following Ref.\cite{iten1} for the $N=2$ case,
is defined by ${\hat {\cal S}}_{c(E\, or\, T)} = {\vec {\cal
S}}_{c(E\, or\, T)} / |{\vec {\cal S}}_{c(E\, or\, T)}|$, ${\hat
{\cal R}}_{c(E\, or\, T)}$, ${\hat {\cal V}}_{c(E\, or\, T)} =
{\hat {\cal R}}_{c(E\, or\, T)} \times {\hat {\cal S}}_{c(E\, or\,
T)}$, with ${\hat {\cal S}}_{c(E\, or\, T)} \cdot {\hat {\cal
R}}_{c(E\, or\, T)} = 0$, ${\hat {\cal S}}_{c(E\, or\, T)}^2 =
{\hat {\cal R}}_{c(E\, or\, T)}^2 = {\hat {\cal V}}_{c(E\, or\,
T)}^2 = 1$. Then we get the following decomposition ((${\cal
S}_{c(E\, or\, T)} = |{\vec {\cal S}}_{c(E\, or\, T)}|$)

\bea
 {\vec \rho}_{c(E\, or\, T)} &=& \rho_{c(E\, or\, T)}\,
 {\hat {\cal R}}_{c(E\, or\, T)},\qquad \rho_{c(E\, or\, T)} = |
 {\vec \eta}_1 - {\vec \eta}_2|,\nonumber \\
 &&{}\nonumber \\
 {\vec \pi}_{c(E\, or\, T)} &=& {\tilde \pi}_{c(E\, or\, T)}\,
 {\hat {\cal R}}_{c(E\, or\, T)} - {{ {\cal S}_{c(E\, or\, T)}}\over
 {\rho_{c(E\, or\, T)}}}\, {\hat {\cal V}}_{c(E\, or\, T)},\qquad
{\tilde \pi}_{c(E\, or\, T)} = {\vec \pi}_{c(E\, or\, T)} \cdot
{{{\vec \rho}_{c(E\, or\, T)}}\over {\rho_{c(E\, or\,
T)}}},\nonumber \\
 &&{}
 \label{V25}
 \eea

\noindent where $\rho_{c(E\, or\, T)}$ is just the relative
variable appearing in the Coulomb potential. Eqs.(\ref{V25}) show
that only the three variables ${\tilde \pi}_{c(E\, or\, T)}$ and $
{\hat {\cal R}}_{c(E\, or\, T)} = {\vec \rho}_{c(E\, or\, T)}/
\rho_{c(E\, or\, T)}$ are still unknown. Then from Eqs.(\ref{V23})
and (\ref{V24}) we get

\bea
 \rho^r_{c(E\, or\, T)}\, \pi^u_{c(E\, or\, T)} +
 \rho_{c(E\, or\, T)}^u\, \pi^r_{c(E\, or\, T)}
 &=& 2\, \rho_{c(E\, or\, T)}\, {\tilde
 \pi}_{c(E\, or\, T)}\, {\hat {\cal R}}^r_{c(E\, or\, T)}\,
  {\hat {\cal R}}^u_{c(E\, or\, T)} -\nonumber \\
  &-& {\cal S}_{c(E\, or\, T)}\, \Big( {\hat {\cal R}}^r_{c(E\, or\, T)}\,
  {\hat {\cal V}}^u_{c(E\, or\, T)} + {\hat {\cal
 R}}^u_{c(E\, or\, T)}\, {\hat {\cal V}}^r_{c(E\, or\, T)}\Big) =\nonumber \\
 &=& p^{ru}_c + p^{ur}_c - (\zeta^r_{c(E\, or\, T)}\, {\cal P}^u_c +
 \zeta^u_{c(E\, or\, T)}\, {\cal P}^r_c) +\nonumber \\
 &+&  2\, \sum_{i=1}^2\, Q_i\, \int d^3\sigma\, c(\vec
 \sigma - {\vec \eta}_i(\tau ))\, [\partial^r\, A^u_{\perp} + \partial^u\, A^r_{\perp}]
 (\tau ,\vec \sigma ) =\nonumber \\
 &{\buildrel {def}\over =}& F^{ru}_{c(E\, or\, T)},\nonumber \\
 &&{}\nonumber \\
 &&\qquad F^{ru}_{c(E\, or\, T)}\, {\cal S}^u_{c(E\, or\, T)} \equiv 0.
 \label{V26}
 \eea

But these are three independent equations for  ${\tilde
\pi}_{c(E\, or\, T)}$ and for the two degrees of freedom in the
unit vector ${\hat {\cal R}}_{c(E\, or\, T)}$ in terms of the
known quantities $F^{ru}_{c(E\, or\, T)}$, ${\cal S}_{c(E\, or\,
T)}$, $\rho_{c(E\, or\, T)} = |{\vec \eta}_1 - {\vec \eta}_2|$.
For instance we get ${\tilde \pi}_{c(E\, or\, T)} = (\sum_r\,
F^{rr}_{c(E\, or\, T)})/ 2\, \rho_{c(E\, or\, T)}$: due to the
transversality of the vector potential ${\tilde \pi}_{c(E\, or\,
T)}$ does not depend on it. In conclusion the external
electro-megnetic potential ${\vec A}_{\perp}$ enters only in the
determination of the axis ${\hat {\cal R}}_{c(E\, or\, T)}$ of the
spin frame.
\medskip

This completes the construction of the effective relative
variables and of the effective spin frame using the extra input of
the 3-momentum dipole. In this way we get a description of the
two-body cluster as an effective two-body system instead of a
pole-dipole system. However, the weak point of this description of
the open system as an extended object is that, whatever definition
of effective center of motion one uses, the symmetrized momentum
dipole $p_c^{ru} + p_c^{ur}$ {\it does not depend only} on the
cluster properties but {\it also} on the external electro-magnetic
transverse vector potential at the particle positions, as shown by
Eq.(\ref{V8}). As a consequence the spin frame, or equivalently
the 3 Euler angles associated with the internal spin, depends upon
the external fields.

\medskip

If we accept this drawback, it is reasonable that, by taking into
account higher multipoles, it is possible to give a description of
a cluster of $n \geq 3$ particles in terms of as many  effective
$n$-body systems as effective dynamical body frames following the
scheme of Ref.\cite{iten1}.
\bigskip

This would open the possibility to have effective descriptions of
two clusters of $n_1$ and $n_2$ particles, respectively, and to
compare it with the effective description of the cluster composed
by the same $n_1 + n_2$ particles to find the relation between the
three centers of motion  of the $(n_1 + n_2)$-, $n_1$- and $n_2$-
clusters and the relative motion of the two $n_1$- and $n_2$-
clusters. To this end the use of {\it the center of energy as
center of motion} seems unavoidable due to the simple composition
law (\ref{V14}). Whatever choice we adopt, however, it turns out
that the relative motion of the two clusters depends on the
external fields besides the effective parameters of the clusters.

These techniques can be extended to relativistic perfect fluids,
if described in the rest-frame instant form as done in Refs.
\cite{92}. Moreover, they are needed for the determination of the
post-Minkowskian approximation to the quadrupole formula for the
emission of gravitational waves (re-summation of the
post-Newtonian approximations) in the background-independent
Hamiltonian linearization of tetrad gravity \cite{93} plus a
perfect fluid \cite{80}.

\vfill\eject

\section{Conclusions.}

A relativistic description of open systems like binary stars
embedded in the gravitational field would be an important
achievement nowadays in view of the construction of templates for
the gravitational radiation. Even by approximating such
description by means of a multipolar expansion in a way suitable
for doing actual calculations, either analytical or numerical, a
big amount of kinematical technical preliminaries is needed
anyway. With this in view, we had in mind to develop methods which
could be useful in general relativity with relativistic perfect
fluids as matter, where single or binary stars could be described
by open fluid subsystems of the isolated system formed by the
gravitational field plus the fluid in the rest-frame instant form
of either metric or tetrad gravity \cite{93,80}.

To pursue our program, in the present paper we have first of all
completed the study of the relativistic kinematics of the system
of N free scalar positive-energy particles in the rest-frame
instant form of dynamics on Wigner hyper-planes, initiated in
Ref.\cite{iten1}.

Then, we have evaluated the energy momentum tensor of the system
on the Wigner hyperplane and then determined Dixon's multipoles
for the N-body problem with respect to the {\it internal} 3-center
of mass located at the origin of the Wigner hyperplane \cite{c34}.
For an isolated system most of the existing definitions of a
collective centroid identify a unique world-line, associated with
the internal canonical 3-center of mass. In the rest-frame instant
form these multipoles are {\it Cartesian (Wigner-covariant)
Euclidean tensors}. While the study of the {\it monopole} and {\it
dipole} moments in the rest frame gives information on the mass,
the spin and the {\it internal} center of mass, the {\it
quadrupole} moment provides the only (though not unique) way of
introducing the concept of {\it barycentric tensor of inertia} for
extended systems in special relativity.

By exploiting the {\it canonical spin bases} of
Refs.\cite{iten2,iten1}, after the elimination of the {internal}
3-center of mass (${\vec q}_{+}={\vec \kappa}_{+}=0$), the
Cartesian multipoles $q_T^{r_1...r_nAB}$ can be expressed in terms
of 6 {\it orientational} variables (the {\it spin vector} and the
three {\it Euler angles} identifying the {\it dynamical body
frame}) and of $6N-6$ ({\it rotational scalar}) {\it shape}
variables, i.e. in terms of the canonical pairs of a canonical
spin basis.
\bigskip

Having completed the discussion of the isolated system of $N$
positive energy free scalar particles the previous formalism has
been applied to an isolated system of $N$ positive-energy
particles with {\it mutual action-at-a-distance interactions}.
Here again we find a unique world-line describing the collective
motion of the system.
\medskip

On the other hand, in the case of an {\it open $n < N$ particle
subsystem} of an isolated system consisting of $N$ charged
positive-energy particles  plus the electro-magnetic field a more
complex description appears. In the rest frame of the isolated
system a suitable definition of the energy-momentum tensor of the
open subsystem allows to define its {\it effective mass,
3-momentum and angular momentum}. However, unlike the case of
isolated systems, each centroid putatively describing the {\it
collective centers of motion}, gives rise to a different
world-line. Starting from the evaluation of the rest-frame Dixon
multipoles of the energy-momentum tensor of the open subsystem
with respect to various centroids we are given therefore {\it many
candidates for an effective center of motion and for an effective
intrinsic spin}. Two centroids (the center of energy and Tulczyjew
centroid) seems to be preferred because of their specific
properties. In the case $n = 2$ it is possible to replace the
pole-dipole description of the 2-particle cluster with a
description of the cluster as an extended system (whose effective
spin frame can be evaluated) at the price of introducing an
explicit dependence on the action of the external electro-magnetic
field upon the cluster.

Finally, by comparing the effective parameters of an open cluster
of $n_1 + n_2$ particles with the effective parameters of the two
clusters with $n_1$ and $n_2$ particles, it is shown that only the
effective {\it center of energy} can in fact play the role of a
useful center of motion,

The kinematical concepts we have defined for closed and open
N-body systems are enough for the treatment of relativistic
continua like relativistic fluids. In \cite{92} a preliminary
extension to closed relativistic fluids is given.

\vfill\eject

\appendix

\section{Non-Relativistic Multipolar Expansions for N Free  Particles.}

In the review paper of Ref.\cite{dixon1} it can be found a study
of the Newtonian multipolar expansions for a continuum isentropic
distribution of matter characterized by a mass density $\rho
(t,\vec \sigma )$,  a velocity field $U^r(t,\vec \sigma )$, and a
stress tensor $\sigma^{rs}(t,\vec \sigma )$, with $\rho (t,\vec
\sigma ) \vec U(t,\vec \sigma )$ the momentum density. In case the
system is isolated, the only dynamical equations are the mass
conservation and the continuum equations of motion, respectively
\medskip

\bea &&{{\partial \rho (t,\vec \sigma )}\over {\partial
t}}-{{\partial \rho (t,\vec \sigma ) U^r(t,\vec \sigma )}\over
{\partial \sigma^r}}=0,\nonumber  \\
 &&{{\partial \rho (t,\vec \sigma )U^r(t,\vec \sigma )}\over {\partial
t}}-{{\partial [\rho U^r U^s -\sigma^{rs}](t,\vec \sigma )}\over
{\partial \sigma^s}}\, {\buildrel \circ \over =}\, 0.
 \label{a1}
\eea

We can adapt this description to an isolated system of N particles
in the following way. The mass density

\begin{equation}
\rho (t,\vec \sigma )=\sum_{i=1}^N m_i \delta^3(\vec \sigma -{\vec
\eta}_i(t)),
 \label{a2}
\end{equation}

\noindent satisfies

\begin{equation}
{{\partial \rho (t,\vec \sigma )}\over {\partial
t}}=-\sum_{i=1}^Nm_i {\dot {\vec \eta}}_i(t)\cdot {\vec
\partial}_{{\vec \eta}_i} \delta^3(\vec \sigma -{\vec
\eta}_i(t))\, {\buildrel {def} \over =}\, {{\partial}\over
{\partial \sigma^r}}[\rho U^r](t,\vec \sigma ),
 \label{a3}
\end{equation}

\noindent while the momentum density \cite{c35} is

\begin{equation}
\rho (t,\vec \sigma ) U^r(t,\vec \sigma )= \sum_{i=1}^N m_i {\dot
{\vec \eta}}_i(t) \delta^3(\vec \sigma -{\vec \eta}_i(t )),
 \label{a4}
\end{equation}

\noindent The associated constant of motion is the total mass
$m=\sum_{i=1}^N$.
\medskip

If we define a function $\zeta (\vec \sigma ,{\vec \eta}_i)$
concentrated in the N points ${\vec \eta}_i$, i=1,..,N, such that
$\zeta (\vec \sigma ,{\vec \eta}_i)=0$ for $\vec \sigma \not=
{\vec \eta}_i$ and $\zeta ({\vec \eta}_i,{\vec
\eta}_j)=\delta_{ij}$ \cite{c36}, the velocity field associated to
N particles becomes

\begin{equation}
\vec U(t,\vec \sigma )= \sum_{i=1}^N {\dot {\vec \eta}}_i(t) \zeta
(\vec \sigma ,{\vec \eta}_i(t)).
 \label{a5}
\end{equation}

The continuum equations of motion are replaced by

\begin{eqnarray}
&&{{\partial}\over {\partial t}} [\rho (t,\vec \sigma ) U^r(t,\vec
\sigma )]\, {\buildrel \circ \over =}\, {{\partial}\over {\partial
\sigma^s}} \sum_{i=1}^N m_i {\dot \eta}^r_i(t) {\dot \eta}^s_i(t)
\delta^3(\vec \sigma -{\vec \eta}_i(t)) +\sum_{i=1}^N m_i {\ddot
\eta}^r_i(t)=\nonumber \\
 &&{\buildrel {def} \over =}\,
 {{\partial [\rho U^r U^s -\sigma^{rs}](t,\vec \sigma )}\over
{\partial \sigma^s}}.
 \label{a6}
\end{eqnarray}

For a system of free particles we have ${\ddot {\vec \eta}}_i(t)\,
{\buildrel \circ \over =}\, 0$ so that $\sigma^{rs}(t,\vec \sigma
)=0$. If there are interparticle interactions, they will determine
the effective stress tensor.

\bigskip
Let us consider an arbitrary point $\vec \eta (t)$. The {\it
multipole moments} of the mass density $\rho$ and momentum density
$\rho \vec U$ and of the stress-like density $\rho U^rU^s$  with
respect to the point $\vec \eta (t)$ are defined by setting ($N
\geq 0$)

\begin{eqnarray}
 m^{r_1...r_n}[\vec \eta (t) ]&=& \int d^3\sigma
[\sigma^{r_1}-\eta^{r_1}(t)]...[\sigma^{r_n}-\eta^{r_n}(t)]\rho
(\tau ,\vec \sigma )=\nonumber \\
 &=&\sum_{i=1}^N m_i
[\eta^{r_1}_i(\tau
)-\eta^{r_1}(t)]...[\eta^{r_n}_i(t)-\eta^{r_n}(t)],
 \nonumber \\
 &&n=0\quad\quad m[\vec \eta (t)]=m=\sum_{i=1}^N m_i,\nonumber \\
 &&{}\nonumber \\
 p^{r_1...r_nr}[\vec \eta (t) ]&=& \int d^3\sigma
[\sigma^{r_1}-\eta^{r_1}(t)]...[\sigma^{r_n}-\eta^{r_n}(t)]\rho
(t,\vec \sigma )U^r(t,\vec \sigma )=\nonumber \\
 &=&\sum_{i=1}^N m_i {\dot \eta}^r_i(t)
[\eta^{r_1}_i(t)-\eta^{r_1}(t)]...[\eta^{r_n}_i(t)-\eta^{r_n}(t)],
\nonumber \\
 && n=0\quad\quad p^r[\vec \eta (t)]=\sum_{i=1}^Nm_i{\dot \eta}_i(t)
 =\sum_{i=1}^N\kappa_i^r=\kappa^r_{+}\approx 0,\nonumber \\
  &&{}\nonumber \\
  p^{r_1...r_nrs}[\vec \eta (t) ]&=& \int d^3\sigma
[\sigma^{r_1}-\eta^{r_1}(t)]...[\sigma^{r_n}-\eta^{r_n}(t)]\rho
(t,\vec \sigma )U^r(t,\vec \sigma )U^s(t,\vec \sigma )=\nonumber
\\
 &=&\sum_{i=1}^N m_i {\dot \eta}^r_i(t){\dot \eta}^s_i(t)
[\eta^{r_1}_i(t)-\eta^{r_1}(t)]...[\eta^{r_n}_i(t)-\eta^{r_n}(t)].
 \label{a7}
 \end{eqnarray}

The {\it mass monopole} is the conserved mass, while the {\it
momentum monopole} is the total 3-momentum, vanishing in the rest
frame.
\medskip

If the {\it mass dipole} vanishes, the point $\vec \eta (t)$ is
the {\it center of mass}:

\begin{equation}
m^r[\vec \eta (t)]=\sum_{i=1}^Nm_i[\eta^r_i(t)-\eta^r(t)]=0
\Rightarrow \vec \eta (t)={\vec q}_{nr}.
 \label{a8}
\end{equation}

The time derivative of the mass dipole is

\begin{equation}
{{d m^r[\vec \eta (t)]}\over {dt}} = p^r[\vec \eta (t)]-m {\dot
\eta}^r(t)=\kappa_{+}^r-m{\dot \eta}^r(t).
 \label{a9}
\end{equation}

When $\vec \eta (t)={\vec q}_{nr}$, from the vanishing of this
time derivative we get the {\it momentum-velocity relation for the
center of mass}

\begin{equation}
p^r[{\vec q}_{nr}]=\kappa^r_{+} = m {\dot q}_{+}^r \quad [\approx
0\, in\, the\, rest\, frame].
 \label{a10}
\end{equation}

The {\it mass quadrupole} is

\begin{equation}
m^{rs}[\vec \eta
(t)]=\sum_{i=1}^Nm_i\eta^r_i(t)\eta^s_i(t)-m\eta^r(t)\eta^s(t)-
\Big( \eta^r(t)m^s[\vec \eta (t)]+\eta^s(t)m^r[\vec \eta
(t)]\Big),
 \label{a11}
\end{equation}

\noindent so that the {\it barycentric mass quadrupole and  tensor
of inertia} are, respectively

\begin{eqnarray}
m^{rs}[{\vec q}_{nr}]&=&\sum_{i=1}^N m_i\eta^r_i(t)\eta^s_i(t)-m
q^r_{nr}q^s_{nr},\nonumber \\
 &&{}\nonumber \\
 I^{rs}[{\vec q}_{nr}]&=&\delta^{rs} \sum_um^{uu}[{\vec q}_{nr}]-m^{rs}[{\vec
 q}_{nr}]=\nonumber \\
 &=&\sum_{i=1}m_i[\delta^{rs} {\vec \eta}_i^2(t)-\eta^r_i(t)\eta^s_i(t)]-
 m[\delta^{rs}{\vec q}^2_{nr}-q^r_{nr}q^s_{nr}]=\nonumber \\
 &=&\sum_{a,b}^{1...N-1}k_{ab}({\vec \rho}_a\cdot {\vec
\rho}_b\delta^{rs}-\rho^r_a\rho^s_b) ,\nonumber \\ \Rightarrow&&
m^{rs}[{\vec q}_{nr}]=\delta^{rs}\sum_{a,b=1}^{N-1}k_{ab}\, {\vec
\rho}_a\cdot {\vec \rho}_b-I^{rs}[{\vec q}_{nr}].
 \label{a12}
\end{eqnarray}

The antisymmetric part of the barycentric momentum dipole gives
rise to the {\it spin vector} in the following way

\begin{eqnarray}
p^{rs}[{\vec q}_{nr}]&=&\sum_{i=1}^Nm_i\eta^r_i(t){\dot
\eta}^s_i(t)-q^r_{nr}p^s[{\vec
q}_{nr}]=\sum_{i=1}^N\eta^r_i(t)\kappa_i^s(t)-
q^r_{+}\kappa^s_{+},\nonumber \\
 &&{}\nonumber \\
 S^u&=&{1\over 2}\epsilon^{urs} p^{rs}[{\vec q}_{nr}]=\sum_{a=1}^{N-1}
 ({\vec \rho}_a\times {\vec \pi}_{qa})^u.
 \label{a13}
\end{eqnarray}

The {\it multipolar expansions} of the mass and momentum densities
around the point  $\vec \eta (t)$ are

\begin{eqnarray}
\rho (t,\vec \sigma )&=& \sum_{n=0}^{\infty} {{m^{r_1....r_n}[\vec
\eta ]}\over {n!}} {{\partial^n}\over {\partial
\sigma^{r_1}...\partial \sigma^{r_n}}} \delta^3(\vec \sigma -{\vec
\eta }(t)),\nonumber \\ &&{}\nonumber \\ \rho (t,\vec \sigma )
U^r(t,\vec \sigma )
 &=&\sum_{n=0}^{\infty}{{p^{r_1....r_nr}[\vec \eta ]}\over {n!}}
{{\partial^n}\over {\partial \sigma^{r_1}...\partial
\sigma^{r_n}}} \delta^3(\vec \sigma -{\vec \eta }(t)).
 \label{a14}
\end{eqnarray}

Finally, for the {\it barycentric multipolar expansions} we have

\begin{eqnarray}
\rho (t,\vec \sigma )&=&m \delta^3(\vec \sigma -{\vec q}_{nr})-
{1\over 2}(I^{rs}[{\vec q}_{nr}]-{1\over
2}\delta^{rs}\sum_uI^{uu}[{\vec q}_{nr}]){{\partial^2}\over
{\partial \sigma^r\partial \sigma^s}} \delta^3(\vec \sigma -{\vec
q}_{nr})+ \nonumber \\ &+&\sum_{n=3}^{\infty}
{{m^{r_1....r_n}[{\vec q}_{nr}]}\over {n!}} {{\partial^n}\over
{\partial \sigma^{r_1}...\partial \sigma^{r_n}}} \delta^3(\vec
\sigma -{\vec q}_{nr}),\nonumber \\ &&{}\nonumber \\ \rho (t,\vec
\sigma ) U^r(t,\vec \sigma )&=&\kappa^r_{+} \delta^3(\vec \sigma
-{\vec q}_{nr})+\Big[ {1\over 2}\epsilon^{rsu}S^u+p^{(sr)}[{\vec
q}_{nr}] \Big] {{\partial}\over {\partial \sigma^s}} \delta^3(\vec
\sigma - {\vec q}_{nr})+\nonumber \\
 &+&\sum_{n=2}^{\infty}{{p^{r_1....r_nr}[{\vec q}_{nr}]}\over {n!}}
{{\partial^n}\over {\partial \sigma^{r_1}...\partial
\sigma^{r_n}}} \delta^3(\vec \sigma -{\vec q}_{nr}).
 \label{a15}
\end{eqnarray}

\vfill\eject

\section{Symmetric Trace-Free Tensors.}

In the applications to gravitational radiation, {\it irreducible
symmetric trace-free Cartesian tensors (STF tensors)}
\cite{sachs,pir,thorne,luc} are needed instead of {\it Cartesian
tensors}. While a {\it Cartesian multipole tensor of rank $l$}
(like the rest-frame Dixon multipoles) on $R^3$ has $3^l$
components, ${1\over 2}(l+1)(l+2)$ of which are in general
independent, a {\it spherical multipole moment of order $l$} has
only $2l+1$ independent components. Even if spherical multipole
moments are preferred in calculations of molecular interactions,
spherical harmonics have various disadvantages in numerical
calculations: for analytical and numerical calculations Cartesian
moments are often more convenient (see for instance
Ref.\cite{hinsen} for the case of the electrostatic potential). It
is therefore preferable using the {irreducible Cartesian STF
tensors}\cite{gel,coope} (having $2l+1$ independent components if
of rank $l$), which are obtained by using {\it Cartesian spherical
(or solid) harmonic tensors} in place of spherical harmonics.

Given an Euclidean tensor $A_{k_1...k_I}$ on $R^3$, one defines
the completely symmetrized tensor $S_{k_1..k_I} \equiv
A_{(k_1..k_I)} = {1\over {I!}} \sum_{\pi} A_{k_{\pi (1)}...k_{\pi
(I)}}$. Then, the associated STF tensor is obtained by removing
all traces ($[I/2]=$ largest integer $\leq I/2$)\medskip

\bea
 A_{k_1...k_I}^{(STF)} &=&\sum_{n=0}^{[I/2]} a_n\, \delta_{(k_1k_2} ...
\delta_{k_{2n-1}k_{2n}} S_{k_{2n+1}...k_I)
i_1i_1...j_nj_n},\nonumber \\
 &&{}\nonumber \\
 a_n &\equiv& (-1)^n {{ l! (2l-2n-1)!!}\over {(l-2n)! (2l-1)!!(2n)!!}}.
 \label{b1}
 \eea

\noindent For instance $(T_{abc})^{STF} \equiv T_{(abc)}- {1\over
5}\Big[\delta_{ab} T_{(iic)}+
\delta_{ac}T_{(ibi)}+\delta_{bc}T_{(aii)}\Big]$.

\vfill\eject

\section{The Gartenahus-Schwartz Transformation.}

In Ref.\cite{iten1} we defined canonical internal relative
variables with respect to the internal 3-center of mass
$\vec{q}_+$ by exploiting a {\it Gartenhaus-Schwartz} canonical
transformation. The canonical generator of this transformation is

\begin{equation}
G=\vec{q}_+\cdot\vec{\kappa}_+,
 \label{c1}
 \end{equation}

\noindent so that the finite transformation, depending on a
parameter $\alpha$, on a generic function $F$ on the phase space
is

\begin{equation}
F(\alpha)=F+\int^\alpha_0
d\overline{\alpha}\;\{F(\overline{\alpha}),G(\overline{\alpha})\}.
 \label{c2}
 \end{equation}

In particular we have

\begin{equation}
\lim_{\alpha\rightarrow\infty}\vec{\kappa}_+(\alpha)=0,\;\;\;
\lim_{\alpha\rightarrow\infty}\vec{q}_+(\alpha)=\infty .
 \label{c3}
 \end{equation}

\medskip

As said in Section II, if we define the  canonical transformation
(\ref{II19}), then the quantities

\begin{equation}
\vec{\pi}_{qa}=\lim_{\alpha\rightarrow\infty}\vec{\pi}_a(\alpha),\;\;\;
\vec{\rho}_{qa}=\lim_{\alpha\rightarrow\infty}\vec{\rho}_a(\alpha),
 \label{c4}
 \end{equation}

\noindent are well defined and the transformation

\beq
 \vec{\eta}_i,\vec{\kappa}_i \rightarrow
\vec{\kappa}_+,\vec{q}_+,\vec{\rho}_{qa},\vec{\pi}_{qa},
 \label{c5}
 \eeq

\noindent is a canonical transformation

\begin{equation}
\{q^r_+,\kappa^s_+\}=\delta^{rs},\;\;\{\rho_{qa}^r,\pi_{qb}^s\}=\delta^{rs}\delta_{ab},
 \label{c6}
 \end{equation}

\noindent as said in Eq.(\ref{II18}).

\medskip

The quantities $\vec{\rho}_{qa},\vec{\pi}_{qa}$ are the searched
internal relative variables: they describe the system after the
gauge fixing $\vec{q}_+\approx 0,\;\; \vec{\kappa}_+\approx 0$. We
have also

\begin{equation}
\vec{\kappa}_+\approx 0\Rightarrow
\vec{\rho}_{qa}\approx\vec{\rho}_a, \;\;
\vec{\pi}_{qa}\approx\vec{\pi}_a.
 \label{c7}
 \end{equation}

\hfill

Thanks to these results, we can calculate a function $F$
independent of $\vec{q}_+$ on the phase space, under the
constraint $\vec{\kappa}_+\approx 0$, by simply performing the
limit

\begin{equation}
F\mid_{\vec{\kappa}_+\approx 0}(\vec{\rho}_{qa},\vec{\pi}_{qa})
=\lim_{\alpha\rightarrow\infty}F(\alpha).
 \label{c8}
 \end{equation}

\medskip

This method is applied in Section IV for calculating the
multipoles after the gauge fixing $\vec{q}_+\approx 0,\;\;
\vec{\kappa}_+\approx 0$. These multipoles depend on
$\vec{\kappa}_i$, so that (see Ref.\cite{iten1})

\begin{equation}
\lim_{\alpha\rightarrow\infty}\vec{\kappa}_i(\alpha)
=\sqrt{N}\sum_{a=1}^{N-1}\gamma_{ai}\vec{\pi}_{qa},
 \label{c9}
 \end{equation}

\noindent and on $(\vec{\eta}_i-\vec{R}_+)$, so that

\begin{eqnarray}
(\vec{\eta}_i-\vec{R}_+)&=& \sum_j(\vec{\eta}_i-\vec{\eta}_j)
\frac{\sqrt{m_j^2+\vec{\kappa}_j^2}}{\sum_k\sqrt{m_k^2+\vec{\kappa}_k^2}}=\nonumber\\
&=&\sum_j\sum_a\sqrt{N}(\gamma_{ai}-\gamma_{aj})\vec{\rho}_a
\frac{\sqrt{m_j^2+\vec{\kappa}_j^2}}{\sum_k\sqrt{m_k^2+\vec{\kappa}_k^2}}.
 \label{c10}
 \end{eqnarray}

Then using the (\ref{c9}) and (\ref{c4}) we have

\begin{equation}
\lim_{\alpha\rightarrow\infty}(\vec{\eta}_i(\alpha)-\vec{R}_+(\alpha))=
\sum_j\sum_a\sqrt{N}(\gamma_{ai}-\gamma_{aj})\vec{\rho}_{qa}
\frac{\sqrt{m_j^2+N\sum_{ab}\vec{\pi}_{qa}\cdot\vec{\pi}_{qb}\gamma_{aj}\gamma_{bj}}}
{\sum_k\sqrt{m_k^2+N\sum_{ab}\vec{\pi}_{qa}\cdot\vec{\pi}_{qb}\gamma_{ak}\gamma_{bk}}}.
\label{c11}
\end{equation}

The following notation is used to denote the limits (\ref{c8})

\beq
 F\rightarrow_{\alpha\rightarrow\infty}
F\mid_{\vec{\kappa}_+\approx 0}(\vec{\rho}_{qa},\vec{\pi}_{qa}).
 \label{c12}
 \eeq

For example Eqs. (\ref{c9}) and (\ref{c11}) become

\begin{eqnarray}
\vec{\kappa}_i&\rightarrow_{\alpha\rightarrow\infty}&
\sqrt{N}\sum_{a=1}^{N-1}\gamma_{ai}\vec{\pi}_{qa},\nonumber \\
 (\vec{\eta}_i-\vec{R}_+)&\rightarrow_{\alpha\rightarrow\infty}&
\sum_j\sum_a\sqrt{N}(\gamma_{ai}-\gamma_{aj})\vec{\rho}_{qa}
\frac{\sqrt{m_j^2+N\sum_{ab}\vec{\pi}_{qa}\cdot\vec{\pi}_{qb}\gamma_{aj}\gamma_{bj}}}
{\sum_k\sqrt{m_k^2+N\sum_{ab}\vec{\pi}_{qa}\cdot\vec{\pi}_{qb}\gamma_{ak}\gamma_{bk}}}.
 \label{c13}
 \end{eqnarray}

\medskip

The closed form (\ref{II20}) - (\ref{II22}) of the canonical
transformation (\ref{c5}) was not given in Ref.\cite{iten1}, but
it can derived from the following two equations of that paper [its
Eq.(5.13) and (5.24)]

\begin{eqnarray}
{\vec \pi}_a(\alpha ) &=& {1\over {\sqrt{N}}} \sum_{i=1}^N
\gamma_{ai} {\vec \kappa}_i(\alpha ),\nonumber \\
 &&{}\nonumber \\
 {\vec \pi}_{qa}&{\buildrel {def}\over =}&{\vec \pi}_a(\infty )
={1\over {\sqrt{N}}} \sum_{i=1}^N {\vec \kappa}_i(\infty
)=\nonumber \\
 &=&{\vec \pi}_a+ {{{\vec n}_{+}}\over {{\cal M}}}
[( M_{sys}- {\cal M}) {\vec n} _{+}\cdot {\vec \pi}_a-|{\vec
\kappa}_{+}| H_a]=\nonumber \\
 &=& {\vec \pi}_a -{{{\vec
\kappa}_{+}}\over {\sqrt{ M_{sys}^2-{\vec \kappa}_{+}^2}}} [H_a-{{
M_{sys} -\sqrt{ M_{sys}^2-{\vec \kappa}^2_{+}}}\over {{\vec
\kappa}_{+}^2}} {\vec \kappa}_{+}\cdot {\vec \pi}_a] \approx {\vec
\pi}_a,\nonumber \\
 &&{}\nonumber \\
 H_a&=& {1\over {\sqrt{N}}} \sum_{i=1}^N \gamma_{ai} H_i =
 {1\over {\sqrt{N}}} \sum_{i=1}^N \gamma_{ai}\, \sqrt{m_i^2 + {\vec \kappa}_i^2},\nonumber \\
 &&{}\nonumber \\
 {\vec \kappa}_i(\infty )&=& \sqrt{N} \sum_{a=1}^{N-1} \gamma_{ai} {\vec \pi}_{qa},\nonumber \\
 H_{(rel) i} &=& H_i(\infty ) =  \sqrt{m_i^2+N
\sum_{ab}^{1..N-1}\gamma_{ai}\gamma_{bi}{\vec \pi}_{qa}\cdot {\vec
\pi}_{qb}}, \nonumber \\
 M_{sys}&=& \sum_{i=1}^NH_i =\sqrt{{\cal M}^2 +{\vec \kappa}_{+}^2}
 \approx  H_{(rel)} = H_M(\infty ) = {\cal M} =\nonumber \\
 &=& \sum_{i=1}^N H_i(\infty ) = \sum_{i=1}^N \sqrt{m_i^2+N
\sum_{ab}^{1..N-1}\gamma_{ai}\gamma_{bi}{\vec \pi}_{qa}\cdot {\vec
\pi}_{qb}}.
 \label{c14}
 \end{eqnarray}

\begin{eqnarray}
{\vec \rho}_{qa}&{\buildrel {def}\over =}& {\vec \rho}_a(\infty
)={\vec \rho}_a-\nonumber \\
 &-&\sum_{i,j=1}^N\sum_{b=1}^{N-1}
\gamma_{aj}(\gamma_{bi}-\gamma_{bj}) {{H_i}\over { M_{sys}}} \Big[
{{|{\vec \kappa}_{+}| {\vec \kappa}_j(\infty )}\over {H_j(\infty )
\sqrt{\Pi}}}+({{ M_{sys}}\over {\sqrt{\Pi}}}-1) {\vec n}_{+} \Big]
{\vec n}_{+}\cdot {\vec \rho}_b=\nonumber \\
 &=& {\vec \rho}_a- \sum_{i,j=1}^N\sum_{b=1}^{N-1}
\gamma_{aj}(\gamma_{bi}-\gamma_{bj}) {{H_i}\over { M_{sys}}}
 {{ {\vec \kappa}_j(\infty )}\over {H_j(\infty )
\sqrt{\Pi}}} {\vec \kappa}_{+}\cdot {\vec \rho}_b\, \approx {\vec
\rho}_a.
 \label{c15}
 \end{eqnarray}

\vfill\eject

\section{More on Dixon's Multipoles.}

In this Appendix we shall consider multipoles with respect to the
origin, i.e. with $\vec \eta =0$ [we use the notation
$t_T^{\mu_1...\mu_n\mu\nu}(T_s)\, {\buildrel {def}\over =}\,
t_T^{\mu_1...\mu_n\mu\nu}(T_s, 0)$] and we shall give some of
their properties following Ref.\cite{dixon}. The proofs of these
results are identical to those given in Ref.\cite{dixon} after the
following kinematical modifications of Eqs.(3.6) and (3.7) of that
paper:

i) let ${\cal W}$ be the world-tube containing the compact support
of the isolated system and $w^{\mu}(\tau ) = z^{\mu}(\tau ,\vec
\eta (\tau ))$ a time-like world-line inside it, with tangent
vector ${\dot w}^{\mu}(\tau ) = z^{\mu}_{\tau}(\tau ,\vec \eta
(\tau )) + z^{\mu}_r(\tau ,\vec \eta (\tau ))\, \eta^r(\tau )$,
used to evaluate the multipoles (see Section IV);

ii) the Wigner hyper-planes $\Sigma_{W\tau}$ of the rest-frame
instant form are in general {\it not orthogonal} to the world-line
(differently from the hyper-surfaces $\Sigma (s)$ of
Ref.\cite{dixon});

iii) Eq.(3.6) is replaced with $\int\, d^4z\, f(z) = \int d\tau\,
\int_{\Sigma_{\tau}} d^3\sigma\, \sqrt{g(\tau ,\vec \sigma )}\,
f(z(\tau ,\vec \sigma )) = \int d\tau\, \int_{\Sigma_{\tau}}\,
d^3\Sigma_{\mu}\, z^{\mu}_{\tau}(\tau ,\vec \sigma )\, f(z(\tau
,\vec \sigma ))$ (see before Eq.(\ref{II1}) for the notations);

iv) since $l_{\mu}(\tau ,\vec \sigma ) = [l_{\rho}\, z^{\rho}_A\,
z^A_{\mu}](\tau ,\vec \sigma ) = [\sqrt{{g\over {\gamma}}}\,
z^{\tau}_{\mu}](\tau ,\vec \sigma )$, we get
$\int_{\Sigma_{\tau}}\, d^3\Sigma_{\mu}\, f(z(\tau ,\vec \sigma ))
= \int_{\Sigma_{\tau}}\, d^3\sigma\, \sqrt{\gamma (\tau ,\vec
\sigma )}\, l_{\mu}(\tau ,\vec \sigma )\, f(z(\tau ,\vec \sigma ))
= \int_{\Sigma_{\tau}}\, d^3\sigma\, \sqrt{g(\tau ,\vec \sigma
)}\, z^{\tau}_{\mu}(\tau ,\vec \sigma )\, f(z(\tau ,\vec \sigma
))$, so that we have ${d\over {d\tau}}\, \int_{\Sigma_{\tau}}\,
d^3\Sigma_{\mu}\, f(z(\tau ,\vec \sigma )) =
\int_{\Sigma_{\tau}}\, d^3\sigma\, {{\partial}\over {\partial
\tau}}\, [\sqrt{g(\tau ,\vec \sigma )}\, z^{\tau}_{\mu}(\tau ,\vec
\sigma )\, f(z(\tau ,\vec \sigma ))] = \int_{\Sigma_{\tau}}\,
d^3\sigma\, \partial_A\, [\sqrt{g(\tau ,\vec \sigma )}\,
z^A_{\mu}(\tau ,\vec \sigma )\, f(z(\tau ,\vec \sigma ))]$ after
having added $\int_{\Sigma_{\tau}}\, d^3\sigma\, \partial_r\,
[\sqrt{g(\tau ,\vec \sigma )}\, z^r_{\mu}(\tau ,\vec \sigma )\,
f(z(\tau ,\vec \sigma ))] = 0$;

v) since ${{\partial}\over {\partial \sigma^A}}\, [\sqrt{g}\,
z^A_{\mu}](\tau ,\vec \sigma ) = {{\partial}\over {\partial
\sigma^A}}\, [{1\over {3!}}\, \epsilon^{ABCD}\,
\epsilon_{\mu\beta\gamma\delta}\, {{\partial z^{\beta}}\over
{\partial \sigma^B}}\, {{\partial z^{\gamma}}\over {\partial
\sigma^C}}\, {{\partial z^{\delta}}\over {\partial
\sigma^D}}](\tau ,\vec \sigma ) = 0$, we get that Eq.(3.7) is
replaced by ${d\over {d\tau}}\, \int_{\Sigma_{\tau}}\,
d^3\Sigma_{\mu}\, f(z(\tau ,\vec \sigma )) =
\int_{\Sigma_{\tau}}\, d^3\sigma\, \sqrt{g(\tau ,\vec \sigma )}\,
z^A_{\mu}(\tau ,\vec \sigma )\, {{\partial f(z(\tau ,\vec \sigma
))}\over {\partial \sigma^A}} = \int_{\Sigma_{\tau}}\, d^3\sigma\,
\sqrt{\gamma (\tau ,\vec \sigma )}\, l_{\nu}(\tau ,\vec \sigma )\,
z^{\nu}_{\tau}(\tau ,\vec \sigma )\, {{\partial f(z(\tau ,\vec
\sigma ))}\over {\partial z^{\mu}}} = \int_{\Sigma_{\tau}}\,
d^3\Sigma_{\nu}\, z^{\nu}_{\tau}(\tau ,\vec \sigma )\, {{\partial
f(z(\tau ,\vec \sigma ))}\over {\partial z^{\mu}}} $;

vi) as a consequence the results quoted in this Appendix hold if
the multipoles are taken with respect to world-lines $w^{\mu}(\tau
) = z^{\mu}(\tau ,\vec \eta )$ such that $\vec \eta (\tau ) = \vec
\eta = const.$ (they are the integral lines of the vector field
$z^{\mu}_{\tau}(\tau ,\vec \sigma )\, \partial_{\mu}$), so that
${\dot w}^{\mu}(\tau ) = z^{\mu}_{\tau}(\tau ,\vec \eta )$.

\bigskip

As shown in Ref.\cite{dixon}, if a field has a compact support W
on the Wigner hyper-planes $\Sigma_{W\tau}$ and if $f(x)$ is a
$C^{\infty}$ complex-valued scalar function on Minkowski
space-time with compact support \cite{c37}, we have
\medskip

\begin{eqnarray}
< T^{\mu\nu},f >&=& \int d^4x T^{\mu\nu}(x) f(x)=\nonumber \\
 &=&\int dT_s\int d^3\sigma f(x_s+\delta
 x_s)T^{\mu\nu}[x_s(T_s)+\delta x_s(\vec \sigma )][\phi ]=\nonumber \\
 &=&\int dT_s \int d^3\sigma \int {{d^4k}\over {(2\pi )^4}}
\tilde f(k) e^{-ik\cdot [x_s(T_s)+\delta x_s(\vec \sigma
 )]}T^{\mu\nu} [x_s(T_s)+\delta x_s(\vec \sigma )][\phi ]=\nonumber \\
 &=&\int dT_s\int {{d^4k}\over {(2\pi )^4}} \tilde f(k)
e^{-ik\cdot x_s(T_s)} \int d^3\sigma T^{\mu\nu}[x_s(T_s)+\delta
x_s(\vec \sigma )][\phi ]\nonumber \\
 &&\sum_{n=0}^{\infty}
{{(-i)^n}\over {n!}} [k_{\mu}\epsilon^{\mu}_u(u(p_s))
\sigma^u]^n=\nonumber \\
 &=&\int\, dT_s\int {{d^4k}\over {(2\pi
)^4}} \tilde f(k) e^{-ik\cdot x_s(T_s)} \sum_{n=0}^{\infty}
{{(-i)^n}\over {n!}} k_{\mu_1}...k_{\mu_n} t_T^{\mu_1...
\mu_n\mu\nu}(T_s),
 \label{d1}
\end{eqnarray}

\noindent If, in particular, $f(x)$ is analytic on W \cite{dixon}
\cite{c38}, we get

\begin{eqnarray}
< T^{\mu\nu},f >&=& \int dT_s \sum_{n=0}^{\infty}{1\over {n!}}
t_T^{\mu_1... \mu_n\mu\nu}(T_s) {{\partial^nf(x)}\over {\partial
x^{\mu_1}...\partial x^{\mu _n}}}{|}_{x=x_s(T_s)},\nonumber \\
&&\Downarrow \nonumber \\ T^{\mu\nu}(x)&=&
\sum_{n=0}^{\infty}{{(-1)^n}\over {n!}} {{\partial^n}\over
{\partial x^{\mu_1}...\partial x^{\mu_n}}} \int dT_s
\delta^4(x-x_s(T_s)) t_T^{\mu_1...\mu_n\mu\nu}(T_s).
 \label{d2}
\end{eqnarray}

\noindent For a N particle system this equation may be rewritten
as Eq.(\ref{IV4}). \bigskip

On the other hand, for non-analytic functions $f(x)$ we have

\begin{eqnarray}
< T^{\mu\nu},f > &=& \int dT_s \sum^N_{n=0} {1\over {n!}}
t_T^{\mu_1...\mu_n \mu\nu}(T_s) {{\partial^nf(x)}\over {\partial
x^{\mu_1}...\partial x^{\mu _n}}}{|}_{x=x_s(T_s)}+\nonumber \\
&+&\int dT_s \int {{d^4k}\over {(2\pi )^4}} \tilde f(k)
e^{-ik\cdot x_s(T_s)} \sum_{n=N+1}^{\infty} {{(-i)^n}\over {n!}}
k_{\mu_1}...k_{\mu_n} t_T^{\mu_1... \mu_n\mu\nu}(T_s),
 \label{d3}
\end{eqnarray}

\noindent and, as shown in Ref.\cite{dixon}, from the knowledge of
the moments $t_T^{\mu_1...\mu_n\mu}(T_s)$ for all $n > N$, we can
get $T^{\mu\nu}(x)$ and, therefore, all the moments with $n\leq
N$.\bigskip

In the case of N free particles and more in general for isolated
systems, the Hamilton equations \cite{c39} for the multipoles
(\ref{IV3}), with $p_T^{\mu_1...\mu_n\mu}(T_s)\, {\buildrel
{def}\over =}\, p_T^{\mu_1...\mu_n\mu}(T_s, \vec 0)$, imply
\medskip

\begin{eqnarray}
{{dp_T^{\mu}(T_s)}\over {dT_s}}\, &{\buildrel \circ \over =}\,&
0,\quad for\, n=0,\nonumber \\ {{d
p_T^{\mu_1...\mu_n\mu}(T_s)}\over {dT_s}}\, &{\buildrel \circ
\over =}\,& -nu^{(\mu_1}(p_s) p_T^{\mu_2...\mu_n)\mu}(T_s)+n
t_T^{(\mu_1...\mu_n)\mu}(T_s), \quad n\geq 1.
 \label{d4}
\end{eqnarray}

\bigskip

If we define for $n \geq 1$
\medskip

\begin{eqnarray}
b_T^{\mu_1...\mu_n\mu}(T_s)&{\buildrel {def}\over
=}&p_T^{(\mu_1...\mu_n\mu )}(T_s)=\nonumber \\
 &=& \epsilon^{(\mu_1}_{r_1}(u(p_s))....
\epsilon^{\mu_n}_{r_n}(u(p_s))\epsilon^{\mu )}_A(u(p_s))
\,\,q_T^{r_1..r_nA\tau}(T_s) ,\nonumber \\
 &&{}\nonumber \\
c_T^{\mu_1...\mu_n\mu}(T_s)&{\buildrel {def}\over
=}&c_T^{(\mu_1...\mu_n )\mu}(T_s)=p_T^{\mu_1...
\mu_n\mu}(T_s)-p_T^{(\mu_1...\mu_n\mu )}(T_s)=\nonumber \\
&=&[\epsilon^{\mu_1}_{r_1}(u(p_s))...\epsilon^{\mu_n}_{r_n}\epsilon^{\mu}_A(u(p_s))-
    \nonumber \\
 &-&\epsilon^{(\mu_1}_{r_1}(u(p_s))...\epsilon^{\mu_n}_{r_n}(u(p_s))
\epsilon_A^{\mu )}(u(p_s))] q_T^{r_1..r_nA\tau}(T_s),\nonumber \\
 &&{}\nonumber \\
&&c_T^{(\mu_1...\mu_n\mu )}(T_s)=0,\qquad S^{\mu\nu}_T=2
p_T^{[\mu\nu ]}=2 c_T^{\mu\nu}, \nonumber \\
 &&{}\nonumber \\
 &&{}\nonumber \\
 \epsilon^{r_1}_{\mu_1}(u(p_s))....\epsilon^{r_n}_{\mu_n}(u(p_s)) b_T^{\mu_1...\mu_n\mu}(T_s)
 &=&{1\over {n+1}} u^{\mu}(p_s) q_T^{r_1...r_n\tau\tau}(T_s)+\epsilon^{\mu}_r(u(p_s))
 q_T^{(r_1...r_nr)\tau}(T_s),\nonumber \\
 &&{}\nonumber \\
 \epsilon^{r_1}_{\mu_1}(u(p_s))....\epsilon^{r_n}_{\mu_n}(u(p_s)) c_T^{\mu_1...\mu_n\mu}(T_s)
 &=& {n\over {n+1}}u^{\mu}(p_s) q_T^{r_1...r_n\tau\tau}(T_s) +\nonumber \\
 &+&\epsilon^{\mu}_r(u(p_s)) [ q_T^{r_1...r_nr\tau}(T_s) - q_T^{(r_1...r_nr)\tau}(T_s)],
 \label{d5}
\end{eqnarray}

\noindent and  for $n\geq 2$
\medskip

\begin{eqnarray}
d_T^{\mu_1...\mu_n\mu\nu}(T_s)&=&d_T^{(\mu_1...\mu_n)(\mu\nu
)}(T_s)\, {\buildrel {def}\over =}\,
 t_T^{\mu_1...\mu_n\mu\nu}(T_s)-\nonumber \\
 &-&{{n+1}\over
n}[t_T^{(\mu_1...\mu_n\mu )\nu}(T_s)+t_T^{(\mu_1...\mu_n\nu )\mu}
(T_s)]+\nonumber \\
 &+&{{n+2}\over n}t_T^{(\mu_1...\mu_n\mu\nu
)}(T_s)=\nonumber
\\ &=&\Big[ \epsilon^{\mu_1}_{r_1} ... \epsilon^{\mu_n}_{r_n}
\epsilon^{\mu}_A \epsilon^{\nu}_B-
 {{n+1}\over n}\Big( \epsilon^{(\mu_1}_{r_1} ... \epsilon^{\mu_n}_{r_n} \epsilon^{\mu )}_A
 \epsilon^{\nu}_B+\nonumber \\
 &+&\epsilon^{(\mu_1}_{r_1} ... \epsilon^{\mu_n}_{r_n} \epsilon^{\nu )}_B
 \epsilon^{\mu}_A\Big) +{{n+2}\over n} \epsilon^{(\mu_1}_{r_1} .. \epsilon^{\mu_n}_{r_n}
 \epsilon^{\mu}_A \epsilon^{\nu )}_B\Big] (u(p_s))\nonumber \\
 && q_T^{r_1..r_nAB}(T_s),\nonumber \\
 &&{}\nonumber \\
 &&d_T^{(\mu_1...\mu_n\mu )\nu}(T_s)=0,\nonumber \\
 &&{}\nonumber \\
 \epsilon^{r_1}_{\mu_1}(u(p_s))....\epsilon^{r_n}_{\mu_n}(u(p_s)) d_T^{\mu_1...\mu_n\mu\nu}(T_s)
 &=& {{n-1}\over {n+1}} u^{\mu}(p_s)u^{\nu}(p_s) q_T^{r_1...r_n\tau\tau}(T_s)+\nonumber \\
 &+&{1\over n} [u^{\mu}(p_s)\epsilon^{\nu}_r(u(p_s))+u^{\nu}(p_s)\epsilon^{\mu}_r(u(p_s))]
   \nonumber \\
 &&[(n-1) q_T^{r_1...r_nr\tau}(T_s)+ q_T^{(r_1...r_nr)\tau}(T_s)]+\nonumber \\
 &+&\epsilon^{\mu}_{s_1}(u(p_s))\epsilon^{\nu}_{s_2}(u(p_s)) [q_T^{r_1...r_ns_1s_2}(T_s)
 -\nonumber \\
  &-&{{n+1}\over n}( q_T^{(r_1...r_ns_1)s_2}(T_s)+ q_T^{(r_1...r_ns_2)s_1}(T_s)) +\nonumber \\
 &+& q_T^{(r_1...r_ns_1s_2)}(T_s)],
 \label{d6}
\end{eqnarray}

\medskip

\noindent then Eqs.(\ref{d4}) may be rewritten in the form\medskip

\begin{eqnarray}
&&1)\quad n=1\nonumber \\
 &&{}\nonumber \\
t^{\mu\nu}_T(T_s)&=&t^{(\mu\nu )}_T(T_s)\, {\buildrel \circ \over
=}\, p_T^{\mu}(T_s)u^{\nu }(p_s)+{1\over 2}{d\over
{dT_s}}(S^{\mu\nu}_T(T_s)+2b_T ^{\mu\nu}(T_s)),\nonumber \\
 &&\Downarrow \nonumber \\
 t^{\mu\nu}_T(T_s)\, &{\buildrel \circ \over
=}\,&p_T^{(\mu}(T_s)u^{\nu )}(p_s) +{d\over
{dT_s}}b_T^{\mu\nu}(T_s)=M u^{\mu}(p_s)u^{\nu}(p_s)+\nonumber \\
 &+&\kappa^r_{+} [u^{(\mu}(p_s)\epsilon^{\nu )}_r(u(p_s))+
\epsilon^{(\mu}_r(u(p_s))u^{\nu )}(p_s)]+ \nonumber \\
 &&\epsilon^{(\mu}_r(u(p_s))\epsilon^{\nu )}_s(u(p_s))\sum_{i=1}^N
 {{\kappa^u_i \kappa^v_i}\over {\sqrt{m_i^2+{\vec \kappa}_i^2}}},\nonumber \\
 {d\over {dT_s}}S^{\mu\nu}_T(T_s)\, &{\buildrel \circ \over
=}\,&2p_T^{[\mu}(T_s) u^{\nu
]}(p_s)=2\kappa^r_{+}\epsilon^{[\mu}_r(u(p_s))u^{\nu ]}(p_s)
 \approx 0,\nonumber \\
 &&{}\nonumber \\
 &&{}\nonumber \\
 &&2)\quad n=2\quad [identity\,\,
t_T^{\rho\mu\nu}=t_T^{(\rho\mu )\nu}+t_T^{(\rho \nu
)\mu}+t_T^{(\mu\nu )\rho}]\nonumber \\
 &&{}\nonumber \\
 2t_T^{(\rho\mu )\nu}(T_s)\, &{\buildrel \circ \over =}\,&
2u^{(\rho}(p_s)b_T ^{\mu )\nu}(T_s)+u^{(\rho}(p_s)S_T^{\mu
)\nu}(T_s)+{d\over {dT_s}}(b_T^{\rho\mu\nu}
(T_s)+c_T^{\rho\mu\nu}(T_s)),\nonumber \\
 &&\Downarrow \nonumber \\
 t_T^{\rho\mu\nu}(T_s)\, &{\buildrel \circ \over
=}\,&u^{\rho}(p_s)b_T ^{\mu\nu}(T_s)+S_T^{\rho (\mu}(T_s)u^{\nu
)}(p_s)+{d\over {dT_s}}({1\over 2}b_T
^{\rho\mu\nu}(T_s)-c_T^{\rho\mu\nu}(T_s)),\nonumber \\
 &&{}\nonumber \\
 &&{}\nonumber \\
  &&3) \quad n \geq 3 \nonumber \\
   &&{}\nonumber \\
t_T^{\mu_1...\mu_n\mu\nu}(T_s)\, &{\buildrel \circ \over =}\,&
d_T^{\mu_1...
\mu_n\mu\nu}(T_s)+u^{(\mu_1}(p_s)b_T^{\mu_2...\mu_n)\mu\nu}(T_s)+2u^{(\mu
_1}(p_s)c_T^{\mu_2...\mu_n)(\mu\nu )}(T_s)+\nonumber \\
 &=&{2\over
n}c_T^{\mu_1...\mu_n(\mu}(T_s)u^{\nu )}(p_s)+{d\over {dT_s}}
[{1\over {n+1}}b_T^{\mu_1...\mu_n\mu\nu}(T_s)+{2\over
n}c_T^{\mu_1...\mu _n(\mu\nu )}(T_s)].
 \label{d7}
\end{eqnarray}

\bigskip

This allows  to rewrite $< T^{\mu\nu},f >$ in the form\cite{dixon}
\medskip

\begin{eqnarray}
< T^{\mu\nu},f > &=&\int dT_s \int {{d^4k}\over {(2\pi )^4}}
\tilde f(k) e^{-ik\cdot x_s(T_s)} \Big[ u^{(\mu}(p_s)p_T^{\nu
)}(T_s)-ik_{\rho}S^{\rho (\mu}_T(T_s) u^{\nu )}(p_s)+\nonumber \\
&+&\sum_{n=2}^{\infty}{{(-i)^n}\over {n!}} k_{\rho_1}...k_{\rho_n}
I_T^{\rho_1 ...\rho_n\mu\nu}(T_s)\Big],
 \label{d8}
\end{eqnarray}
\medskip

\noindent with

\begin{eqnarray}
I_T^{\mu_1...\mu_n\mu\nu}(T_s)&=&I_T^{(\mu_1...\mu_n )(\mu\nu
)}(T_s)\, {\buildrel {def}\over =}\,
d_T^{\mu_1...\mu_n\mu\nu}(T_s)-\nonumber
\\
 &-&{2\over {n-1}}u^{(\mu_1}(p_s)c_T^{\mu_2...\mu_n )(\mu\nu )}(T_s)+\nonumber \\
 &+&{2\over n}
c_T^{\mu_1...\mu_n(\mu}(T_s)u^{\nu )}(p_s)=\nonumber \\
  &=&\Big[
\epsilon^{\mu_1}_{r_1}...\epsilon^{\mu_n}_{r_n} \epsilon^{\mu}_A
\epsilon^{\nu}_B-
 {{n+1}\over n}\Big( \epsilon^{(\mu_1}_{r_1}...\epsilon^{\mu_n}_{r_n} \epsilon^{\mu )}_A
 \epsilon^{\nu}_B+\nonumber \\
 &+&\epsilon^{(\mu_1}_{r_1}...\epsilon^{\mu_n}_{r_n} \epsilon^{\nu )}_B \epsilon^{\mu}_A\Big) +
 {{n+2}\over n} \epsilon^{(\mu_1}_{r_1}...\epsilon^{\mu_n}_{r_n}
 \epsilon^{\mu}_A \epsilon^{\nu )}_B\Big] (u(p_s))\nonumber \\
 && q_T^{r_1..r_nAB}(T_s)-\nonumber \\
 &-&\Big[ {2\over {n-1}} u^{(\mu_1}(p_s) \Big( \epsilon^{\mu_2}_{r_1}...
 \epsilon^{\mu_n)}_{r_{n-1}} \epsilon^{(\mu}_{r_n} \epsilon^{\nu )}_A-
 \epsilon^{(\mu_2}_{r_1}...\epsilon^{\mu_n)}_{r_{n-1}} \epsilon^{(\mu}_{r_n}
 \epsilon^{\nu ))}_A\Big)-\nonumber \\
 &-&{2\over n} \Big( \epsilon^{\mu_1}_{r_1}...\epsilon^{\mu_n}_{r_n} \epsilon^{(\mu}_A-
 \epsilon^{(\mu_1}_{r_1}... \epsilon^{\mu_n}_{r_n} \epsilon^{(\mu )}_A u^{\nu )}(p_s)
 \Big] (u(p_s))\nonumber \\
 && q_T^{r_1..r_nA\tau}(T_s),\nonumber \\
 &&{}\nonumber \\
 &&I_T^{(\mu_1...\mu_n\mu )\nu}(T_s)=0,\nonumber \\
 &&{}\nonumber \\
 \epsilon^{r_1}_{\mu_1}(u(p_s))....\epsilon^{r_n}_{\mu_n}(u(p_s)) I_T^{\mu_1...\mu_n\mu\nu}(T_s)
 &=& {{n+3}\over {n+1}} u^{\mu}(p_s)u^{\nu}(p_s) q_T^{r_1...r_n\tau\tau}(T_s)+\nonumber \\
 &+&{1\over n}[u^{\mu}(p_s)\epsilon^{\nu}_r(u(p_s))+u^{\nu}(p_s)\epsilon^{\mu}_r(u(p_s))]
 q_T^{r_1...r_nr\tau}(T_s)+\nonumber \\
 &+&\epsilon^{\mu}_{s_1}(u(p_s))\epsilon^{\nu}_{s_2}(u(p_s)) [q_T^{r_1...r_ns_1s_2}(T_s)
 -\nonumber \\
 &-&{{n+1}\over n}( q_T^{(r_1...r_ns_1)s_2}(T_s)+ q_T^{(r_1...r_ns_2)s_1}(T_s)) +
    \nonumber \\
 &+&q_T^{(r_1...r_ns_1s_2)}(T_s)].
 \label{d9}
\end{eqnarray}

For a N particle system, Eq.(\ref{d8})implies Eq.(\ref{IV15}).

\bigskip

Finally, a set of multipoles equivalent to the
$I_T^{\mu_1..\mu_n\mu\nu}$ is \cite{c40}:

\begin{eqnarray}
&&for\quad n \geq 0\nonumber \\
 &&{}\nonumber \\
J_T^{\mu_1...\mu_n\mu\nu\rho\sigma}(T_s)&=&J_T^{(\mu_1...\mu_n)[\mu\nu
][\rho \sigma ]}(T_s)\, {\buildrel {def}\over =}\,
I_T^{\mu_1...\mu_n[\mu [\rho\nu ]\sigma ]}(T_s)=\nonumber \\
 &&{}\nonumber \\
 &=&t_T^{\mu_1...\mu_n[\mu [\rho\nu ]\sigma ]}(T_s)-\nonumber \\
 &-&{1\over {n+1}}\Big[ u^{[\mu}(p_s)p_T^{\nu ]\mu_1...\mu_n[\rho\sigma
]}(T_s)+\nonumber \\
 &+&u^{[\rho}(p_s)p_T^{\sigma ]\mu_1...\mu_n[\mu\nu ]}(T_s)\Big]=\nonumber\\
 &&{}\nonumber \\
&=&\Big[ \epsilon^{\mu_1}_{r_1} .. \epsilon^{\mu_n}_{r_n}
\epsilon^{[\mu}_r
 \epsilon^{[\rho}_s \epsilon^{\nu ]}_A \epsilon^{\sigma ]}_B\Big] (u(p_s))
 q_T^{r_1..r_nAB}(T_s)-\nonumber \\
  &-&{1\over {n+1}}\Big[ u^{[\mu}(p_s) \epsilon^{\nu ]}_r(u(p_s)) \epsilon^{[\rho}_s(u(p_s))
  \epsilon^{\sigma ]}_A(u(p_s))+\nonumber \\
   &+&u^{[\rho}(p_s) \epsilon^{\sigma ]}_r(u(p_s))
  \epsilon^{[\mu}_s(u(p_s)) \epsilon^{\nu ]}_A(u(p_s))\Big]\nonumber \\
  &&\epsilon^{\mu_1}_{r_1}(u(p_s))...\epsilon^{\mu_n}_{r_n}(u(p_s))
  q_T^{rr_1..r_nsA\tau}(T_s),\nonumber \\
 &&{}\nonumber \\
 (n+4)(3n+5) && linearly\, independent\, components,\nonumber \\
&&{}\nonumber \\
u_{\mu_1}(p_s)J_T^{\mu_1...\mu_n\mu\nu\rho\sigma}(T_s)&=&
J_T^{\mu_1...\mu_{n-1}(\mu_n\mu\nu )\rho\sigma}(T_s)=0,\quad\quad
for\, n \geq 1,\nonumber \\
 &&{}\nonumber \\
 I_T^{\mu_1...\mu_n\mu\nu}(T_s)&=&{{4(n-1)}\over {n+1}}
J_T^{(\mu_1...\mu_{n-1}| \mu |\mu_n)\nu}(T_s),\quad\quad for\, n
\geq 2,\nonumber \\
 &&{}\nonumber \\
  \epsilon^{r_1}_{\mu_1}(u(p_s))....\epsilon^{r_n}_{\mu_n}(u(p_s))
  J_T^{\mu_1...\mu_n\mu\nu\rho\sigma}(T_s) &=& \Big[ \epsilon^{[\mu}_r
 \epsilon^{[\rho}_s \epsilon^{\nu ]}_A \epsilon^{\sigma ]}_B\Big] (u(p_s))
 q_T^{r_1..r_nAB}(T_s)-\nonumber \\
  &-&{1\over {n+1}}\Big[ u^{[\mu}(p_s) \epsilon^{\nu ]}_r(u(p_s)) \epsilon^{[\rho}_s(u(p_s))
  \epsilon^{\sigma ]}_A(u(p_s))+\nonumber \\
   &+&u^{[\rho}(p_s) \epsilon^{\sigma ]}_r(u(p_s))
  \epsilon^{[\mu}_s(u(p_s)) \epsilon^{\nu ]}_A(u(p_s))\Big]
  q_T^{rr_1..r_nsA\tau}(T_s).\nonumber \\
  &&{}
 \label{d10}
\end{eqnarray}

The $J_T^{\mu_1..\mu_n\mu\nu\rho\sigma}$ are the Dixon {\it
$2^{n+2}$-pole inertial moment tensors} of the extended system:
they (or equivalently the $I_T^{\mu_1...\mu_n\mu\nu}$'s) determine
the energy-momentum tensor together with the {\it monopole}
$P^{\mu}_T$ and the {\it spin dipole} $S^{\mu\nu}_T$.
\medskip

As shown in Section 5 of Ref.\cite{dixon}, the equations of motion
$\partial_{\mu} T^{\mu\nu}\, {\buildrel \circ \over =}\, 0$ do not
imply equations of motion for the multipoles
$I_T^{\mu_1...\mu_n\mu\nu}$ ($n \geq 2$) or
$J_T^{\mu_1...\mu_n\mu\nu\rho\sigma}$ ($n \geq 0$), but only
Eqs.(\ref{IV16}) for the multipoles $P^{\mu}_T$ and $S^{\mu\nu}_T$
\cite{c41}. Instead the multipoles $t_T^{\mu_1...\mu_n\mu\nu}$ and
$p^{\mu_1...\mu_n\mu}_T$ have non trivial equations of motion.

\medskip

When all the multipoles $J_T^{\mu_1..\mu_n\mu\nu\rho\sigma}$ are
zero (or negligible) one speaks of a {\it pole-dipole} system.
\bigskip

On the Wigner hyperplane, the content  of these {\it
$2^{n+2}$-pole inertial moment tensors} is replaced by the {\it
Euclidean Cartesian tensors} $q_T^{r_1...r_n\tau\tau}$,
$q_T^{r_1...r_nr\tau}$, $q_T^{r_1...r_nrs}$. As shown in Appendix
B, we can decompose these Cartesian tensors in their irreducible
STF (symmetric trace-free) parts (the {\it STF tensors}).

\bigskip

Thus the {\it multipolar expansion} (\ref{IV4}) may be rewritten
as

\bea &&T^{\mu\nu}[ x_s^{({\vec q}_{+})
\beta}(T_s)+\epsilon^{\beta}_r(u(p_s))\sigma^r ]= T^{\mu\nu}[
w^{\beta}(T_s)+\epsilon^{\beta}_r(u(p_s))
(\sigma^r-\eta^r(T_s))]=\nonumber \\
 &&{}\nonumber \\
 &=& u^{(\mu}(p_s) \epsilon^{\nu )}_A(u(p_s)) [\delta^A_{\tau}
  M+\delta^A_u \kappa^u_{+}] \delta^3(\vec \sigma -\vec \eta
  (T_s))+\nonumber \\
 &&{}\nonumber \\
  &+&{1\over 2} S_T^{\rho (\mu}(T_s)[\vec \eta ] u^{\nu )}(p_s)
   \epsilon^r_{\rho}(u(p_s)) {{\partial}\over {\partial \sigma^r}}
   \delta^3(\vec \sigma -\vec \eta (T_s))+\nonumber \\
 &&{}\nonumber \\
   &+&\sum_{n=2}^{\infty} {{(-1)^n}\over {n!}} \Big[
 {{n+3}\over {n+1}} u^{\mu}(p_s)u^{\nu}(p_s) q_T^{r_1...r_n\tau\tau}(T_s,\vec \eta )+\nonumber \\
 &+&{1\over n}[u^{\mu}(p_s)\epsilon^{\nu}_r(u(p_s))+u^{\nu}(p_s)\epsilon^{\mu}_r(u(p_s))]
 q_T^{r_1...r_nr\tau}(T_s,\vec \eta )+\nonumber \\
 &+&\epsilon^{\mu}_{s_1}(u(p_s))\epsilon^{\nu}_{s_2}(u(p_s))
 [q_T^{r_1...r_ns_1s_2}(T_s,\vec \eta )-\nonumber \\
 &-&{{n+1}\over n}( q_T^{(r_1...r_ns_1)s_2}(T_s,\vec \eta )+
 q_T^{(r_1...r_ns_2)s_1}(T_s,\vec \eta )) +
 q_T^{(r_1...r_ns_1s_2)}(T_s, \vec \eta )]\Big] \nonumber \\
 &&{{\partial^n}\over {\partial \sigma^{r_1}..\partial \sigma^{r_n}}}
   \delta^3(\vec \sigma -\vec \eta (T_s)),
 \label{d11}
  \eea

\noindent leading to Eq.(\ref{IV15}).

\bigskip

For open systems, subsystems of an isolated system like in Section
V, we have $\partial_{\nu}\, T^{\mu\nu}\, \cir\, F^{\mu} \not= 0$,
with $F^{\mu}$ an external force. As shown in Ref.\cite{dixon} for
the case in which $F^{\mu} = - F^{\mu\nu}\, J_{\nu}$
($\partial_{\mu}\, J^{\mu}\, \cir 0$) is the Lorentz force, in
this case the multipolar expansion (\ref{IV15}) is still valid,
while the equations of motion (\ref{IV16}) become ($P^{\mu}$ and
$T_s$ are the conserved 4-momentum and the rest-frame time of the
global isolated system)

\bea
 {{d P^{\mu}_c(T_s)}\over {dT_s}} &\cir& \int d^3\sigma\,
 F^{\mu}(T_s, \vec \sigma ),\nonumber \\
 {{d S^{\mu\nu}_c(T_s)[\vec \eta = 0]}\over {dT_s}} &\cir& 2\,
 p_c^{[\mu}(T_s)\, u^{\nu ]}(P) - \int d^3\sigma\,
 \sigma^r\, [\epsilon^{\mu}_r(u(P))\, F^{\nu}(T_s, \vec \sigma ) -
 \epsilon^{\nu}_r(u(P))\, F^{\mu}(T_s, \vec \sigma )].
 \label{d12}
 \eea

\vfill\eject


\begin{thebibliography}{}
\bibitem{c1}Such hyper-surfaces are the leaves
of a foliation of Minkowski space-time, associated with one among
its 3+1 splittings, and are equivalent to a congruence of
time-like accelerated observers.
\bibitem{c2}${\vec z}_s/\epsilon_s$ is
the classical analogue of the Newton-Wigner 3-position operator,
only covariant under the Euclidean subgroup of the Poincar\'e
group.
\bibitem{c6}Their generators are the center-of-mass angular
momentum Noether constants of motion.
\bibitem{c7}Their generators are not constants of motion.
\bibitem{c9}On it we use
the notation $A=(\tau ,r)$. The 3-vectors $\vec B=\{ b^r\}$ on the
Wigner hyper-planes are Wigner spin-1 3-vectors.
$\epsilon^{\mu}_r(u(p_s))=L^{\mu}{}_r(p_s, {\buildrel \circ \over
p}_s)$ and
$\epsilon^{\mu}_{\tau}(u(p_s))=u^{\mu}(p_s)=L^{\mu}{}_o(p_s,{\buildrel
\circ \over p}_s)$ are the columns of the standard Wigner boost
for time-like Poincar\'e orbits. See Appendix B of
Ref.\cite{iten1}.
\bibitem{c11}See Refs.\cite{israel,ehlers} for the definition of this
concept in general relativity. By using the interpretation of
Ref.\cite{israel}, also the special relativistic limit of the
general relativistic Dixon centroid of Ref.\cite{dixon1} gives the
centroid $x^{\mu}_s(\tau )$: it coincides with  the special
relativistic Dixon centroid of Ref.\cite{dixon} defined by using
the conserved energy momentum tensor, as we shall see  in Section
IV.
\bibitem{c12}It describes a point living on the Wigner
hyper-planes and has the covariance of the little group O(3) of
time-like Poincar\'e orbits, like the Newton-Wigner position
operator.
\bibitem{c14}It implies $\lambda (\tau )=-1$ and
identifies the time parameter $\tau$ with the Lorentz scalar time
of the center of mass in the rest frame, $T_s=p_s\cdot {\tilde
x}_s/M_{sys}$; $M_{sys}$  generates the evolution in this time.
\bibitem{c16}As in every instant
form of dynamics, there are four independent Hamiltonians $p^o_s$
and $J^{oi}_s$, functions of the invariant mass $M_{sys}$; we give
also the expression in the basis $T_s$, $\epsilon_s$, ${\vec
z}_s$, ${\vec k}_s$.
\bibitem{c17}This {\it internal Poincar\'e algebra realization} should not
be confused with the previous {\it external} one based on ${\tilde
S}^{\mu\nu}_s$; $\Pi$ and $W^2$ are the two non-fixed invariants
of this realization.
\bibitem{c19}As we shall see in the next
Section, $\vec K\approx 0$ is implied by the natural gauge fixing
${\vec q}_{+}\approx 0$.
\bibitem{c22}In this way we are sure to have imposed the
restriction to positive energy particles and  to have excluded the
other $2^N-1$ branches of the total mass spectrum.
\bibitem{c24}Previous studies of multipole theories of
point particles can be found in
Refs.\cite{mul4,mul8,mul2,mul10,mul9,mul11,mul12,mul13}.
\bibitem{c27}They are $\vec \eta$
independent; see Appendix C of Ref.\cite{iten1} and Appendix A for
the non-relativistic limit.
\bibitem{c34}So that the origin is also the Fokker-Pryce center
of inertia and Pirani and Tulczyjew centroids of the system.
\bibitem{c35}This can be taken as the definitory equation for the velocity field,
even if strictly speaking we do not need it in what follows.
\bibitem{c36}It is a limiting concept deriving from the characteristic
function of a manifold.
\bibitem{c37}So that its Fourier
transform $\tilde f(k)=\int d^4x f(x) e^{ik\cdot x}$ is a slowly
increasing entire analytic function on Minkowski space-time
($|(x^o+iy^o) ^{q_o}...(x^3+iy^3)^{q_3}f(x^{\mu}+iy^{\mu})| <
C_{q_o...q_3} e^{a_o|y^o|+...+ a_3|y^3|}$, $a_{\mu} > 0$,
$q_{\mu}$ positive integers for every $\mu$ and $C_{q_o...q_3} >
0$), whose inverse is $f(x)=\int {{d^4k}\over {(2\pi )^4}} \tilde
f(k)e^{-ik\cdot x}$.
\bibitem{c39}In
Ref.\cite{dixon} this is a consequence of
$\partial_{\mu}T^{\mu\nu}\, {\buildrel \circ \over =}\, 0$.
\end{thebibliography}
\end{document}